\newif\ifmainmode         \mainmodetrue
\newif\ifsuppmode         \suppmodetrue
\newif\ifcombinedmode \combinedmodetrue

\documentclass[
reprint,
aps,
amsmath,
amssymb,
prx,
twocolumn,
superscriptaddress,
longbibliography
]{revtex4-2}

\usepackage[T1]{fontenc}
\usepackage{newtxtext}
\usepackage{newtxmath}
\usepackage{bm}

\usepackage{enumitem}

\usepackage{physics}
\usepackage{amsmath}
\usepackage{mathtools}
\usepackage{dsfont}

\usepackage{appendix}

\usepackage{placeins}

\usepackage{color}
\usepackage[dvipsnames]{xcolor}

\definecolor{nqdcolor}{rgb}{0.5586, 0.0586, 0.4219}
\newcommand*{\nqdcolor}{\color{nqdcolor}}

\usepackage{hyperref}
\hypersetup{colorlinks,citecolor=nqdcolor,linkcolor=nqdcolor,urlcolor=nqdcolor}

\usepackage{makecell}
\usepackage{colortbl} 

\usepackage{graphicx}
\usepackage[all]{hypcap}

\usepackage{comment}

\newcommand{\nocontentsline}[3]{}
\let\origcontentsline\addcontentsline
\newcommand\stoptoc{\let\addcontentsline\nocontentsline}
\newcommand\resumetoc{\let\addcontentsline\origcontentsline}

\makeatletter
\@ifundefined{ifmainmode}{\newif\ifmainmode\mainmodetrue}{}%
\@ifundefined{ifsuppmode}{\newif\ifsuppmode\suppmodetrue}{}%
\@ifundefined{ifcombinedmode}{\newif\ifcombinedmode\combinedmodetrue}{}%
\makeatother

\begin{document}

\author{Yu-Bo Shi}
\affiliation{Max Planck Institute for the Physics of Complex Systems, N\"othnitzer Str.~38, 01187 Dresden, Germany}
\affiliation{School of Physics, Nankai University, Tianjin 300071, China}

\author{Roderich Moessner}
\affiliation{Max Planck Institute for the Physics of Complex Systems and Cluster of Excellence ctd.qmat, N\"othnitzer Str.~38, 01187 Dresden, Germany}

\author{Ricard Alert}
\thanks{Equal contribution, ralert@pks.mpg.de, mgbukov@pks.mpg.de}
\affiliation{Max Planck Institute for the Physics of Complex Systems, N\"othnitzer Str.~38, 01187 Dresden, Germany}
\affiliation{Center for Systems Biology Dresden, Pfotenhauer Str.~108, 01307 Dresden, Germany}
\affiliation{Cluster of Excellence Physics of Life, TU Dresden, 01062 Dresden, Germany}
\affiliation{Departament de F\'{i}sica de la Mat\`{e}ria Condensada, Universitat de Barcelona, Barcelona, Spain}
\affiliation{Universitat de Barcelona Institute of Complex Systems (UBICS), Barcelona, Spain}
\affiliation{Instituci\'{o} Catalana de Recerca i Estudis Avan\c{c}ats (ICREA), Barcelona, Spain}

\author{Marin Bukov}
\thanks{Equal contribution, ralert@pks.mpg.de, mgbukov@pks.mpg.de}
\affiliation{Max Planck Institute for the Physics of Complex Systems, N\"othnitzer Str.~38, 01187 Dresden, Germany}

\clearpage
\title{\nqdcolor{Hamiltonian description of nonreciprocal interactions}} 

\ifmainmode

\begin{abstract}
In many systems, including sedimenting particles and bird flocks, interactions do not derive from a potential and are generally non-reciprocal, meaning that they do not obey the action–reaction principle. 
As a result, one cannot define a conventional energy function or use analytical and numerical tools that rely on it. 
Here we address this limitation by constructing a Hamiltonian with auxiliary degrees of freedom that, under a constraint, generates the original non-reciprocal dynamics. 
We show that Monte Carlo simulations based on the constrained Hamiltonian reproduce both stationary and non-stationary states of the original Langevin dynamics, as we illustrate for dissipative XY spins with vision-cone interactions.
The symplectic structure inherent to the construction also lets us apply established ideas from Hamiltonian engineering, which we demonstrate by varying the amplitude of a periodic (Floquet) drive to tune the spin interactions between square- and chain-lattice geometries. 
Overall, our construction paves the way towards extending statistical mechanics and Hamiltonian dynamics to non-reciprocal systems.
\end{abstract}

\maketitle
	
\stoptoc

\section{\label{sec:intro}Introduction}

Newton's third law states that every action has a reaction equal in magnitude and opposite in direction. For two particles interacting via a potential, the interaction forces are reciprocal: $\mathbf{F}_{2\to 1} = -\mathbf{F}_{1\to 2}$. However, for many mesoscopic and macroscopic particles, interactions do not arise from a potential but are instead mediated by a nonequilibrium field~\cite{bowick2022symmetry,fruchart2026nonreciprocal}. For example, interactions between colloids can be mediated by hydrodynamic flows, by a dissolved chemical, or by a temperature field~\cite{dzubiella2003depletion,golestanian2012collective,soto2014self,stark2018artificial,liebchen2019interactions,agudo2019active}. Respectively, interactions between animals, such as birds, wildebeest or humans, can be mediated by their vision field~\cite{romanczuk2009collective,barberis2016large,durve2018active}. In such systems, interactions are in general nonreciprocal: $\mathbf{F}_{2\to 1}\neq - \mathbf{F}_{1\to 2}$. Nonreciprocal interactions have been reported in systems including sedimenting particles~\cite{ramaswamy2001issues,fruchart2026nonreciprocal,smoluchowski1913stokes,smoluchowski1911wechselwirkung}, spinning starfish embryos~\cite{tan2022odd} as well as active colloids and droplets~\cite{schmidt2019light,soni2019odd,meredith2020predator,zhang2021active,wu2021ion,gupta2022active,bililign2022motile,maity2023spontaneous,das2024flocking}; they have also been programmed in colloids~\cite{bauerle2018self,lavergne2019group} and in robots~\cite{brandenbourger2019nonreciprocal,ghatak2020observation,fruchart2021non}. Despite recent advances~\cite{ivlev2015statistical,kryuchkov2018dissipative,fruchart2021non,dinelli2023non,loos2023long,avni2023non,huang2024active}, a general statistical-mechanics framework for nonreciprocal systems is still missing~\cite{ivlev2015statistical}.

To appreciate the reason for this, note that in classical and statistical mechanics, one may distinguish between Hamiltonian and non-Hamiltonian systems, see Fig.~\ref{fig:classes_of_systems}. 
Hamiltonian systems possess conjugate pairs of variables that span phase space (e.g., position and momentum). Mathematically, this is expressed by the Poisson bracket $\{\cdot,\cdot\}$ which endows phase space with a so-called symplectic structure~\cite{arnol2013mathematical}. 
This enables the use of canonical transformations~\cite{chan2004exact,birkhoff1927dynamical}, which preserve the Poisson bracket, to find integrals of motion or derive simplified effective models that approximate the dynamics~\cite{birkhoff1927dynamical,gustavson1966on,schrieffer1966relation}.

\begin{figure}[t!]
	\centering
	\includegraphics[width=0.48\textwidth]{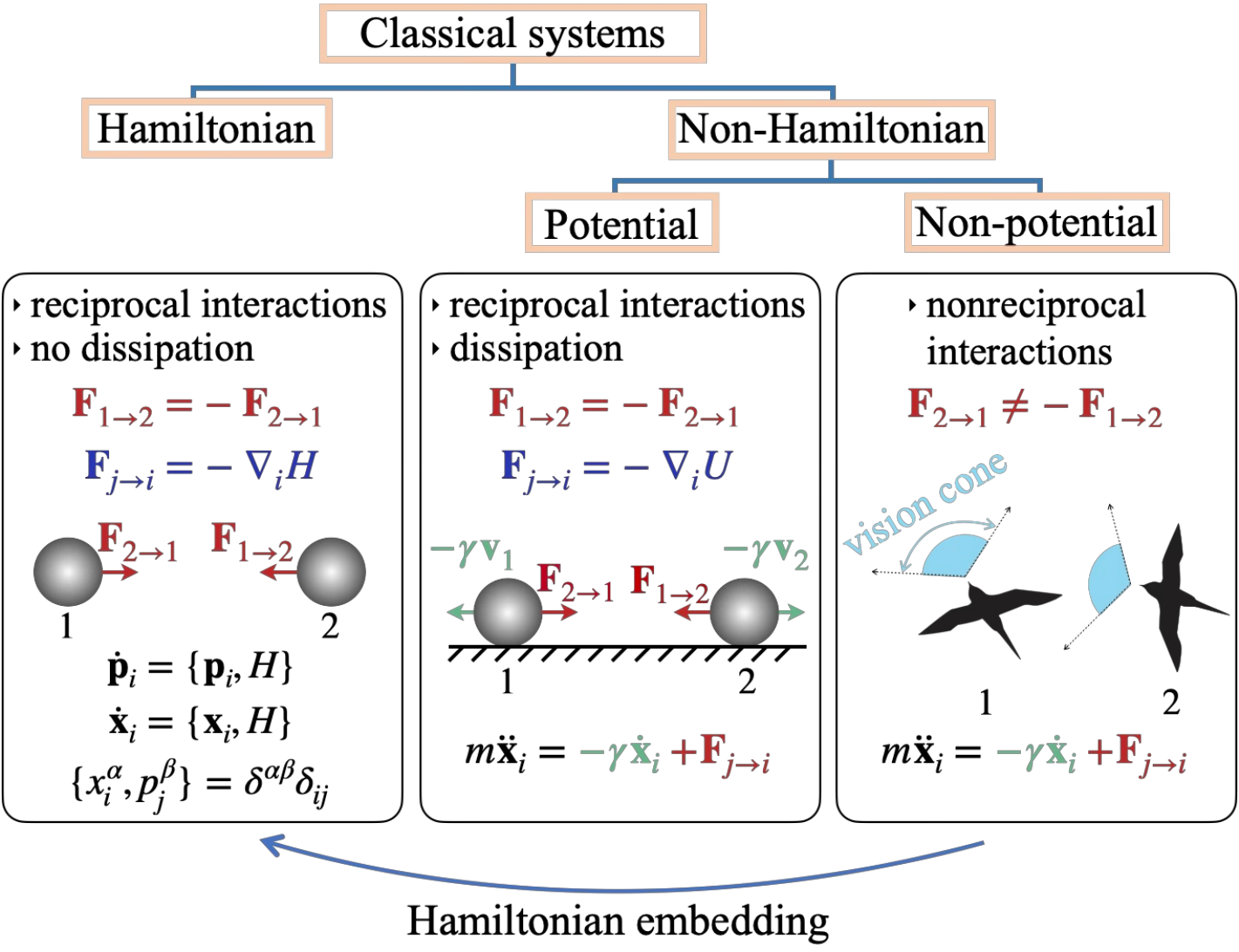}
	\caption{
		Classical systems can be classified as either Hamiltonian (e.g., two electric charges in vacuum, left panel) or non-Hamiltonian. The latter can have interactions that may be derived from a potential (e.g., two electric charges on a frictional substrate, middle panel) or not (e.g., birds with vision-cone interactions, right panel). In this work, we construct a Hamiltonian embedding for nonreciprocal systems.}
	\label{fig:classes_of_systems}
\end{figure}

Non-Hamiltonian systems, on the other hand, lack conjugate variables, and hence do not enjoy any of the advantages offered by the symplectic structure. Broadly speaking, they fall into two categories: systems whose equations of motion are governed by conservative forces that can be derived from a potential, and those that cannot, cf.~Fig.~\ref{fig:classes_of_systems}. 
A simple example of a non-Hamiltonian system with interactions arising from a potential is a pair of particles subject to dissipation, for example through friction with a substrate (Fig.~\ref{fig:classes_of_systems}, middle panel)~\footnote{We define dissipation by lack of conservation of phase-space volume under the dynamics.}. A notable advantage of having a potential is the ability to use Monte-Carlo methods~\cite{landau2021guide} to investigate the properties of the system coupled to a thermal bath. 

Conversely, in the simplest setting, nonreciprocal systems are modeled by forces that cannot be derived from a single mutual interaction potential~\cite{ivlev2015statistical}. These nonconservative forces do not admit an energy function and are, furthermore, non-Hamiltonian in general. Physically, this means that nonreciprocal systems are intrinsically out of thermal equilibrium, which substantially complicates their study beyond the bare simulation of the underlying equations of motion. For instance, due to the absence of a potential, it is a priori unclear how to apply Monte-Carlo methods to simulate and analyze properties of their nonequilibrium steady states. Similarly, due to the lack of conjugate variables, deriving effective descriptions for nonreciprocal models cannot be done using canonical transformations. 

In this work, we remove these obstacles by designing a Hamiltonian embedding for nonreciprocal systems. Specifically, for any equation of motion with pairwise interactions described by \textit{nonreciprocal} forces, we present a construction of a \textit{reciprocal} Hamiltonian, comprised of the original system and an auxiliary degree of freedom for each original one [Fig.~\ref{fig:schematic}]. By imposing a constraint on each (original, auxiliary)-pair, the equation of motion derived from this Hamiltonian reproduces the original nonreciprocal dynamics. Our results apply equally to inertial and dissipative nonreciprocal dynamics; for dissipative nonreciprocal systems in particular, the original and auxiliary degrees of freedom form canonically conjugate variables. 

We exemplify our construction using dissipative XY spins on a square lattice with vision-cone and chase-and-run interactions -- paradigmatic examples of nonreciprocal interactions~\cite{Dadhichi2020nonmutual,loos2023long,Popli2025dont,dopierala2025inescapable}. To demonstrate the utility of our construction, we first show (both analytically and numerically) the equivalence of the states, either steady or nonstationary, obtained from Monte-Carlo simulations with Glauber dynamics based on the Hamiltonian embedding on the one hand, and the original equations of motion subject to Langevin noise on the other. We then use the emergent symplectic structure of the Hamiltonian embedding to analyze the vision-cone XY system subject to a periodic drive using Floquet theory~\cite{bukov2015universal}, and derive effective equations of motion for the driven system. 
In the SM~\cite{SI}, we provide a general proof for the construction, accompanied by applications to five experimental systems featuring nonreciprocal interactions: metal-dielectric Janus particles~\cite{zhang2021active,das2024flocking}, walker robots~\cite{baconnier2022selective}, robotic metamaterials~\cite{brandenbourger2019nonreciprocal}, feedback-controlled active particles~\cite{bauerle2018self}, and sedimenting particles~\cite{ramaswamy2001issues}.

\begin{figure*}[t!]
	\centering\includegraphics[width=0.98\textwidth]{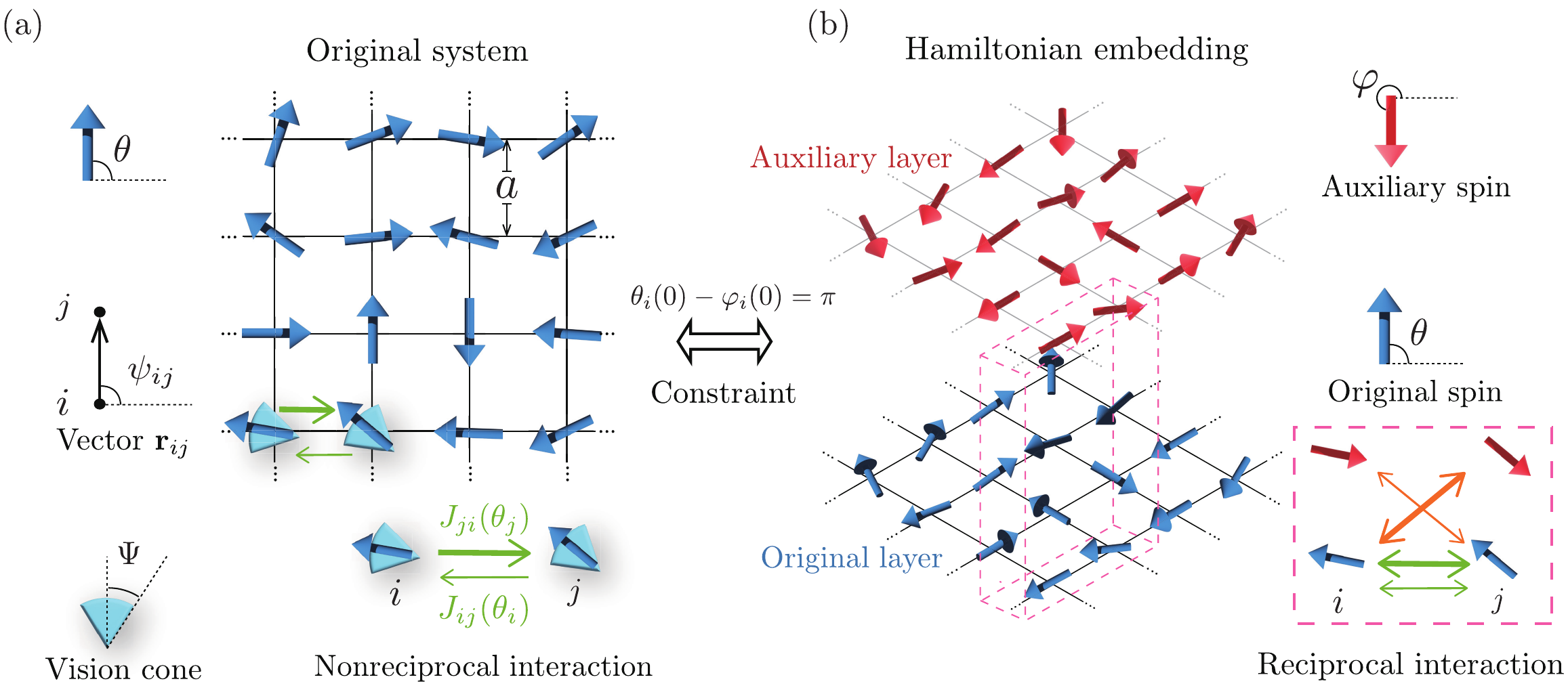}
	\caption{
		(a) Schematic illustration of XY spins with vision-cone interactions on a square lattice: each spin (blue arrow) described by the angle $\theta_j$ interacts only with nearest neighbors within its vision cone of angle $\Psi$ (cyan); a pair of unidirectional thin/thick green arrows show the nonreciprocal interaction strength. 
		The vector from site $i$ to site $j$ is denoted by  $\mathbf{r}_{ij}{=}a(\cos\psi_{ij},\sin\psi_{ij})$. (b) Corresponding Hamiltonian embedding: A layer of auxiliary spins $\varphi_j$ (red arrows) is coupled to the original spins (blue) reciprocally (cf.~inset and Eq.~\eqref{eq:H_tot}).
		Imposing the constraint that auxiliary spins mirror the original ones, $\theta_i(0){-}\varphi_i(i){=}\pi$, we recover the equations of motion for the nonreciprocal system from the equations of motion of its Hamiltonian embedding.
		\label{fig:schematic}
	}
\end{figure*}

\section{\label{sec:model}XY spins with nonreciprocal Vision-Cone Interactions}

Consider a square lattice of linear dimension $L$ and lattice constant $a$ with periodic boundary conditions. Each lattice site $i$ contains a classical spin whose orientation, $\mathbf{S}_{i}{=}(\cos \theta _{i},\sin \theta _{i})$, can rotate continuously within the lattice plane, and is parametrized by the angle $\theta _{i}$. The spins interact via pairwise vision-cone interactions of strength
\begin{equation*}
	J_{ij}(\theta _{i})=\left\{ 
	\begin{array}{cc}
		J, & \text{for}\;  \min \left\{ 2\pi -\left\vert \theta _{i}-\psi _{ij}\right\vert
		,\left\vert \theta _{i}-\psi _{ij}\right\vert \right\} \leq \Psi \\ 
		0, & \mathrm{else,}
	\end{array}
	\right. 
\end{equation*}
where $\Psi $ denotes half the vision cone angle, and $\psi _{ij}$ is the angle between the line connecting sites $i$ and $j$ and the horizontal [Fig.~\ref{fig:schematic}(a), inset]. The reciprocal XY model is recovered for $\Psi{=}\pi$.
We couple the system to an external bath of temperature $T$ through a Gaussian (white) noise term $\eta_i(t)$ with zero mean and unit variance; this bath defines the notion of temperature used throughout this work. The interacting spins undergo Langevin dynamics, following the equation of motion
\begin{equation}
	\label{eq:VC-XY_Langevin_EOM}
	\dot{\theta}_{i}\left( t\right) =-\sum_{j \in \{\langle ij \rangle\} }J_{ij}(\theta _{i})\sin (\theta_{i}-\theta _{j})+\sqrt{2 T}\eta _{i}(t),
\end{equation}
where the Boltzmann constant $k_B{=}1$, and the damping coefficient $\gamma{=}1$.
The corresponding vision-cone XY interaction [Fig.~\ref{fig:schematic}(a)] gives rise to a minimal model for nonreciprocal interactions, which has recently gained significant attention. Previous work has studied the role of nonreciprocity on the transition from the disordered to the ferromagnetic phase \cite{Dadhichi2020nonmutual,loos2023long,Popli2025dont,dopierala2025inescapable}, as well as on the creation and annihilation of topological defects \cite{rouzaire2025nonreciprocal,Popli2025dont}.

Since a spin may be in the vision cone of its neighbor, but not the other way around [see Fig.~\ref{fig:schematic}(a)], vision-cone interactions are nonreciprocal: $J_{ij}(\theta_i){\neq}J_{ji}(\theta_j)$; hence, the magnitudes of the torques acting on spins $i$ and $j$, as given by the right-hand-side of Eq.~\eqref{eq:VC-XY_Langevin_EOM}, differ. 
Thus, one cannot assign an unambiguous interaction energy to each lattice bond [Fig.~\ref{fig:schematic}(a), inset], and therefore vision-cone interactions cannot be derived from a potential.

\section{\label{sec:H_ext}Constrained Hamiltonian Embedding}

We now present a procedure that enables us to obtain nonreciprocal interactions in the equations of motion from an effective Hamiltonian function, in the absence of the noise term $\eta_i$. This is achieved by first extending the model by adding an auxiliary set of degrees of freedom, i.e., doubling the configuration space. Introducing suitable couplings between original and auxiliary degrees of freedom using \textit{reciprocal} interactions allows us to construct a well-defined Hamiltonian. 
The original nonreciprocal dynamics is generated from Hamilton's equations of motion for states obeying one constraint on each pair of (original, auxiliary) degrees of freedom, as explained below. This constraint reduces the size of the configuration space back to that of the original nonreciprocal system. Crucially, the constraint is preserved by the dynamics, so that it suffices to impose it once, e.g. on the initial state, of the time evolution.

Let us demonstrate the procedure using the vision-cone dynamics from Eq.~\eqref{eq:VC-XY_Langevin_EOM}, see SI~\cite{SI} for more examples. We introduce a set of auxiliary spins, $\mathbf{a}_{i}{=}(\cos \varphi _{i},\sin \varphi _{i})$, on an auxiliary lattice layer [red spins in Fig.~\ref{fig:schematic}(b)], which point opposite to the spins in the original system: $\mathbf{a}_{i}{=}{-}\mathbf{S}_{i}$. 
To grasp the essence of the idea, consider first two neighboring spins $\mathbf{S}_i, \mathbf{S}_j$ which interact nonreciprocally. The key insight is to symmetrize the nonreciprocal interaction between $\mathbf{S}_i$ and $\mathbf{S}_j$, but then use the interaction with the neighboring auxiliary spins $\mathbf{a}_{i}, \mathbf{a}_{j}$ to each cancel one part of the symmetric coupling between the $\mathbf{S}$-spins, cf.~Fig.~\ref{fig:schematic}(b) inset. Note that this cancellation only works if the auxiliary spins are antiparallel to the original, $\mathbf{a}_i{=}{-}\mathbf{S}_i$. This is the origin of the constraint that we need to enforce, which we call the mirror constraint. There is no interaction between the auxiliary spins $\mathbf{a}_i$. 

Following this procedure, we write a Hamiltonian
\begin{equation}
	\label{eq:H_tot}
	H(\theta_i,\varphi_i)=H_{SS}(\theta_i)+H_{Sa}(\theta_i,\varphi_i),
\end{equation}%
with conjugate degrees of freedom $\{\theta_i, \varphi_j\}{=}\delta_{ij}$, where $\{\cdot,\cdot\}$ is the Poisson bracket; hence, $\varphi_i$ acts as the canonical momentum for $\theta_i$, see SI~\cite{SI} for details.

The first term, $H_{SS}$, describes the symmetrized interactions within the original system,
\begin{equation}
	\label{eq:H_SS}
	H_{SS}=-\sum_{ \langle ij \rangle }\left[ J_{ij}(\theta _{i})+J_{ji}(\theta _{j})\right] \cos (\theta _{i}-\theta _{j}) . 
\end{equation}%
Here, the energy of bond $\langle ij\rangle$ is given by the sum of the vision-cone interactions along that bond, $J_{\left\langle ij\right\rangle }\left( \theta _{i},\theta _{j}\right) {=}J_{ij}(\theta_{i}){+}J_{ji}(\theta _{j})$. Although this symmetrized interaction is reciprocal, it depends on the orientations $\theta _{i}$ and $\theta _{j}$. 

The second term, $H_{Sa}$, represents the coupling between the two
layers, where the auxiliary spins $\mathbf{a}_{i}$ interact with the original spins $\mathbf{S}_{j}
$, following
\begin{equation*}
	H_{Sa}=\sum_{ \langle ij \rangle } \left[-J_{ij}(\theta _{i})\cos (\theta _{j}-\varphi_{i}) - J_{ji}(\theta _{j})\cos (\theta _{i}-\varphi _{j}) \right].
\end{equation*}%
The magnitude of each of the two terms in $H_{Sa}$ above directly corresponds to that in the symmetrized interaction, see Fig.~\ref{fig:schematic}(b) inset; as we show next, this is essential to recover the original nonreciprocal dynamics. 

The equations of motion generated by the Hamiltonian~\eqref{eq:H_tot}, $\dot\theta_i {=} \left\{ \theta_i,H\right\}$ and $\dot\varphi_i {=} \left \{\varphi_i , H \right \}$, form a set of equations that couple the auxiliary and original degrees of freedom:
\begin{eqnarray}
	\label{eq:H_theta_EOM}
	\dot{\theta}_i \left( t \right) 
	&=&\partial_{\varphi_i}H = 
	\sum_{j \in \{\langle ij \rangle\}} J_{ij}\left(\theta_i\right) \sin \left( \varphi_i-\theta_j \right) , \\
	\dot{\varphi}_{i}\left( t\right) 
	&=&-\partial_{\theta_i}H = 
	\sum_{j \in \{\langle ij \rangle\}} -J_{ij}(\theta _{i})\sin(\theta _{i}-\theta _{j})  \notag \\
	&&- J_{ji}(\theta _{j})\left[ \sin (\theta_{i}-\varphi _{j})+\sin (\theta _{i}-\theta _{j})\right] \notag \\
	&&+\partial _{\theta _{i}}J_{ij}(\theta _{i})\left[ \cos (\theta_{j}-\varphi _{i})+\cos (\theta _{j}-\theta _{i})\right] . \nonumber
\end{eqnarray}
The $\partial _{\theta _{i}}J_{ij}(\theta _{i})$ term appears since the coupling strength $J_{ij}(\theta _{i})$ depends on the spin degree of freedom $\theta_i$; this implicit dependence is at the origin of the discrepancy between the dynamics generated by the reciprocal textbook XY model with the replacement $J_{ij}\mapsto J_{ij}(\theta_i)$ and the dynamics of Eq.~\eqref{eq:VC-XY_Langevin_EOM}~\cite{rouzaire2025nonreciprocal}.

If the mirror constraint is imposed on the initial conditions, 
\begin{equation}
	\label{eq:constraint}
	\mathbf{a}_i(0)=-\mathbf{S}_i(0) \quad \Longleftrightarrow \quad \theta_i(0) -\varphi_i(0) = \pi, 
\end{equation}
it is preserved under the equations of motion, i.e., $\theta_i(t) {-}\varphi_i(t) {=} \pi$ for all times, as we show in the SI~\cite{SI}. 
Because, under the constraint, the dynamics of the $\varphi_i(t)$ degree of freedom is tied to that of the $\theta_i(t)$ spins, the equations of motion~\eqref{eq:H_theta_EOM} uncouple. Moreover, each degree of freedom recovers precisely the nonreciprocal vision-cone interaction term from Eq.~\eqref{eq:VC-XY_Langevin_EOM}. Formally, the constraint~\eqref{eq:constraint} causes the unwanted terms in the last two lines of Eq.~\eqref{eq:H_theta_EOM} to cancel, and makes the remaining equations identical to one another~\footnote{Notice that, after applying the constraint, the vision cones for $\varphi_i$ and $\theta_i$ point in the same directions, even though $\theta_i-\varphi_i=\pi$.}. 

The procedure illustrated above is not limited to vision-cone XY interactions. In the SI~\cite{SI}, we first present a general proof for arbitrary nonreciprocal interactions. To illustrate the wide applicability of our framework, we construct the Hamiltonian for five specific examples: metal-dielectric Janus particles~\cite{das2024flocking,zhang2021active}, self-aligning active particles such as walker robots~\cite{baconnier2022selective,baconnier2025self}, robotic metamaterials~\cite{brandenbourger2019nonreciprocal}, feedback-controlled active particles~\cite{bauerle2018self}, and sedimenting particles~\cite{ramaswamy2001issues}.
In these examples, the dynamics are overdamped, i.e., dissipation dominates over inertia. However, using suitable kinetic-energy terms, the Hamiltonian embedding also applies to inertial (i.e., non-dissipative) nonreciprocal dynamics, suggesting a broad applicability, see SI~\cite{SI}.

According to Liouville's theorem, without constraint, the Hamiltonian embedding gives rise to nondissipative dynamics in the embedding phase space, even if the original equations of motion are dissipative. This feature is brought about by the symplectic structure inherent to the Hamiltonian embedding (e.g., in the vision-cone XY spins that we discussed, $\theta$ and $\varphi$ are conjugate variables). By contrast, on the submanifold defined by the constraint, the dynamics are dissipative. This apparent contradiction is resolved by noticing that the constrained submanifold is a set of measure zero in the embedding phase space which does not contribute to phase-space volume.

As a direct consequence of the constraint, trajectories with initial conditions on the constraint manifold remain confined to it for all time, and hence never intersect with trajectories that start outside. Therefore, dynamics outside the manifold, which are reciprocal, do not affect the dynamics on the constraint manifold, which are nonreciprocal. 
Thence, the phase-space probability distribution evolves independently on the constraint manifold and outside it. It is therefore possible for an initial ensemble on the constraint manifold to evade thermalization, even when the embedding Hamiltonian dynamics outside the manifold are ergodic. This provides a way to embed nonequilibrium configuration-space distributions of nonreciprocal systems into the larger Hamiltonian phase space. 
Because the nonreciprocal manifold is a set of measure zero within the total phase space, it does not affect expectation values of observables defined on the total phase space. Hence, observables on the embedding phase space are insensitive to non-stationary behaviors on the nonreciprocal manifold, such as the breaking of continuous time-translation symmetry or persistent configuration-space currents.

Furthermore, notice that, when evaluated on the constraint~\eqref{eq:constraint}, the total Hamiltonian $H$ vanishes identically; therefore, $H$ does not describe the energy of the original nonreciprocal system, but merely provides a generator for its dynamics. 
Moreover, the ground-state manifold of $H$ need not obey the constraint.
These unusual features reflect the inherent out-of-equilibrium character of the original nonreciprocal system.

Finally, we note that the form of the Hamiltonian embedding is not unique. For instance, one can distribute the interaction symmetrically among the two types of degrees of freedom without changing the resulting equations of motion upon imposing the constraint. It is currently an open question whether this freedom can be used to imprint additional properties or constraints into the embedding Hamiltonian.

In what follows, we illustrate the utility of the proposed Hamiltonian description by 
(i) demonstrating that the long-time states of the
original Langevin dynamics ~\eqref{eq:VC-XY_Langevin_EOM}, either steady or nonstationary, coincides with those of Glauber dynamics generated by the Hamiltonian~\eqref{eq:H_tot}, thus enabling the use of Monte-Carlo methods to study nonreciprocal dynamics.
In addition, (ii) we engineer the strength of vision cone interactions along different lattice directions with the help of high-frequency periodic drives, explicitly using the symplectic structure of the Hamiltonian formulation; this enables the application of Floquet engineering to control the phase transitions of nonreciprocal systems.

\section{\label{sec:Langevin_vs_MC}Equivalence between Hamiltonian Monte-Carlo and Langevin Dynamics Subject to White Noise}

The absence of a well-defined interaction energy per bond precludes the direct application of Monte-Carlo methods to study nonreciprocal systems.
To circumvent this problem, recent works have introduced new concepts and constructions~\cite{weiderpass2025solving,rajeev2024ising,seara2023non} to perform Monte-Carlo simulations, such as the `selfish energy' of a spin on the lattice site $i$, $E_i {=} 
{-} \sum_{j \in \langle ij \rangle} J_{ij}(\theta_i) \cos(\theta_i{-}\theta_j)$~\cite{avni2023non,loos2023long,blom2025local}, or modified acceptance rates that include nonreciprocal interactions through a Legendre-like transform similar to a grand-canonical ensemble~\cite{osat2023non}. 
A notable common disadvantage of these tools is that it is a priori unclear whether the resulting steady states correspond to those of the microscopic Langevin dynamics; moreover, many nonreciprocal systems evolve into nonstationary states. Additionally, these methods do not enjoy formal mathematical guarantees for convergence, unlike Monte-Carlo methods for reciprocal systems~\cite{landau2021guide}.

Starting from the reciprocal Hamiltonian embedding, we propose a Monte-Carlo algorithm to sample nonequilibrium steady states and nonstationary states of nonreciprocal systems. 
Specifically, we apply constrained Glauber dynamics~\cite{glauber1963time} at temperature $T$ defined by the transition rates $w \left( \theta_{i}\rightarrow \theta_{i}^{\prime }\right) ={1}/{2}\left[1-\tanh(\Delta E(\theta_i,\theta'_i)/(2T))\right]$ with
\begin{equation}
	\label{eq:Glauber_rate}
	\Delta E(\theta_i,\theta'_i) = \frac{1}{2}\big\{H(\theta',\varphi)-H(\theta,\varphi')\big\}\big|_\text{constr.}\, .
\end{equation}
Here, the notation $\{\cdot\}|_\text{constr.}$ means that the operations in the bracket are carried out first, and only then is the constraint applied. As a result, the auxiliary variable $\varphi_i$ drops out of the expression for the energy difference $\Delta E(\theta_i,\theta'_i)$. Consequently, we only need to simulate the dynamics of the original $\mathbf{S}$-spin layer, thus keeping the number of simulated degrees of freedom equal to that of the original system.  

For large system sizes, $L{\gg} 1$, these Glauber dynamics give rise to a Fokker-Planck equation for the configuration-space distribution $\rho(\bm{\theta},t_\text{MC})$ (see Methods):
\begin{eqnarray}
	\label{eq:rho_vs_time}
	\frac{\partial \rho(\bm{\theta}, t_{\textrm{MC}})}{\partial t_{\textrm{MC}}} 
	= \alpha \sum_{i=1}^{L^2}
	\partial_{\theta_i}
	\Big[
	T \partial_{\theta_i} \rho(\bm{\theta}, t_{\textrm{MC}})
	{-}F_i(\bm{\theta})\rho(\bm{\theta}, t_{\textrm{MC}})
	\Big], \nonumber \\
\end{eqnarray}
where $F_i(\bm{\theta}){=}{-}\partial_{\theta_i}H(\bm{\theta},\bm{\varphi})\big|_\text{constr.}$ is the (nonreciprocal) force acting on spin $\mathbf{S}_i$.
Here $\alpha$ is a scaling factor relating the physical time and Monte-Carlo time: $t=\alpha t_\text{MC}$, set by temperature $T$ and the support size $\Delta$ of the uniform distribution according to which the Glauber updates are performed (see Methods).

In the Methods section, we prove that Eq.~\eqref{eq:rho_vs_time} coincides with the Fokker-Planck equation obtained from the original Langevin equations of motion. Therefore, the Hamiltonian embedding allows us to define transition rates that enable the use of the Glauber algorithm in nonreciprocal systems. Our Monte-Carlo scheme allows us to sample nonequilibrium states since detailed balance is broken by the constraint, see Eq.~\eqref{eq:Glauber_rate} and SM~\cite{SI}. Hence, we can bypass brute-force Langevin dynamics and directly sample both nonequilibrium steady states as well as oscillatory nonstationary states, as we now demonstrate in numerical simulations.

\begin{figure}[t!]
	\centering\includegraphics[width=0.48\textwidth]{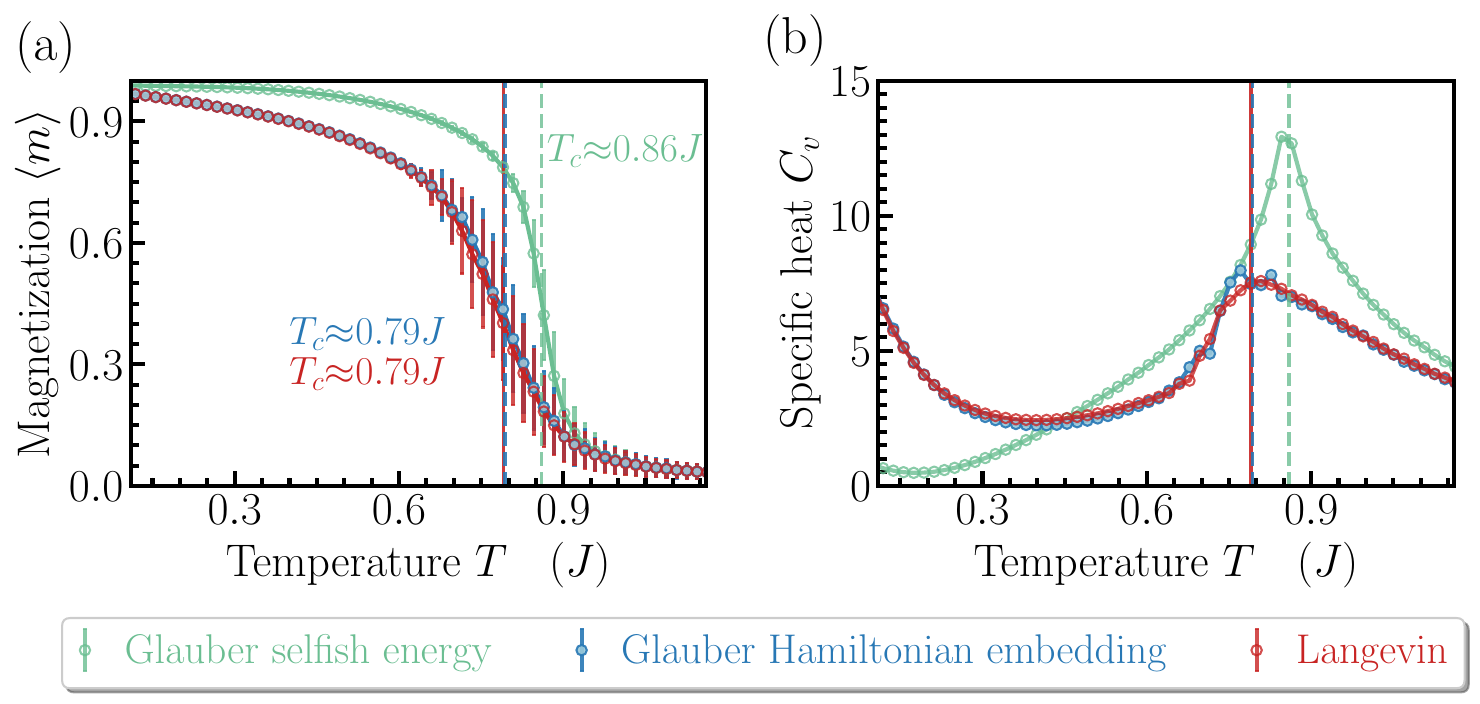}
	\caption{
		Nonequilibrium steady state in XY spins with vision-cone interactions. 
		(a) Average magnetization $\langle m \rangle$ with standard deviation $\sigma_m$ (errorbars), and (b) specific heat $C_v$ as functions of temperature $T$, obtained from Monte-Carlo simulations with Glauber dynamics based on either the selfish energy (green) or the constrained Hamiltonian embedding (blue), as well as simulations of the original Langevin dynamics (red).
		Whereas the constrained Glauber and Langevin simulations show the same critical temperature within the statistical uncertainty: $T_c/J =0.79 \pm 0.04$ (Langevin) and $T_c/J =0.79 \pm 0.04$ (constrained Glauber), the selfish-energy Glauber dynamics predicts $T_c/J = 0.86\pm0.02J$.
		The square lattice has linear dimension $L{=}100$, the vision cone angle is $\Psi {=} 0.85\pi$.
		For the Langevin dynamics, we set a time step $J\delta t{=}0.01$ and use $100 L^2$ iterations. 
		In the constrained Glauber algorithm~\cite{SI}, we set a support size $\Delta{=}0.1$, and use use $200 L^4$ iterations.
	A total of $n_\textrm{eq}=5\times 10^4$ data points are uniformly sampled from the last $50L^2$ ($50L^4$) iterations for Langevin (constrained Glauber) dynamics of $\mathcal{N}=100$ independent trajectories, each initialized with a different random seed, at each temperature. The total number of samples is therefore $n=n_\textrm{eq}\times\mathcal{N}=5\times 10^6$. For selfish-energy Glauber dynamics, $n_\textrm{eq}=10^4$ and $\mathcal{N}=100$ (ref.~\cite{zenodo_repo}).
	}\label{fig:Langevin_Hamiltonian_correspondence}
\end{figure}

\subsection{Nonequilibrium steady states}
\label{subsec:VC_steady_state}

For the vision-cone spins from Eq.~\eqref{eq:VC-XY_Langevin_EOM}, using the Hamiltonian embedding~\eqref{eq:H_tot} we compute the energy difference entering the constrained Glauber transition rates as $\Delta E(\theta_i,\theta'_i){=}
{-}\frac{1}{2}\sum_{j\in\langle ij\rangle}[J_{ij}(\theta_i'){+}J_{ij}(\theta_i)]\times \qquad\qquad[\cos(\theta_i'{-}\theta_j){-}\cos(\theta_i{-}\theta_j)]$.

In Fig.~\ref{fig:Langevin_Hamiltonian_correspondence}, we compare (a) the average magnetization $\langle m \rangle {=}\left\langle\sqrt{(\sum_i \cos \theta
	_{i})^{2}{+}\left( \sum_i \sin \theta _{i}\right) ^{2}}/L^{2}\right\rangle$ and (b) the specific heat, $C_{v}{=}\left( \left\langle H_{SS}^{2}\right\rangle {-}\left\langle
H_{SS} \right\rangle ^{2}\right) /\left( LT\right)^{2}$, obtained as the variance of the energy in the original spin layer~\cite{galley2013classical}, as a function of temperature, for constrained Glauber dynamics (blue) vs.~Langevin dynamics (green). We initialize the system in an ordered state, $\theta_i {=}0$, with initial energy of the original system $H_{SS}{=}3L^2$. Consistent with previous works~\cite{Dadhichi2020nonmutual,loos2023long,Popli2025dont,dopierala2025inescapable}, our data indicates the existence of a phase transition between an ordered phase at low temperature and a disordered phase at high temperature [Fig.~\ref{fig:Langevin_Hamiltonian_correspondence}(a)].
Accordingly, the specific heat $C_v$ exhibits a peak at the phase transition point [see Fig.~\ref{fig:Langevin_Hamiltonian_correspondence}(b)].
Importantly, both the Langevin and the constrained Monte-Carlo simulations exhibit the same critical temperature, $T_c/J=0.79\pm 0.04$, for the phase transition, within statistical uncertainty, see SI~\cite{SI}. By contrast, a Monte-Carlo simulation using the Hamiltonian~\eqref{eq:H_SS} without the effect of the auxiliary spins $\mathbf{a}_i$ produces a critical temperature of $T_c/J{\approx} 1.70$~\cite{SI,bandini2024xy} (not shown). 

Unlike Glauber dynamics based on the selfish energy~\cite{avni2023non,loos2023long}, our constrained-Hamiltonian Monte-Carlo update rule results in the same configuration-space density for the nonequilibrium steady state as the Langevin equation. Indeed, the critical temperatures for the phase transition estimated from the Langevin and the selfish-energy approach do not coincide in general [compare green and red curves in Fig.~\ref{fig:Langevin_Hamiltonian_correspondence}], although for some parameters the values may agree within statistical uncertainty. By ensuring the equivalence of the Langevin and the constrained Monte-Carlo dynamics, our Hamiltonian embedding resolves this long-standing issue in the study of vision-cone interactions~\cite{rouzaire2025nonreciprocal}.

\subsection{Nonstationary states}
\label{subsec:run-chase_nonstationary_state}

\begin{figure}[t!]
	\centering\includegraphics[width=0.48\textwidth]{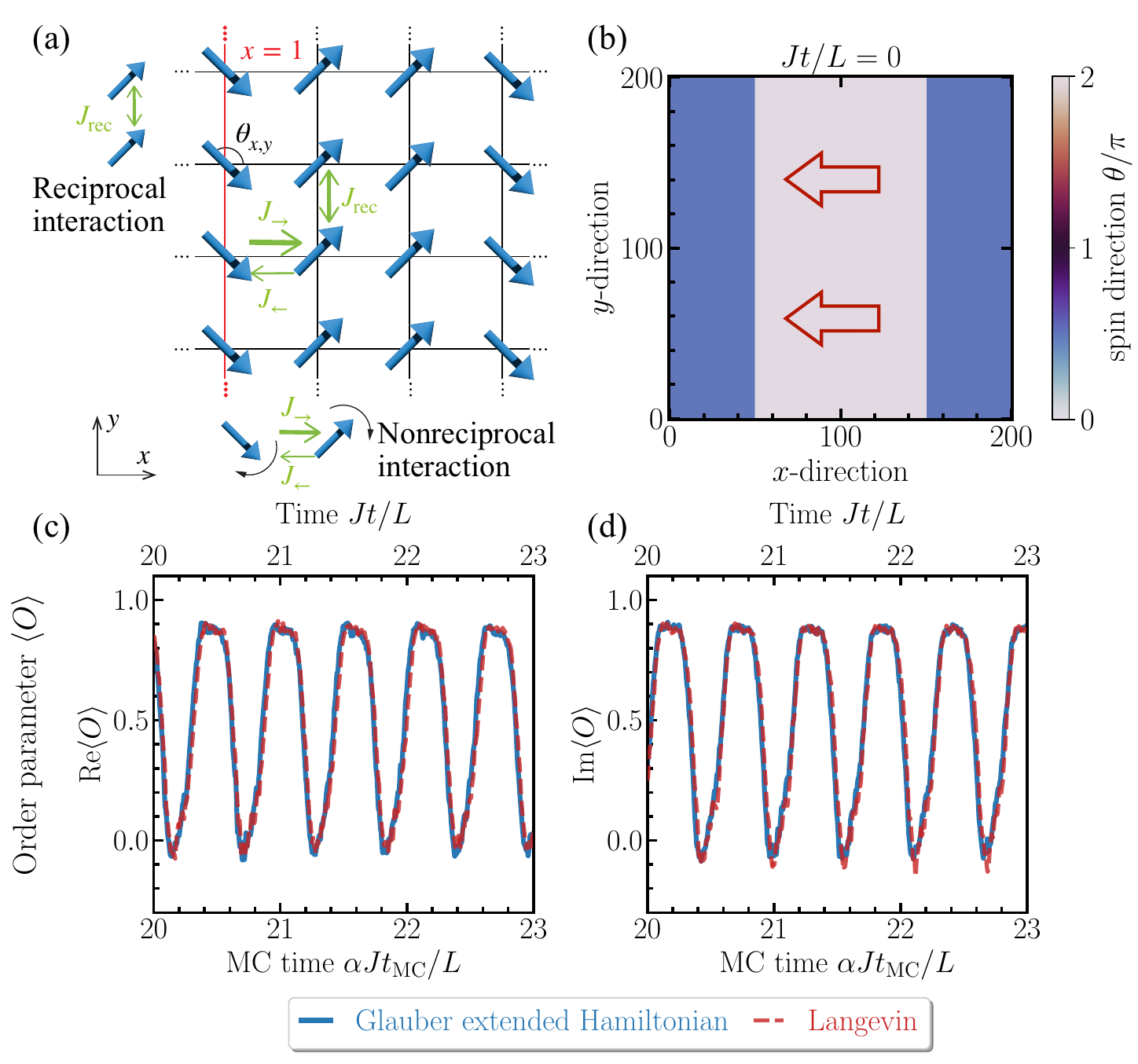}
	\caption{
		Nonstationary state of XY spins with chase-and-run interactions. 
		(a) XY spins on a square lattice with nonreciprocal interactions along the $x$-direction, and reciprocal interactions along the $y$-direction. The $x{=}1$ column used to define the order parameter $O(t)$ is marked in red. 
		(b) Stripe pattern used as an initial steady state, which moves perpetually to the left due to the nonreciprocal interactions (see Video).  
		(c)/(d) Long-time evolution of the real/imaginary part of the stripe order parameter $\langle O(t)\rangle$ averaged over an ensemble of $\mathcal{N}{=}100$ independent realizations of the noisy dynamics. 
		The data show excellent agreement between the Langevin [dashed line, top horizontal axis] and the constrained Glauber simulations [solid line, bottom horizontal axis] with transition rates obtained from the Hamiltonian embedding. The time axis is rescaled by the linear system size $L$.  
		The parameters are $J_{\rightarrow}/J{=}1$, $J_{\leftarrow}/J {=} {-}0.99$, $J_\text{rec}/J {=}10$, $T/J{=} 0.08$, $\Delta{=}0.05$, and $L=200$, with the scaling factor $\alpha(\Delta,T){=}2.37{\times}10^{-3}$ extracted numerically~\cite{SI}.
	}\label{fig:run_chase}
\end{figure}

We now show that the Hamiltonian embedding also describes nonreciprocal systems that exhibit long-time nonstationary states, such as sustained oscillations, corresponding to the breaking of time-translation symmetry.

Consider the dynamics of XY spins on a square lattice following Eq.~\eqref{eq:VC-XY_Langevin_EOM}, but with nonreciprocal interactions $J_{\leftarrow}{\neq} J_{\rightarrow}$ along the $x$-direction (independent of the spin orientation), and reciprocal interactions $J_\text{rec}$ along the $y$-direction, see Fig.~\ref{fig:run_chase}(a). The initial state is the vertical stripe pattern shown in Fig.~\ref{fig:run_chase}(b). Whenever $J_{\leftarrow} {\approx} {-}J_{\rightarrow}$, the spins exhibit a chase-and-run dynamics~\cite{fruchart2021non} and the pattern moves to the left (for $J_\rightarrow{>}0$) at a constant speed (see Video). In the thermodynamic limit, this nonstationary state propagates perpetually due to an interplay between noise and many-body interactions~\cite{SI}, breaking continuous time-translation symmetry. 
These sustained dynamics do not have an equilibrium analogue, and hence present an ideal testbed for the validity of our constrained Glauber dynamics.  

Following the procedure in Sec.~\ref{sec:H_ext}, we calculate a Hamiltonian embedding for this model (Methods); then, using Eq.~\eqref{eq:Glauber_rate}, we find the expression for the energy difference $\Delta E(\theta,\theta')$, cf.~Eq.~\eqref{eq:DeltaE_pattern}.

Figure~\ref{fig:run_chase}(c-d) shows the dynamics of the complex-valued order parameter
$
O(t)=L^{-1} \sum_{y = 1}^{L} e^{i \theta_{1,y}(t)},
$
which characterizes the average orientation of the spins within the $x{=}1$ column of the lattice [shown in red in Fig.~\ref{fig:run_chase}(a)]. 
Then, with periodic boundary conditions, the real and imaginary parts of $\langle O(t) \rangle$ exhibit temporal oscillations as a result of the collective spin rotation at the interface between the stripes, which moves due to the nonreciprocal interactions.
As anticipated from our analytical theory, we find an excellent agreement between the Langevin and Monte-Carlo dynamics at all times, including transients~\cite{SI}. This provides numerical evidence that Glauber dynamics with transition rates determined from the Hamiltonian embedding capture correctly the nonstationary-state behavior of nonreciprocal systems.

We emphasize that, while above we used two specific models to demonstrate the equivalence between Monte-Carlo dynamics and the original nonreciprocal Langevin equation, the Hamiltonian embedding applies to generic nonreciprocal systems, as we show in~\cite{SI}.

\section{\label{sec:Floquet}Hamiltonian engineering for nonreciprocal systems}

Whereas the Glauber dynamics in the previous section rely on the embedding Hamiltonian to define transition rates, they do not make use of the underlying symplectic (or phase-space) structure, as Monte-Carlo simulations only require an energy function. 
The symplectic structure of the embedding Hamiltonian enables the use of canonical transformations to analyze the physics of nonreciprocal systems, e.g., in identifying sets of strongly and weakly interacting degrees of freedom which define fast and slow variables, or in determining conserved quantities. Its existence is also inherently related to the existence of a Hamiltonian as a generator of dynamics.

As a concrete example of leveraging the symplectic structure provided by the embedding Hamiltonian, here we demonstrate the application of Hamiltonian engineering to modify the properties of nonreciprocal systems using high-frequency periodic driving~\cite{bukov2015universal,goldman2014periodically,eckardt2017colloquium}. 
To this end, we consider a hypothetical experiment on a system of XY spins interacting nonreciprocally on a square lattice (Fig.~\ref{fig:periodic_drive}a). Hamiltonian engineering allows one to tune the interaction strength independently along the different lattice directions, which may enable one to explore regimes not accessible otherwise. As an example, below we show how to use high-frequency periodic driving to modify the interactions and achieve a dimensional crossover, whereby a square lattice of spins with vision-cone interactions is effectively turned into a one-dimensional chain (Fig.~\ref{fig:periodic_drive}a).

To define the drive, note that the XY model arises naturally in the description of Josephson junction arrays; when subject to an external AC magnetic field, the field can be modeled by a time-dependent vector potential $A_{ij}$ which couples to the phase difference $\theta_i{-}\theta_j$ of the superconducting order parameter~\cite{newrock2000two}.
While the discussion here is not bound to a specific experimental system, this setting illustrates concretely how a periodic drive can modify the interaction term when it is not possible to directly tune its coupling strength $J_{ij}$. In the presence of the drive, the equations of motion read as

\begin{equation}
	\label{eq:periodic_Langevin}
	\dot{\theta}_{i}\left( t\right) {=}{-}\!\!\sum_{ j\in \{\langle ij \rangle \}}\!\!\!\! J_{ij}(\theta _{i})\sin (\theta
	_{i}-\theta _{j}-
	A_{ij}\cos\omega t
	)
	+\sqrt{2 T}%
	\eta _{i}(t). 
\end{equation}
We model the external periodic drive along the $x$-direction with frequency $\omega$ and amplitude $A_{ij}{=}aA_0\cos\psi_{ij}$, where the angle $\psi_{ij}$ is defined in Fig.~\ref{fig:schematic}a, and $a$ is the lattice constant~\cite{dunlap1986dynamic,lignier2007dynamical,arimondo2012kilohertz}. 

We now demonstrate that the periodic drive enables us to effectively suppress the spin interactions along the $x$-direction only [Fig.~\ref{fig:periodic_drive}a] while keeping the vision-cone interaction along the $y$-direction~
\footnote{ 
	In the literature on quantum systems~\cite{bukov2015universal,goldman2014periodically,eckardt2017colloquium}, this phenomenon is known as dynamical localization, since an external drive is used to suppress the transfer of spin-state population between neighboring lattice sites, controlled by the corresponding spin-spin interaction strength. In the case of nonreciprocal interactions, the periodic drive leads to weakening and eventually complete decoupling along the $x$-bonds of the lattice.
}
-- an instance of Floquet engineering~\cite{bukov2015universal,goldman2014periodically,eckardt2017colloquium}.
In systems with limited control over the interaction strength, this strategy would allow experiments to study the fate of the phase transition, as the dimensionality of the system is effectively changed from $d=2$ (a square lattice) to $d=1$ (a collection of uncoupled chains). 
However, the Floquet engineering framework uses a Hamiltonian, which is not directly available in nonreciprocal systems.

The Hamiltonian formalism from Sec.~\ref{sec:H_ext} proves convenient to analyze the driven system due to its symplectic structure. Adding the auxiliary spin $\mathbf{a}_{j}$, we construct the time-periodic Hamiltonian embedding
\begin{eqnarray}
	\label{eq:periodic_drive_Hamiltonian}
	H\left( t\right)  
	=-\sum_{\langle ij\rangle}&&J_{ij}(\theta _{i})[\cos (\theta _{i}-\theta
	_{j}-A_{ij}\cos\omega t )   \\
	&&+\cos (\varphi _{i}-\theta _{j}-A_{ij}\cos\omega t)] +(i\leftrightarrow j) , \notag
\end{eqnarray}
with $\left\{ \theta _{j},\varphi _{i}\right\} {=}\delta _{ij}$. 
To derive an effective description at high driving frequencies, $J_{ij}/\omega{\ll}1$, we apply the Floquet-Magnus expansion $H_{F} {=}H_{F}^{(0)}{+}\mathcal{O}(\omega^{-1})$~\cite{bukov2015universal}; the leading-order Floquet Hamiltonian is the time-average
\begin{eqnarray}
	\label{eq:HF_0}
	H_{F}^{(0)} &{=}&{-}\sum_{\langle ij \rangle}
	\mathcal{J}_{0}\left( A_{ij} \right)  J_{ij}(\theta _{i}) \left[ \cos \left( \theta _{i}{-}\theta _{j}\right) {+}\cos \left( \varphi_{i}{-}\theta _{j}\right) \right] \nonumber\\
	&& +(i\leftrightarrow j),
\end{eqnarray}
where $\mathcal{J}_{0}\left( A_{ij} \right)$ is the Bessel function of the first kind.
Higher-order terms in the Floquet-Magnus expansion can be readily obtained in a straightforward way using this formalism; they control the validity of the approximation.

\begin{figure}[t!]
	\centering\includegraphics[width=0.48\textwidth]{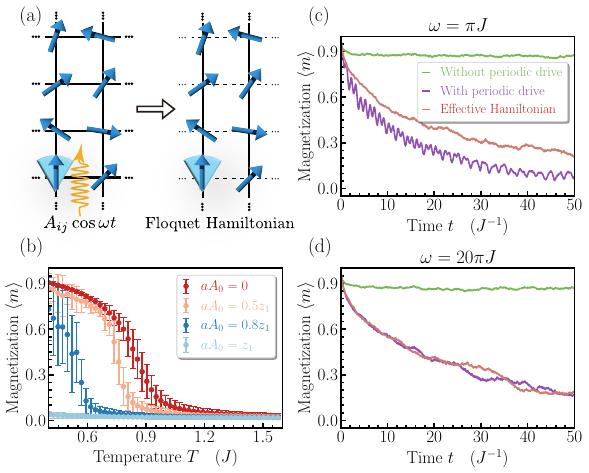}
	\caption{
		The Hamiltonian embedding enables the use of Floquet techniques to analyze driven nonreciprocal interacting systems.
		(a) Schematic illustration of the periodically driven XY spins (left panel, Eq.~\eqref{eq:periodic_drive_Hamiltonian}) and the effective time-averaged interactions (right panel, Eq.~\eqref{eq:HF_0}). Thin black dashed lines indicate suppression of interactions in the $x$-direction. 
		(b) Average magnetization $\langle m \rangle$ standard deviation $\sigma_m$ (error bars) obtained from Langevin dynamics as a function of temperature $T$ for different drive amplitude strengths $A_0$. As $A_0$ increases, the transition temperature decreases; as predicted by the time-averaged Floquet Hamiltonian, when $A_0 {=} z_\alpha a^{-1}$ (see text), the system no longer exhibits a long-range ordered phase. We use $100L^2$ iterations with a timestep of $J \delta t=0.01$; $\omega{=}20 \pi J$, and $\Psi=7\pi/8$. 
		A total of $n_\textrm{eq}{=}10^4$ data points are uniformly sampled from the last $50L^2$ timesteps of each of $\mathcal{N}=100$ independent trajectories, each initialized with a different random seed, at every temperature. The total number of samples used to compute averages and standard error bars is therefore $n{=}n_\textrm{eq}\times\mathcal{N}{=}10^6$ (ref.~\cite{zenodo_repo}).
		(c-d) Average magnetization $\langle m \rangle$ obtained from Langevin dynamics as a function of time $t$ for two drive frequencies: $\omega {=} \pi J$ (c) and $\omega {=} 20\pi J$ (d), for $T{=}0.5J$. As the frequency increases, the dynamics corresponding to the equations of motion derived from $H_F^{(0)}$ (orange) become closer to the exact periodically driven Langevin dynamics (purple). Both show a significant deviation from the Langevin dynamics of the non-driven system from Eq.~\eqref{eq:VC-XY_Langevin_EOM} (green). 
		Parameters: $A_0 {=} 2a^{-1}$, and $L{=}64$.
		No ensemble averaging is performed here.
	}
	\label{fig:periodic_drive}
\end{figure}

To verify the validity of the Floquet Hamiltonian description in the nonreciprocal setting, we study the average magnetization as a function of time $t$ for two drive frequencies in Fig.~\ref{fig:periodic_drive}(c,d). The deviation between the results from the exact Langevin equations of motion (purple) and the effective equations of motion derived from $H_F^{(0)}$ (orange) decreases as the frequency increases [from Fig.~\ref{fig:periodic_drive}(c) to Fig.~\ref{fig:periodic_drive}(d)]; moreover, the drive modifies the behavior significantly (compare orange and green curves).

By tuning the amplitude $A_{0}$ of the drive close to a zero of the Bessel function, $\mathcal{J}_{0}(z_\alpha){=}0$, one can suppress spin interactions in the $x$-direction, thereby realizing a dimensional crossover from a two-dimensional lattice to a set of one-dimensional chains [Fig.~\ref{fig:periodic_drive}a]. 
Figure~\ref{fig:periodic_drive}(b) shows the average magnetization obtained from simulations of the spin dynamics~\eqref{eq:periodic_Langevin} in the high-frequency regime as a function of temperature $T$. As we gradually increase the drive amplitude $A_{0}$, the transition temperature $T_c$ decreases. To leading order in $\omega^{-1}$, when $A_{0}{=}z_\alpha a^{-1}$, the system decouples into a set of independent 1D XY-chains and no longer exhibits an ordered phase at any finite reservoir temperature. The behavior in Fig.~\ref{fig:periodic_drive}(b) is consistent with our analysis using the Floquet Hamiltonian $H_F^{(0)}$, since the dimensional crossover occurs precisely at the zeroes of the Bessel function.
Finally, note that, even though periodic drives typically lead to enhanced energy absorption, the heating rates are either power-law or exponentially suppressed with increasing the drive frequency $\omega$~\cite{mcroberts2023prethermalization}.

In summary, Floquet-Hamiltonian engineering based on the framework from Sec.~\ref{sec:H_ext} provides enhanced control over model parameters in nonreciprocal systems, enabling access to wider regions of their phase diagrams and allowing one to probe dimensional crossovers. 
More broadly, the Hamiltonian embedding -- of which computing the Floquet Hamiltonian is one application -- unlocks generic Hamiltonian methods for nonreciprocal systems. We anticipate its symplectic structure to similarly enable the transfer of other techniques to nonreciprocal systems.

\section{\label{sec:outro}Discussion and Outlook}

In the past, isolated examples of Hamiltonian descriptions of nonreciprocal interactions have been reported, e.g.\ for the Lotka-Volterra predator-prey model~\cite{nutku1990hamiltonian}. 
In this work, we have constructed a general Hamiltonian embedding for nonreciprocal systems by introducing auxiliary degrees of freedom that couple to the original ones in a reciprocal way. By imposing a constraint on the phase-space dynamics of the embedding, we preserve the total number of independent degrees of freedom and recover the original nonreciprocal interactions. 
As a direct consequence of our construction, it follows that the phase space of the reciprocal Hamiltonian $H$ from Eq.~\eqref{eq:H_tot} contains a nonreciprocal submanifold defined by the constraint, which is closed under the dynamics.

Crucially, our construction works for pairwise interactions, and may not apply to arbitrary non-potential systems [cf.~Fig.~\ref{fig:classes_of_systems}]; a generalization to velocity-dependent (e.g., Lorentz-like) forces is presented in~\cite{SI}.
Unlike in systems where nonreciprocity arises as a result of projecting/tracing out bath degrees of freedom that mediate the interactions (e.g., a chemical concentration field)~\cite{bowick2022symmetry}, we emphasize that here the auxiliary degree of freedom does not play the role of a bath for the original one; instead, the projecting/tracing out procedure is replaced by the constraint. 
The framework is generic and applies to both inertial and dissipative dynamics featuring nonreciprocal pairwise interactions.

Our Hamiltonian embedding advances a century-old effort of using auxiliary degrees of freedom to describe generic nonconservative forces. First attempts date back to Bateman~\cite{bateman1931dissipative}, where a Lagrangian was constructed for a dissipative oscillator; however, these early results cannot be generalized beyond harmonic oscillators as they neglect the dynamical constraint from Eq.~\eqref{eq:constraint}.
Markovian embeddings have also been frequently used to describe and simulate non-Markovian systems with memory kernels~\cite{Siegle2010markovian,jung2018generalized,doerries2021correlation}.
More recently, Galley presented a general formalism to describe nonconservative forces by formulating Hamilton's principle as an initial value problem~\cite{galley2013classical}: it requires doubling the degrees of freedom and relies on an educated guess for the interaction term that couples them. This approach produces the correct nonreciprocal equations of motion, when supplemented with our Hamiltonian~\eqref{eq:H_tot}. A key contribution of our work is the generic construction of the interaction energy between the original and the auxiliary degrees of freedom for any pairwise nonreciprocal interaction; previous literature has not discussed nonreciprocal interactions within a Hamiltonian framework. 
Critically, our approach differs from Galley’s in the implementation: Galley imposes an equality condition on the degrees of freedom at the \textit{final} time, from which the so-called physical limit~\cite{galley2013classical} arises by setting the boundary terms in the variational principle to vanish; the conceptual difference in our approach is the \textit{constraint} that we impose at the \textit{initial} time, which is then automatically preserved by the dynamics at all times.

We discussed two applications using XY spins on a square lattice with vision-cone interactions:
(i) We proved the equivalence between Glauber dynamics based on the Hamiltonian embedding and the original nonreciprocal Langevin dynamics. In particular, we showed that both procedures yield the same critical temperature for the transition between the ferromagnetic and non-magnetized phases. These results open the door to investigating the nature of this and similar phase transitions in large systems using fast Monte-Carlo algorithms.
(ii) We make use of the inherent symplectic structure of the embedding to derive an effective Hamiltonian for periodically driven vision-cone XY spins using Floquet theory; this allows us to interpret the disappearance of ferromagnetic order when varying the drive amplitude as a dimensional crossover. These results suggest exciting possibilities for experiments to apply Floquet engineering techniques~\cite{bukov2015universal, goldman2014periodically, eckardt2017colloquium} to investigate regimes in parameter space that are otherwise inaccessible in the nonreciprocal system.

We note that the Hamiltonian embedding does not introduce new physics content per se beyond what is already captured by the equations of motion defining the original nonreciprocal system. However, it enables the study of nonreciprocal systems using analytical and numerical tools from equilibrium or nonequilibrium physics that require a Hamiltonian structure.
For instance, one can now use Monte-Carlo algorithms that shortcut long transients of Langevin simulations and lead directly to the steady state. 
Numerical techniques based on the embedding may require keeping track of twice the number of original degrees of freedom; however, the numerical complexity of any algorithm remains unchanged by adding the auxiliary degrees of freedom (which effectively doubles the system size).
Likewise, one can apply various canonical transformations (e.g., Schrieffer-Wolff-like transformations such as the Birkhoff-Gustavson normal form~\cite{birkhoff1927dynamical,gustavson1966on,schrieffer1966relation}) 
to identify fast and slow effective degrees of freedom in nonreciprocal models.
An illustrative example is the Floquet-Magnus expansion for periodically driven systems: higher-order terms in the inverse drive frequency contain effective spin-spin interactions~\cite{bukov2015universal}. Note that whereas it is possible to derive effective stroboscopic equations of motion in the high-frequency regime directly from the equations of motion to any order in the inverse frequency $\omega^{-1}$ by using two-times perturbation theory~\cite{strogatz2018nonlinear}, the Hamiltonian formulation has the advantage of providing a systematic model-agnostic algorithmic procedure. This formalism can prove useful to analyze the behavior of time-crystalline nonreciprocal states~\cite{hanai2024nonreciprocal}.

The Hamiltonian framework we propose has the potential to expand the scope of research on nonreciprocal systems and enable novel applications. In experiments that simulate classical and quantum models~\cite{georgescu2014quantum,mcdonald2018phase,brandenbourger2019nonreciprocal,Wang2019non}, our construction can offer a new way to engineer nonreciprocal dynamics using reciprocal Hamiltonian systems subject to a constraint. 
Promising future directions include developing kinetic Monte-Carlo schemes to capture non-steady states \cite{landau2021guide}, canonically quantizing the Hamiltonian embedding, or using it to construct
statistical mechanics and thermodynamic descriptions~\cite{zhang2023entropy,loos2020irreversibility} of nonreciprocal systems.

\section*{Methods}

\subsection*{Equivalence between Constrained Glauber Monte-Carlo and Langevin dynamics}

The constrained Glauber Monte-Carlo algorithm with transition rates defined using the Hamiltonian embedding and the Langevin equation with generic nonreciprocal interactions, share the same Fokker-Planck equation. This equation describes the configuration space density distribution $\rho(\bm{\theta}, t)$, with $\bm{\theta}=(\theta_i, \dots, \theta_{L^2})$. We now demonstrate this equivalence explicitly.

First, we sketch the derivation of the Fokker-Planck equation from the Langevin equation. Denoting the force acting on spin $\theta_i$ by $F_i(\bm{\theta})$, and using It\^o formalism, the equation of motion is 
\begin{equation}
	\label{eq:Ito-EOM}
	\theta_i(t+\Delta t)-\theta_i(t)
	=
	F_i(\bm{\theta})\,\Delta t + \sqrt{2T}\,\Delta W_i(t)\; ,
\end{equation}
where $\Delta W_i(t)$ is a Wiener increment obeying $\langle \Delta W_i\rangle=0$ and
$\langle \Delta W_i\,\Delta W_j\rangle=\delta_{ij}\Delta t$.
Following the standard steps~\cite{SI}, we arrive at the Fokker-Planck equation for the  configuration-space probability density $\rho(\bm{\theta},t)$:
\begin{equation}
	\label{eq:Fokker-Planck}
	\partial_t \rho(\bm \theta ,t)
	=
	-\sum_i \partial_{\theta_i}\left[F_i(\bm\theta)\,\rho(\bm\theta ,t)\right]
	+T\sum_i \partial_{\theta_i}^2 \rho(\bm\theta ,t).
\end{equation}

In what follows, we independently derive Eq.~\eqref{eq:Fokker-Planck} from the Glauber dynamics with transition probability rates
\begin{eqnarray}
	\label{eq:Glauber_embeddong_rates}
	w(\theta_i\to\theta_i')&=&\frac{1}{2}\left(1-\tanh\frac{\Delta E(\theta,
		\theta')}{2T}\right),\\
	\Delta E(\theta,
	\theta')
	&=& \frac{1}{2}\big(H(\theta',\varphi)-H(\theta,\varphi')\big)\big|_\text{constr.($\theta,\varphi$)}
\end{eqnarray}
defined by eliminating the auxiliary degree of freedom from the Hamiltonian embedding by enforcing the constraint~\cite{SI}.
We note that $T$ here is the temperature of the bath; the nonreciprocal system of interest is never in a thermal state at any point during its evolution.

In each Glauber update step $n$, we pick a spin at random and update it according to $\theta_j\to\theta_j'=\theta_j+\delta$. 
The probability of selecting spin $j$ is uniformly distributed on the square lattice of linear size $L$, and is hence given by $1/L^2$. At the same time, the update angle $\delta$ is drawn uniformly from the interval $[-\Delta,\Delta]$, with probability density $1/(2\Delta)$.

The discrete-time master equation governing these Glauber dynamics is then given by
\begin{eqnarray}
	&&\rho (\bm{\theta}, n+1) - \rho (\bm{\theta}, n) = 
	\frac{1}{2\Delta}
	\int_{-\Delta}^{\Delta} \mathrm d \delta \frac{1}{L^2}\sum_i   \Big[
	\\
	&& \rho(\theta_i+\delta,\{\bm{\theta}\}_{j\neq i},n)\; 
	w (\theta_i {+} \delta \to \theta_i)
	{-} \rho (\bm{\theta}, n) \; 
	w (\theta_i \to \theta_i {+} \delta) \Big]. \nonumber
\end{eqnarray}
Treating $\Delta \ll 1$ as a small parameter, we can expand the integrand to leading order in $\delta$ using
\begin{eqnarray}
	\rho(\theta_i{+}\delta,\{\bm{\theta}\}_{j\neq i},n) 
	&{=}&
	\rho(\bm{\theta},n) {+} \partial_{\theta_i} \rho(\bm{\theta},n) \delta  {+}
	\frac{1}{2} \partial_{\theta_i}^2 \rho(\bm{\theta},n) \delta^2 \nonumber \\
	&&{+} \mathcal{O}(\delta^3)\; , \nonumber\\
	w(\theta_i {\to} \theta_i {+} \delta)&{=}& \frac{1}{2} {+} \frac{1}{4T} F_i(\bm{\theta}) \delta {+} \frac{1}{8T} \partial_{\theta_i} F_i(\bm{\theta}) \delta^2  \nonumber \\
	&&{+} \mathcal{O}(\delta^3)\; , \nonumber\\
	w(\theta_i{+} \delta\to\theta_i)&{=}&\frac{1}{2} {-} \frac{1}{4T} F_i(\bm{\theta}) \delta  {-} \frac{1}{8T} \partial_{\theta_i} F_i(\bm{\theta}) \delta^2 \nonumber \\
	&&+ \mathcal{O}(\delta^3)\; , 
\end{eqnarray} 
where the force acting on spin $i$ is obtained from the embedding Hamiltonian under the constraint as 
\begin{eqnarray}
	F_i(\bm{\theta}) &=& - \frac{1}{2}\left(\partial_{\theta_i} - \partial_{\varphi_i}\right)H(\bm{\theta},\bm{\varphi})\big|_\text{constr.($\theta,\varphi$)} \nonumber\\
	&=& - \partial_{\theta_i}H(\bm{\theta},\bm{\varphi})\big|_\text{constr.($\theta,\varphi$)}\; .
\end{eqnarray}
We emphasize that one first computes the derivatives and only then applies the constraint. 

Noticing that linear terms in $\delta$ vanish since the integral limits are symmetric, we obtain to leading order in $\Delta$:
\begin{eqnarray}
	&&\rho (\bm{\theta}, n+1) {-} \rho (\bm{\theta}, n) \approx   \\
	&& \frac{1}{2\Delta L^2}
	\int_{-\Delta}^{\Delta} \delta^2 \mathrm d \delta  \sum_i  \frac{1}{4T} \partial_{\theta_i}\Big[
	-\rho(\bm{\theta}, n)F_i(\bm{\theta}) 
	+ T \partial_{\theta_i} \rho(\bm{\theta}, n)
	\Big]  \nonumber  \\
	&&= \frac{1}{L^2}\frac{\Delta^2}{12T} \sum_i
	\partial_{\theta_i}\Big[
	-\rho(\bm{\theta}, n)F_i(\bm{\theta}) 
	+ T \partial_{\theta_i} \rho(\bm{\theta}, n)
	\Big] + \mathcal{O}(\Delta^4). \nonumber
\end{eqnarray}

Recalling that in Eq.~\eqref{eq:Ito-EOM} all $L^2$ spins get updated simultaneously at each time step, we define dimensionless Monte-Carlo time as
\begin{equation}
	\label{eq:MC_time}
	t_{\textrm{MC}}=\frac{n}{L^2}.
\end{equation}
In the limit of large system sizes, $L^2\gg 1$, the corresponding time increment $\mathrm{d} t_\mathrm{MC} = 1/L^2$ is small, and Monte-Carlo time becomes continuous; we obtain
\begin{equation}
	\rho (\bm{\theta}, n+1) - \rho (\bm{\theta}, n)
	= \partial_{t_\mathrm{MC}}\rho(\bm{\theta}, t_{\mathrm{MC}})\; \mathrm{d} t_\mathrm{MC}.
\end{equation}
This allows us to formally define a continuous-time limit for the master equation:
\begin{eqnarray}
	\label{eq:MC_rho_vs_time}
	\frac{\partial \rho(\bm{\theta}{,} t_{\textrm{MC}})}{\partial t_{\textrm{MC}}} 
	&{=}& \frac{\Delta^2}{12T} \sum_i
	\partial_{\theta_i}\Big[
	{-}\rho(\bm{\theta}{,} t_\mathrm{MC})F_i(\bm{\theta}) 
	{+} T \partial_{\theta_i} \rho(\bm{\theta}{,}  t_ \mathrm{MC})
	\Big] \nonumber\\
	&& + \mathcal{O}(\Delta^4).
\end{eqnarray}
In the small-update regime $\Delta\ll 1$, Eq.~\eqref{eq:MC_rho_vs_time} is equivalent to the Fokker-Planck equation derived earlier for the Langevin dynamics, cf.~Eq.~\eqref{eq:Fokker-Planck}, except for the prefactor $\Delta^2/(12T)$.

In units where the friction coefficient is set to unity, it follows from the equations of motion that energy has the same unit as frequency (or inverse time) [see, e.g., Eq.~\eqref{eq:VC-XY_Langevin_EOM}]. Since the only model-independent energy scale in the Glauber rate $w$ is the temperature $T$ (which also has units of energy), the physical time unit in Eq.~\eqref{eq:MC_rho_vs_time} is set by the inverse bath temperature $T$. Hence, Monte-Carlo time $t_\text{MC}$ is related to the physical (Langevin) time $t_\text{Lan}$ by
\begin{equation}
	t_\text{Lan} = \frac{\Delta^2}{12 T}t_\text{MC}.
\end{equation}

Therefore, starting with Glauber dynamics using the transition rates obtained from the Hamiltonian embedding, we derived the Fokker-Planck equation~\eqref{eq:Fokker-Planck} which describes the Langevin dynamics of the original nonreciprocal system. This establishes the equivalence between the two approaches. We present a detailed formal derivation that uses the extended embedding variables in the SM~\cite{SI}.

Finally, we note that the derivation above does not make use of the details of a particular model. For as long as one can construct a Hamiltonian embedding, the latter can be used to define transition rates $w$ using Eq.~\eqref{eq:Glauber_embeddong_rates}, and the equivalence of the Langevin and Monte-Carlo descriptions follows. 

\subsection*{Hamiltonian embedding for chase \& run spin model}

Consider XY spins on a square lattice of linear size $L$ with periodic boundary conditions. Denoting the lattice sites by $i{=}(x,y)$, the corresponding Langevin equation of motion for the chase-and-run spin model discussed in the main text reads as
\begin{eqnarray}
	\label{eq:Langevin_chiral}
	\dot{\theta}_{x,y} &=& - J_{\rightarrow} \sin(\theta_{x,y} - \theta_{x+1,y}) - J_{\leftarrow} \sin(\theta_{x,y} - \theta_{x-1,y}) \nonumber \\
	&& - J_\text{rec} [\sin(\theta_{x,y} - \theta_{x,y+1})+ \sin (\theta_{x,y} - \theta_{x,y-1})] \nonumber\\
	&& + \sqrt{2T} \eta_{x,y}(t) \, .
\end{eqnarray}
Here $J_{\leftarrow}{\neq} J_{\rightarrow}$ are the nonreciprocal interactions along the $x$-direction, and $J_\text{rec}$ denotes the reciprocal interactions along the $y$-direction (see main text). Temperature is denoted by $T$, and $\eta_{x,y}(t)$ adds white noise to the dynamics of each spin, cf.~Eq.~\eqref{eq:VC-XY_Langevin_EOM}. 

The spins are initialized in the perfect stripe state defined by
\begin{equation}
	\theta_{x,y} =
	\left\{  
	\begin{array}{ll}
		0, & \text{for}\;  x \in \left(\frac{L}{4},\frac{3L}{4}\right) \\ \frac{\pi}{2}, & \mathrm{else.}
	\end{array}
	\right. \,
\end{equation}
which is perturbed by the noise in the dynamics. 

The Hamiltonian embedding for the nonreciprocal XY spins in Eq.~\eqref{eq:Langevin_chiral} reads as
\begin{eqnarray}
	\label{eq:Hamiltonian_chiral}
	H_\text{tot} &=& H_{SS} + H_{Sa} - H_{aa},  \\
	H_{SS}\!&{=}&\!\! {-}\!\!   \sum_{ x,y}(J_{\rightarrow} {+} J_{\leftarrow})[ \cos(\theta_{x,y} {-} \theta_{x+1,y}) {+}\cos(\theta_{x,y} {-} \theta_{x-1,y})  ] \nonumber \\ 
	&& \qquad + J_\text{rec}[\cos(\theta_{x,y} {-} \theta_{x,y+1}) {+} \cos(\theta_{x,y} - \theta_{x,y-1})] ,  \nonumber \\
	H_{Sa} \!&{=}&\!\! {-}\!\! \sum_{ x,y} J_{\leftarrow} \cos(\theta_{x,y} {-} \varphi_{x+1,y}) {+} J_{\rightarrow} \cos(\theta_{x,y} {-} \varphi_{x-1,y}), \nonumber \\
	H_{aa} \!&{=}&\!\! {-}\!\! \sum_{ x,y} J_\text{rec}[\cos(\varphi_{x,y} {-} \varphi_{x,y+1}) {+} \cos(\varphi_{x,y} {-} \varphi_{x,y-1})] , \nonumber
\end{eqnarray}
where $\{\theta_{x,y}, \varphi_{x',y'} \} = \delta_{x,x'}\delta_{y,y'}$ are conjugate variables. 

A straightforward calculation shows that the Hamilton equations of motion derived from Eq.~\eqref{eq:Hamiltonian_chiral} reduce to the Langevin equation~\eqref{eq:Langevin_chiral}, when subject to the constraint
\begin{equation}
	\label{eq:constraint_2}
	\theta_{x,y} -\varphi_{x,y} = \pi\, . 
\end{equation}
Thus, the energy difference used in the Glauber update rule can be evaluated following Eq.~\eqref{eq:Glauber_rate} as
\begin{eqnarray}
	\label{eq:DeltaE_pattern}
	\Delta E(\theta_{x,y},\theta'_{x,y}) &{=}&
	{-} J_\rightarrow [\cos(\theta_{x,y}'{-} \theta_{x{+}1,y}){-} \cos(\theta_{x,y}{-} \theta_{x{+}1,y})] \notag \\
	&&{-} J_\leftarrow [\cos(\theta_{x,y}'{-} \theta_{x-1,y}){-} \cos(\theta_{x,y}{-} \theta_{x-1,y})] \notag \\
	&&{-} J_\mathrm{rec} \sum_{\sigma = \pm 1} [ \cos(\theta_{x,y}' {-} \theta_{x,y+\sigma } ) \notag \\
	&&\qquad\qquad\quad {-}  \cos(\theta_{x,y} {-} \theta_{x,y+\sigma } ) ] \; .
\end{eqnarray}
We use this expression in the constrained Glauber simulations of the chase-and-run dynamics. 

\emph{data availability.---}%
The data associated with this Article are available via Zenodo at https://doi.org/10.5281/zenodo.19427733 (Ref.~\cite{zenodo_repo})

\emph{code availability.---}%
The codes associated with this manuscript version are available via Zenodo at https://doi.org/10.5281/zenodo.19427733 (Ref.~\cite{zenodo_repo})

\emph{Acknowledgments.---}%
We thank L.~Hahn, A.~McRoberts, M.~Sonner, Y.~Zhang, and H.~Zhao for related discussions. 
MB was funded by the European Union (ERC, QuSimCtrl, 101113633).
RA acknowledges funding from the European Union through the ERC Starting Grant "Living Fluctuations" (No. 101114584).
YS sincerely thanks Hongzheng Zhao for hosting during the visit at Peking University.
Views and opinions expressed are however those of the authors only and do not necessarily reflect those of the European Union or the European Research Council Executive Agency. Neither the European Union nor the granting authority can be held responsible for them. This work was in part supported by the Deutsche
Forschungsgemeinschaft via Research Unit FOR 5522 (project-id 499180199) and the Würzburg-Dresden Cluster of Excellence on Complexity, and Topology and Dynamics in Quantum Matter - ctd.qmat (EXC 2147, project ID 390858490).

\emph{Author contributions.---}
R.A., M.B., and R.M.~conceived the research.
Y.S.~did the analytical calculations and performed the numerical simulations. 
R.A.~and M.B.~supervised the work on the project, and contributed equally to this work. 
All authors contributed to writing the manuscript. 

\emph{Competing Interests Statement.---}
The authors declare no competing interests.

\else 
\fi 

\resumetoc
\ifsuppmode

\cleardoublepage

\renewcommand{\thefigure}{S\arabic{figure}}
\setcounter{figure}{0}
\renewcommand{\theequation}{S.\arabic{equation}}
\setcounter{equation}{0}
\renewcommand{\thesection}{S\arabic{section}}
\setcounter{section}{0}

\onecolumngrid
\begin{center}
	\textbf{\large{\textit{\nqdcolor{Supplemental Material}}}}\\
	\hfill \break
\end{center}
\twocolumngrid

\ifcombinedmode
\else
\maketitle
\fi 

\begin{widetext}
	\setcounter{figure}{0}
	\begingroup
	\makeatletter
	\def\ltxu@dotsep{0.5em}%
	\tableofcontents
	\endgroup
	
	\section{Constrained Hamiltonian embedding of nonreciprocal systems}
	\label{sec:embedding-proof}
	
	In this section, we first outline a general approach for constructing Hamiltonian embeddings, focusing on velocity-independent interaction forces. We then discuss the more general case of velocity-dependent interactions.  
	
	\subsection{Velocity-independent interaction forces}
	\label{sec:velocity_independent_interaction_forces}
	We begin by emphasizing that, in its simplest form, (non-)reciprocity is a property defined for \textit{pairwise interacting} systems. Let us denote by ${\bf F}_{i\rightarrow j}(\mathbf{r}_i, \mathbf{r}_j)$ the force exerted by particle $i$ at position $\mathbf{r}_i$ on particle $j$ at position $\mathbf{r}_j$. We consider translation invariant systems where ${\bf F}_{i\rightarrow j}(\mathbf{r}_i, \mathbf{r}_j) = {\bf F}_{i\rightarrow j}(\mathbf{r}_i - \mathbf{r}_j)$ depends only on the distance between the two particles involved. In addition, in this first section, we only consider momentum-independent forces, and thus leave out of consideration Lorentz-like forces.
	
	Reciprocal (R) interactions obey Newton's third law 
	\begin{equation}
		\label{eq:R-force}
		{\bf F}^\text{R}_{i\rightarrow j}(\mathbf{r}_i - \mathbf{r}_j) = -{\bf F}^\text{R}_{j\rightarrow i}(\mathbf{r}_j - \mathbf{r}_i),
	\end{equation}
	i.e, the force exerted by particle $i$ on particle $j$, ${\bf F}^\text{R}_{i\rightarrow j}(\mathbf{r}_i - \mathbf{r}_j)$, is equal in magnitude and opposite in direction to the force ${\bf F^\text{R}}_{j\rightarrow i}(\mathbf{r}_j - \mathbf{r}_i)$ exerted by particle $j$ on particle $i$. 
	
	We define nonreciprocal interactions via the relation
	\begin{eqnarray}
		\label{eq:NR-force}
		{\bf F}^\text{NR}_{i\rightarrow j}(\mathbf{r}_i - \mathbf{r}_j) \neq  -{\bf F}^\text{NR}_{j\rightarrow i}(\mathbf{r}_j - \mathbf{r}_i).
	\end{eqnarray}
	Nonreciprocal interactions refer to an asymmetry in the interaction between two particles in which the action and reaction are not equal. 
	
	In general, the equations of motion for particle $i$ may be governed by a combination of reciprocal and nonreciprocal interactions. Hence in its most general form, it takes the form:
	\begin{eqnarray}
		\label{Seq:EOM_general}
		m\ddot{\mathbf{r}}_i +\gamma \dot{\mathbf{r}}_i = \sum_{j \neq i}
		\left[\mathbf{F}_{j\rightarrow i}^\mathrm{NR}(\mathbf{r}_i - \mathbf{r}_j) +
		\mathbf{F}_{j\rightarrow i}^\mathrm{R}(\mathbf{r}_i - \mathbf{r}_j)\right],
	\end{eqnarray}
	where we denote the mass by $m$ and the damping factor by $\gamma$.

	We now assume that the interaction forces can be derived from a one-way interaction potential:
	\begin{equation}
		\label{seq:force-potential-independent}
		{\bf F}_{i\rightarrow j}(\mathbf{r}_i - \mathbf{r}_j) = - \frac{\partial}{\partial \mathbf{r}_i} U_{i\rightarrow j}(\mathbf{r}_i - \mathbf{r}_j).
	\end{equation}
	Reciprocal interactions are conservative and share the same (mutual) interaction potential, and hence obey the relation $U^\text{R}_{i\rightarrow j}=U^\text{R}_{j\rightarrow i}$ which, therefore, implies Newton's third law (Eq.~\eqref{eq:R-force}); it is common to use the simplified notation $U^\text{R}_{i\rightarrow j}=U^\text{R}$. 
	By contrast, nonreciprocal interactions obey $U^\text{NR}_{i\rightarrow j}\neq U^\text{NR}_{j\rightarrow i}$, and are hence nonconservative. Since each interaction function $U^\text{NR}_{i\rightarrow j}(\mathbf{r}_i - \mathbf{r}_j)$, $U^\text{NR}_{j\rightarrow i}(\mathbf{r}_i - \mathbf{r}_j)$ depends on the positions of both particles involved, one cannot write down an energy function such that the Hamiltonian equations of motion for the corresponding conjugate variables are nonreciprocal; this is because the resulting force acting on particle $i$ will inevitably contain the gradients of both interaction potentials, which would make the net force reciprocal.

	To resolve this problem, we introduce a copy of the system described by the
	auxiliary degrees of freedom $\{\mathbf{x}_{i}, \mathbf{p}^\mathbf{x}_j\}=\delta_{ij}$. The original degrees of freedom obey the Poisson bracket $\{\mathbf{r}_{i}, \mathbf{p}^\mathbf{r}_j\}=\delta_{ij}$. One can construct a general Hamiltonian embedding for the joint system using only reciprocal interactions, as
	follows: 
	\begin{eqnarray}
		H &=&\phantom{+}\sum_{j}\frac{\left( \mathbf{p}_{j}^{\mathbf{r}}-\frac{%
				\gamma }{2}\mathbf{x}_{j}\right) ^{2}}{2m}-\frac{\left( \mathbf{p}_{j}^{%
				\mathbf{x}}+\frac{\gamma }{2}\mathbf{r}_{j}\right) ^{2}}{2m}  \notag \\
		&&+\sum_{\langle ij\rangle }\left[ U_{i\rightarrow j}^{\text{NR}}(\mathbf{r}%
		_{i}-\mathbf{r}_{j})+U_{j\rightarrow i}^{\text{NR}}(\mathbf{r}_{j}-\mathbf{r}%
		_{i})-U_{i\rightarrow j}^{\text{NR}}(\mathbf{x}_{j}-\mathbf{r}%
		_{i})-U_{j\rightarrow i}^{\text{NR}}(\mathbf{x}_{i}-\mathbf{r}_{j})\right] 
		\notag \\
		&&+\sum_{\langle ij\rangle } \left [ U^{\text{R}}(\mathbf{r}_{i}-\mathbf{r}_{j})-U^{%
			\text{R}}(\mathbf{x}_{i}-\mathbf{x}_{j}) \right ].
	\end{eqnarray}%
	The notation $\langle ij \rangle$ denotes a summation over pairs of particles $i$ and $j$ (each counted once). 
	Here the kinetic energy is quadratic in the momenta $\mathbf{p}_{j}^{\mathbf{%
			r}},\mathbf{p}_{j}^{\mathbf{x}}$, and hence the system is inertial and the
	EOM is second order in time. 
	Note the apparent similarity of the kinetic energy above to that of a particle in a magnetic field; 
	however, a notable difference is that, in the Hamiltonian $H$ above, the minimal coupling is between $(\mathbf{p}_{j}^{\mathbf{r}}, \mathbf{x}_{j})$ [as opposed to $(\mathbf{p}_{j}^{\mathbf{r}}, \mathbf{r}_{j})$], etc.

	As we showed in the Main Text, in general, the equations of motion derived from the
	Hamiltonian $H$ defines a coupled system of equations 
	\begin{eqnarray}
		\frac{\mathrm d \mathbf{r}_{i}}{\mathrm d t} &=&\frac{\mathbf{p}_{i}^{\mathbf{r}}-\frac{\gamma }{2}%
			\mathbf{x}_{i}}{m}\ ,  \notag \\
		\frac{\mathrm d \mathbf{p}_{i}^{\mathbf{r}}}{\mathrm d t}  &=&\frac{\gamma }{2}\frac{\left( \mathbf{p}%
			_{i}^{\mathbf{x}}+\frac{\gamma }{2}\mathbf{r}_{i}\right) }{m}-\frac{\partial}{\partial{
				\mathbf{r}_{i}}}\sum_{j\in \langle ij\rangle }\left[ U_{i\rightarrow j}^{%
			\text{NR}}(\mathbf{r}_{i}-\mathbf{r}_{j})+U_{j\rightarrow i}^{\text{NR}}(%
		\mathbf{r}_{j}-\mathbf{r}_{i})-U_{i\rightarrow j}^{\text{NR}}(\mathbf{x}_{j}-%
		\mathbf{r}_{i})+U^{\text{R}}(\mathbf{r}_{i}-\mathbf{%
			r}_{j})\right]\ ,   \notag \\
		\frac{\mathrm d \mathbf{x}_{i}}{\mathrm d t}  &=&-\frac{\mathbf{p}_{i}^{\mathbf{x}}+\frac{\gamma }{2}%
			\mathbf{r}_{i}}{m}\ ,  \notag \\
		\frac{\mathrm d \mathbf{p}_{i}^{\mathbf{x}}}{\mathrm d t}  &=&\frac{\gamma }{2}\frac{\left( \mathbf{p}%
			_{i}^{\mathbf{r}}-\frac{\gamma }{2}\mathbf{x}_{i}\right) }{m}+\frac{\partial}{\partial{\mathbf{x}_{i}}}\sum_{j\in \langle ij\rangle }\left[ U_{j\rightarrow i}^{%
			\text{NR}}(\mathbf{x}_{i}-\mathbf{r}_{j})+U^{\text{R}}(\mathbf{x}_{i}-%
		\mathbf{x}_{j})\right] .
	\end{eqnarray}%
	However, if we impose the constraint 
	\begin{equation}
		\mathbf{r}_{i}(0)-\mathbf{x}_{i}(0)=0,\qquad \mathbf{p}_{i}^{\mathbf{r}}(0)+%
		\mathbf{p}_{i}^{\mathbf{x}}(0)=0,
	\end{equation}
	at time $t=0$, it follows immediately from the above equations of motion that the constraint is preserved at all times, i.e.,
	\begin{equation}
		\label{eq:constr_t}
		\frac{\mathrm d}{\mathrm d t}\left[{\mathbf{r}}_{i}(t)-{ \mathbf{x}}_{i}(t)\right]\bigg|_\text{constr.}=0,
		\qquad 
		\frac{\mathrm d}{\mathrm d t}\left[{ \mathbf{p}}_{i}^{\mathbf{r}}(t)+{\mathbf{p}}_{i}^{\mathbf{x}}(t)\right]\bigg|_\text{constr.}=0.
	\end{equation}
	Moreover, the joint equations of motion reduce to the original nonreciprocal dynamics in Eq.~%
	\eqref{Seq:EOM_general}. More precisely, each degree of freedom (i.e., both
	the $\mathbf{r}_{j}$ and the $\mathbf{x}_{j}$ degree of freedom) follows the
	same equations of motion; the dynamics of the $\mathbf{x}_{j}$ degree of freedom perfectly mirrors that of $\mathbf{r}_{j}$.
	We note that the negative kinetic energy for the auxiliary $\mathbf{x}_{j}$ is a sufficient condition to preserve the constraint at all times.
	
	In Sec.~\ref{sec:examples} below, we give the explicit form of the constrained Hamiltonian for various models with nonreciprocal equations of motion.

	\subsection{Velocity-dependent interaction forces}
	\label{subsec:velocity_dependent_interactions}
	
	In Sec.~\ref{sec:velocity_independent_interaction_forces}, we restricted our attention to velocity-independent interactions. We now consider pairwise interactions that may depend on particle velocities, while still requiring the equations of motion to remain second order in time. This requirement severely constrains the admissible form of the interaction Lagrangian and will be the starting point for the embedding constructions discussed below.
	
	\subsubsection{General form of velocity-dependent Lagrangians with reciprocal interactions}
	\label{sec:general_form_of_velocity_dependent_Lagrangian}
	
	We begin by reviewing a few important facts about velocity-dependent reciprocal interactions. We will then build on these ideas to define the embedding for their nonreciprocal counterparts. 
	
	Consider an inertial system of $N$ particles with coordinates $\{\mathbf r_i\}$ and Lagrangian
	\begin{equation}
		\mathcal{L}(\{\mathbf r\},\{\dot{\mathbf r}\})
		=
		\sum_{i=1}^N \frac{m_i}{2}\dot{\mathbf r}_i^2
		-
		\mathcal{U}(\{\mathbf r\},\{\dot{\mathbf r}\}),
	\end{equation}
	where $\mathcal{U}$ denotes the interaction energy. The Euler--Lagrange equations read as
	\begin{equation}
		m_i\ddot{\mathbf r}_i
		=
		-\partial_{\mathbf r_i}\mathcal{U}
		+
		\frac{\mathrm d}{\mathrm dt}\partial_{\dot{\mathbf r}_i}\mathcal{U}
		\equiv \mathbf F_i\; ,
	\end{equation}
	which defines the total force $\mathbf F_i$ acting on particle $i$.
	In general, the total-time-derivative term contains accelerations through
	$\left(\partial_{\dot{\mathbf r}_i}\partial_{\dot{\mathbf r}_j}\mathcal{U}\right)\ddot{\mathbf r}_j$, which is seen by carrying out the total time derivative.
	If we require the force $\mathbf F_i$ to depend only on positions and velocities, and not explicitly on accelerations, these terms must vanish identically along arbitrary solutions of the equations of motion. Therefore,
	\begin{equation}
		\label{seq:velocity_rescriction}
		\partial_{\dot{\mathbf r}_i}\partial_{\dot{\mathbf r}_j}\mathcal{U}=0
		\qquad \forall\, i,j.
	\end{equation}
	Hence, the interaction part of the Lagrangian can be at most linear in the velocities, and in its full generality takes the form
	\begin{equation}
		\mathcal{U}(\{\mathbf r\},\{\dot{\mathbf r}\})
		=
		\phi(\{\mathbf r\})
		+
		\sum_j \mathbf A_j(\{\mathbf r\})\cdot \dot{\mathbf r}_j,
	\end{equation}
	where we introduced the scalar potential $\phi$, and $\mathbf A_j$ are generalized vector potentials. Here, the notation and physical interpretation are borrowed from electromagnetism.
	
	Substituting this form into the Euler--Lagrange equations yields
	\begin{equation}
		\label{seq:velocity_force}
		\mathbf F_i
		=
		-\partial_{\mathbf r_i}\phi
		-\sum_j \partial_{\mathbf r_i}\!\left[\mathbf A_j\cdot \dot{\mathbf r}_j\right]
		+\sum_j (\dot{\mathbf r}_j\cdot \partial_{\mathbf r_j})\mathbf A_i .
	\end{equation}
	It is convenient to write the force in the $\alpha$-th direction in the linear form
	\begin{equation}
		\label{seq:decomposed_force}
		(\mathbf F_i)_\alpha
		=
		V_{i\alpha}(\{\mathbf r\})
		+
		\sum_{j,\beta} T_{i\alpha}^{\,j\beta}(\{\mathbf r\})\,\dot {{r}}_{j\beta},
	\end{equation}
	with the definitions
	\begin{equation}
		\label{seq:tensor_and_v}
		V_{i\alpha}=-\partial_{{r}_{i\alpha}}\phi,
		\qquad
		T_{i\alpha}^{\,j\beta}
		=
		-\partial_{{r}_{i\alpha}}(\mathbf A_j)_\beta
		+
		\partial_{{r}_{j\beta}}(\mathbf A_i)_\alpha .
	\end{equation}
	Conversely, if the force has this form and satisfies a set of integrability conditions, one can reconstruct the potentials locally. 
	First, since ${V}_{i\alpha} = - \partial_{{r}_{i\alpha}} \phi$, the field must be a gradient field in order for $\phi$ to be well defined, i.e., $\partial_{{r}_{j\beta} }{V}_{i \alpha}  = \partial_{{r}_{i\alpha} }{V}_{j \beta} $. 
	Second, since the tensor $T_{i\alpha}^{j\beta}$ already satisfies the closure (Bianchi-type) condition, $\partial_{ r_{k\gamma}}T_{i\alpha}^{j\beta}+\partial_{ r_{i\alpha}}T_{k\gamma}^{j\beta}+\partial_{ r_{j\beta}}T_{i\alpha}^{k\gamma}=0$, the vector potential $\mathbf A_i$ can be reconstructed from the force in Eq.~\eqref{seq:decomposed_force} by integrating $T_{i\alpha}^{j\beta}$ along a path $\Gamma$ in configuration space.
	In particular, 
	\begin{equation}
		\phi(\{\mathbf r\})
		=
		-\int_{\Gamma}\sum_i \mathbf V_i(\{\mathbf r\})\cdot \mathrm d\mathbf r_i ,
	\end{equation}
	and, up to a gauge choice,
	
	\begin{equation}
		(\mathbf{A}_i)_\alpha(\{\mathbf{r}\}) = - \int \mathrm{d}^{3N} r'  \left[ \partial_{r'_{j\beta }} G_{3N} \left( \{\mathbf{r}\} - \{\mathbf{r}'\}  \right)  \right] T_{i \alpha}^{j \beta}\left( \{\mathbf{r}'\} \right)
	\end{equation}
	where $G_{3N}$ is a $(3N)$-dimensional Green's function obtained from solving the equation 
	\begin{equation}
		- \sum_{i,\alpha} \nabla ^2 _{r_{i \alpha}} G_{3N}\left( \{\mathbf{r}\} - \{\mathbf{r}'\} \right) = \delta^{(3N)} \left(\{\mathbf{r}\} - \{\mathbf{r}'\}\right).
	\end{equation}
	
	In the translation-invariant case, the tensor $T_{i \alpha}^{j \beta}$ depends only on the relative coordinate $\mathbf{r} = \mathbf{r}_i - \mathbf{r}_j$. Translational invariance then requires the induced vector potential to depend only on relative separations as well, so it can be written as a sum of pairwise contributions,
	\begin{equation}
		(\mathbf{A}_i)_\alpha(\{\mathbf{r}\}) = \sum_{j} \mathcal{A}_\alpha (\mathbf{r}_i - \mathbf{r}_j).
	\end{equation}
	Correspondingly, the tensor structure reduces to $T_{i \alpha}^{j \beta} = \mathcal{T}_{\alpha \beta}(\mathbf{r})$. Once the dependence on the full configuration space is reduced to a dependence on the single relative coordinate $\mathbf{r}$, the original $(3N)$-dimensional Green's-function problem collapses to the ordinary three-dimensional Poisson problem. The relevant Green's function in three dimensions is therefore
	\begin{equation}
		G_3(\mathbf{r}) = \frac{1}{4 \pi |\mathbf{r}|}.
	\end{equation}
	Substituting this form into the solution above gives the pairwise vector potential
	\begin{equation}
		\mathcal{A}_\alpha(\mathbf{r}) = \frac{1}{4 \pi } \int \mathrm{d}^3 r' \, \frac{r_\beta - r'_\beta}{|\mathbf{r} - \mathbf{r}'|^3} \, \mathcal{T}_{\alpha \beta}(\mathbf{r}').
	\end{equation}

	In the case of reciprocal interactions, it is useful to specialize to the translation-invariant two-body case, as we did in the previous section for velocity-independent interactions. In this setting, the interaction between particles $i$ and $j$ can depend only on their relative coordinate and, because of Eq.~\eqref{seq:velocity_rescriction}, depend at most linearly on their velocities. We therefore introduce
	\begin{equation}
		\mathbf r_{ij}=\mathbf r_i-\mathbf r_j,
		\qquad
		\mathbf v_{ij}=\dot{\mathbf r}_i-\dot{\mathbf r}_j,
		\qquad
		\nabla \equiv \nabla_{\mathbf r_{ij}}.
	\end{equation}
	
	We now show that reciprocity imposes a strong simplification. Although one may start from a velocity-dependent interaction written separately in terms of $\dot{\mathbf r}_i$ and $\dot{\mathbf r}_j$, Newton's third law requires that, up to an irrelevant total time derivative, the dependence on velocity enters only through the relative velocity $\mathbf v_{ij}$ since it requires that $\frac{\mathrm d}{\mathrm d t} (\partial_{\dot{\mathbf{r}}_i}\mathcal U_{ij}^\mathrm{R}) = - \frac{\mathrm d}{\mathrm d t} (\partial_{\dot{\mathbf{r}}_j}\mathcal U_{ij}^\mathrm{R})$.
	
	As a result, any reciprocal pair interaction in this class can be written in the generalized-potential form
	\begin{equation}
		\mathcal U_{ij}^{\mathrm R}
		=
		\phi(\mathbf r_{ij})
		+
		\mathbf A(\mathbf r_{ij})\cdot \mathbf v_{ij}.
	\end{equation}
	Here $\phi(\mathbf r_{ij})$ is an ordinary scalar interaction potential, while $\mathbf A(\mathbf r_{ij})$ plays the role of an effective vector interaction potential. For identical particles, exchange symmetry further requires $\phi(-\mathbf r)=\phi(\mathbf r)$ and $\mathbf A(-\mathbf r)=-\mathbf A(\mathbf r)$. The total reciprocal interaction is then obtained by summing this pair contribution over all pairs $\langle ij\rangle$.
	
	This form has a simple physical interpretation. The scalar potential generates the familiar conservative interaction, whereas the vector-potential term produces the most general velocity-dependent reciprocal force that does not introduce accelerations explicitly into the equations of motion. The force exerted by particle $j$ on particle $i$ takes on the compact form
	\begin{equation}
		\mathbf F_{j\to i}^{\mathrm R}
		=
		-\nabla \phi(\mathbf r_{ij})
		+
		\left[
		(\nabla \mathbf A(\mathbf r_{ij}))^T
		-
		\nabla \mathbf A(\mathbf r_{ij})
		\right]\cdot\mathbf v_{ij}.
	\end{equation}
	Thus, the velocity-dependent part is purely antisymmetric in the velocity sector. In three dimensions, this contribution can be recognized as a Lorentz-like term, $\,-\mathbf v_{ij}\times [\nabla\times \mathbf A(\mathbf r_{ij})]$, generated by the effective magnetic field $\mathbf{B}=\nabla\times\mathbf{A}$ associated with $\mathbf A$.
	
	Written in this way, reciprocity is manifest: after using $\mathbf r_{ji}=-\mathbf r_{ij}$ and $\mathbf v_{ji}=-\mathbf v_{ij}$, one immediately finds $\mathbf F_{j\to i}^{\mathrm R}=-\mathbf F_{i\to j}^{\mathrm R}$. The representation is also gauge-dependent in the usual sense: the transformation $\mathbf A(\mathbf r)\to \mathbf A(\mathbf r)+\nabla \chi(\mathbf r)$ changes the generalized potential $\phi(\mathbf r_{ij})$ only by the total derivative $\mathrm d\chi(\mathbf r_{ij})/\mathrm dt$, and therefore leaves the equations of motion unchanged.
	
	In summary, the most general translation-invariant reciprocal pair interaction compatible with Eq.~\eqref{seq:velocity_rescriction} is fully characterized by a scalar interaction potential $\phi$ and a vector interaction potential $\mathbf A$. This provides the natural starting point for discussing velocity-dependent nonreciprocal forces below.
	In the following subsections, we use this structure as the building block for Hamiltonian embeddings of nonreciprocal velocity-dependent forces.

	\subsubsection{A direct Lagrangian embedding for velocity-dependent nonreciprocal forces}
	\label{sec:velocity_dependent_Lagrangian_embedding_1}
	
	Velocity-dependent interactions require more care than their velocity-independent counterparts because the interaction force is no longer obtained from a spatial gradient alone, but from the full Euler--Lagrange operator, as we now discuss. Throughout this subsection, ``potential'' therefore refers to a \emph{generalized} interaction potential in the Lagrangian sense. We first present the most direct extension of the embedding construction from Sec.~\ref{sec:velocity_independent_interaction_forces}, and then identify its regime of validity.
	
	For a pair of particles $(i,j)$, let the reciprocal sector be described by the generalized potential

	\begin{equation}
		U^{\mathrm R}(\mathbf r_i-\mathbf r_j,\dot{\mathbf r}_i-\dot{\mathbf r}_j),
	\end{equation}
	consistent with translation invariance. For the nonreciprocal sector, we allow a directed (i.e., one-way) generalized potential
	\begin{equation}
		U^{\mathrm{NR}}_{j\to i}(\mathbf r_i,\mathbf r_j,\dot{\mathbf r}_i,\dot{\mathbf r}_j),
	\end{equation}
	which generates the force exerted by particle $j$ on particle $i$,
	\begin{equation}
		\label{eq:NR_force_velocity_generalized_potential}
		\mathbf F^{\mathrm{NR}}_{j\to i}
		=
		-\partial_{\mathbf r_i}U^{\mathrm{NR}}_{j\to i}
		+
		\frac{\mathrm d}{\mathrm dt}\,\partial_{\dot{\mathbf r}_i}U^{\mathrm{NR}}_{j\to i}.
	\end{equation}
	The reciprocal force $\mathbf F^{\mathrm R}_{j\to i}$ follows analogously from $U^{\mathrm R}$. The target equations of motion are then
	\begin{equation}
		\label{Seq:EOM_general_velocity}
		m\ddot{\mathbf r}_i+\gamma\dot{\mathbf r}_i
		=
		\sum_{j\in\langle ij\rangle}
		\Big[
		\mathbf F^{\mathrm{NR}}_{j\to i}(\mathbf r_i,\mathbf r_j,\dot{\mathbf r}_i,\dot{\mathbf r}_j)
		+
		\mathbf F^{\mathrm R}_{j\to i}(\mathbf r_i-\mathbf r_j,\dot{\mathbf r}_i-\dot{\mathbf r}_j)
		\Big].
	\end{equation}
	
	To embed Eq.~\eqref{Seq:EOM_general_velocity} into a reciprocal system of original and auxiliary degrees of freedom, we introduce an auxiliary copy $\mathbf x_i$ and consider the Bateman-like Lagrangian~\cite{bateman1931dissipative}
	\begin{align}
		\label{Seq:Lagrangian_velocity}
		\mathcal L
		&=
		\sum_i
		\left[
		\frac{m}{2}\dot{\mathbf r}_i^2
		-\frac{m}{2}\dot{\mathbf x}_i^2
		+\frac{\gamma}{2}\left(\dot{\mathbf r}_i\!\cdot\!\mathbf x_i-\mathbf r_i\!\cdot\!\dot{\mathbf x}_i\right)
		\right]
		\nonumber\\
		&\quad
		-\sum_{\langle ij\rangle}
		\Big[
		U^{\mathrm{NR}}_{j\rightarrow i}(\mathbf r_i,\mathbf r_j,\dot{\mathbf r}_i,\dot{\mathbf r}_j)
		+U^{\mathrm{NR}}_{i\rightarrow j}(\mathbf r_j,\mathbf r_i,\dot{\mathbf r}_j,\dot{\mathbf r}_i)
		-U^{\mathrm{NR}}_{j\rightarrow i}(\mathbf x_i,\mathbf r_j,\dot{\mathbf x}_i,\dot{\mathbf r}_j)
		-U^{\mathrm{NR}}_{i\rightarrow j}(\mathbf x_j,\mathbf r_i,\dot{\mathbf x}_j,\dot{\mathbf r}_i)
		\nonumber\\
		&\qquad\qquad
		+U^{\mathrm R}(\mathbf r_i-\mathbf r_j,\dot{\mathbf r}_i-\dot{\mathbf r}_j)
		-U^{\mathrm R}(\mathbf x_i-\mathbf x_j,\dot{\mathbf x}_i-\dot{\mathbf x}_j)
		\Big].
	\end{align}
	The first line contains the kinetic energy. 
	The mixed term proportional to $\gamma$ is the standard Bateman construction for linear damping~\cite{bateman1931dissipative}, while the negative sign of the auxiliary kinetic energy is what preserves our mirror constraint dynamically, as shown in Eq.~\eqref{seq:constraint_velocity_preserved} below.
	
	The Euler--Lagrange equations,
	\begin{equation}
		\label{Seq:EOM_velocity}
		\frac{\mathrm d}{\mathrm dt}\frac{\partial\mathcal L}{\partial\dot{\mathbf r}_i}
		-\frac{\partial\mathcal L}{\partial\mathbf r_i}=0,
		\qquad
		\frac{\mathrm d}{\mathrm dt}\frac{\partial\mathcal L}{\partial\dot{\mathbf x}_i}
		-\frac{\partial\mathcal L}{\partial\mathbf x_i}=0,
	\end{equation}
	yield for the original variables
	\begin{align}
		\label{seq:EoM_original_velocity}
		m\ddot{\mathbf r}_i+\gamma\dot{\mathbf x}_i
		&=
		-\sum_{j\in\langle ij\rangle}
		\partial_{\mathbf r_i}
		\Big[
		U^{\mathrm{NR}}_{j\to i}(\mathbf r_i,\mathbf r_j,\dot{\mathbf r}_i,\dot{\mathbf r}_j)
		+
		U^{\mathrm R}(\mathbf r_i-\mathbf r_j,\dot{\mathbf r}_i-\dot{\mathbf r}_j)
		\Big]
		\nonumber\\
		&\quad
		+\sum_{j\in\langle ij\rangle}
		\frac{\mathrm d}{\mathrm dt}\,
		\partial_{\dot{\mathbf r}_i}
		U^{\mathrm{NR}}_{j\to i}(\mathbf r_i,\mathbf r_j,\dot{\mathbf r}_i,\dot{\mathbf r}_j)
		\nonumber\\
		&\quad
		+\sum_{j\in\langle ij\rangle}
		\frac{\mathrm d}{\mathrm dt}\,
		\partial_{\dot{\mathbf r}_i}
		U^{\mathrm R}(\mathbf r_i-\mathbf r_j,\dot{\mathbf r}_i-\dot{\mathbf r}_j)
		\nonumber\\
		&\quad +\boldsymbol{\Xi}_i ,
	\end{align}
	where the terms not present in the original nonreciprocal equations of motion are collected into
	\begin{align}
		\label{eq:Xi_velocity_first_embedding}
		\boldsymbol{\Xi}_i
		&=
		-\sum_{j\in\langle ij\rangle}
		\partial_{\mathbf r_i}
		\Big[
		U^{\mathrm{NR}}_{i\to j}(\mathbf r_j,\mathbf r_i,\dot{\mathbf r}_j,\dot{\mathbf r}_i)
		-
		U^{\mathrm{NR}}_{i\to j}(\mathbf x_j,\mathbf r_i,\dot{\mathbf x}_j,\dot{\mathbf r}_i)
		\Big]
		\nonumber\\
		&\quad
		+\sum_{j\in\langle ij\rangle}
		\frac{\mathrm d}{\mathrm dt}\,
		\partial_{\dot{\mathbf r}_i}
		\Big[
		U^{\mathrm{NR}}_{i\to j}(\mathbf r_j,\mathbf r_i,\dot{\mathbf r}_j,\dot{\mathbf r}_i)
		-
		U^{\mathrm{NR}}_{i\to j}(\mathbf x_j,\mathbf r_i,\dot{\mathbf x}_j,\dot{\mathbf r}_i)
		\Big].
	\end{align}
	Accordingly, the auxiliary variables obey
	\begin{align}
		\label{seq:EoM_auxiliary_velocity}
		m\ddot{\mathbf x}_i+\gamma\dot{\mathbf r}_i
		&=
		-\sum_{j\in\langle ij\rangle}
		\partial_{\mathbf x_i}
		\Big[
		U^{\mathrm{NR}}_{j\to i}(\mathbf x_i,\mathbf r_j,\dot{\mathbf x}_i,\dot{\mathbf r}_j)
		+
		U^{\mathrm R}(\mathbf x_i-\mathbf x_j,\dot{\mathbf x}_i-\dot{\mathbf x}_j)
		\Big]
		\nonumber\\
		&\quad
		+\sum_{j\in\langle ij\rangle}
		\frac{\mathrm d}{\mathrm dt}\,
		\partial_{\dot{\mathbf x}_i}
		U^{\mathrm{NR}}_{j\to i}(\mathbf x_i,\mathbf r_j,\dot{\mathbf x}_i,\dot{\mathbf r}_j)
		\nonumber\\
		&\quad
		+\sum_{j\in\langle ij\rangle}
		\frac{\mathrm d}{\mathrm dt}\,
		\partial_{\dot{\mathbf x}_i}
		U^{\mathrm R}(\mathbf x_i-\mathbf x_j,\dot{\mathbf x}_i-\dot{\mathbf x}_j).
	\end{align}
	
	We now impose the mirror constraint at the initial time,
	\begin{equation}
		\label{seq:constraint_velocity_1}
		\mathbf r_i(0)=\mathbf x_i(0),
		\qquad
		\dot{\mathbf r}_i(0)=\dot{\mathbf x}_i(0).
	\end{equation}
	On this manifold, $\boldsymbol{\Xi}_i\equiv 0$ identically, and Eqs.~\eqref{seq:EoM_original_velocity} and \eqref{seq:EoM_auxiliary_velocity} collapse onto the same equation, namely Eq.~\eqref{Seq:EOM_general_velocity} for $\mathbf r_i$ and, equivalently, for $\mathbf x_i$. It is convenient to define the difference variable $\boldsymbol{\Delta}_i\equiv \mathbf r_i-\mathbf x_i$. Subtracting the two equations of motion then gives, on the constraint manifold,
	\begin{equation}
		\label{seq:constraint_velocity_preserved}
		m\ddot{\boldsymbol{\Delta}}_i-\gamma\dot{\boldsymbol{\Delta}}_i=0.
	\end{equation}
	Hence, the initial conditions in Eq.~\eqref{seq:constraint_velocity_1} imply $\boldsymbol{\Delta}_i(t)\equiv 0$ for all times: the constraint manifold is dynamically preserved, and the embedding reproduces the desired nonreciprocal dynamics.
	
	Whenever the velocity-momentum map is non-singular, the corresponding Hamiltonian follows from a Legendre transform,
	\begin{equation}
		\mathbf p_i^{\mathbf r}=\frac{\partial\mathcal L}{\partial\dot{\mathbf r}_i},
		\qquad
		\mathbf p_i^{\mathbf x}=\frac{\partial\mathcal L}{\partial\dot{\mathbf x}_i},
		\qquad
		H=
		\sum_i
		\left(
		\mathbf p_i^{\mathbf r}\!\cdot\!\dot{\mathbf r}_i
		+
		\mathbf p_i^{\mathbf x}\!\cdot\!\dot{\mathbf x}_i
		\right)
		-\mathcal L.
	\end{equation}
	Thus, as usual, a Hamiltonian embedding exists whenever the Lagrangian description is regular. We therefore focus on the Lagrangian formulation here. 
	
	This construction is the closest analogue of the velocity-independent embedding, but it is not fully general. It has a structural limitation: because the nonreciprocal sector is encoded through an ordinary generalized potential that depends on both the original and the auxiliary velocities, the admissible forces are restricted by Eq.~\eqref{seq:velocity_rescriction}. In particular, within this ansatz, the force can be at most linear in the velocities if the equations of motion are to remain second order in time. The present embedding, therefore, captures only the subset of velocity-dependent nonreciprocal interactions that fit into a conventional generalized-potential description. Nevertheless, the similarity of this formulation of the embedding to electromagnetism and the associated gauge structure may provide new insights into the general structure of these embeddings. In Sec.~\ref{sec:general_form}, we introduce a more flexible construction in which the velocity dependence is attached only to the original degrees of freedom, thereby allowing for a broader class of nonreciprocal forces.

	\subsubsection{Velocity-dependent nonreciprocal interactions: General framework}
	\label{sec:general_form}
	
	We now extend our construction to velocity-dependent nonreciprocal interactions. 
	While the direct Lagrangian embedding from Sec.~\ref{sec:velocity_dependent_Lagrangian_embedding_1} works for equations of motions that are linear in the velocity, here we present a more flexible formulation that accommodates a broader class of velocity-dependent forces.
	
	Consider the Lagrangian for the doubled system:
	\begin{align}
		\label{Seq:Lagrangian_velocity_2}
		\mathcal L
		&=
		\sum_i
		\left[
		\frac{m}{2}\dot{\mathbf r}_i^2
		-\frac{m}{2}\dot{\mathbf x}_i^2
		+\frac{\gamma}{2}\left(\dot{\mathbf r}_i\!\cdot\!\mathbf x_i-\mathbf r_i\!\cdot\!\dot{\mathbf x}_i\right)
		\right]
		\nonumber\\
		&\quad
		-\sum_{\langle ij\rangle}
		\Big[
		U^{\mathrm{NR}}_{j\rightarrow i}(\mathbf r_i,\mathbf r_j,\dot{\mathbf r}_i,\dot{\mathbf r}_j)
		+U^{\mathrm{NR}}_{i\rightarrow j}(\mathbf r_j,\mathbf r_i,\dot{\mathbf r}_j,\dot{\mathbf r}_i)
		\nonumber\\
		&\quad
		-U^{\mathrm{NR}}_{j\rightarrow i}(\mathbf x_i,\mathbf r_j,\dot{\mathbf r}_i,\dot{\mathbf r}_j)
		-U^{\mathrm{NR}}_{i\rightarrow j}(\mathbf x_j,\mathbf r_i,\dot{\mathbf r}_j,\dot{\mathbf r}_i)
		\nonumber\\
		&\quad
		+U^{\mathrm{R}}(\mathbf r_i-\mathbf r_j,\dot{\mathbf r}_i-\dot{\mathbf r}_j)
		-U^{\mathrm{R}}(\mathbf x_i-\mathbf x_j,\dot{\mathbf x}_i-\dot{\mathbf x}_j)
		\Big] .
	\end{align}
	
	The key difference from Eq.~\eqref{Seq:Lagrangian_velocity} is that the velocity-dependent nonreciprocal potential couples only to the original degrees of freedom $\dot{\mathbf r}_i$, and not to the auxiliary ones $\dot{\mathbf x}_i$. This modification enables the embedding to capture velocity-dependent forces that are not restricted to being linear in velocity.
	
	From the Euler-Lagrange equations, the original degrees of freedom satisfy
	\begin{align}
		\label{seq:EoM_original_velocity_2}
		m\ddot{\mathbf r}_i+\gamma\dot{\mathbf x}_i
		&=
		- \sum_{j\in\langle ij\rangle}
		\partial_{\mathbf{r}_i} 
		\Big[
		U^{\mathrm{NR}}_{j\to i}(\mathbf r_i,\mathbf r_j,\dot{\mathbf r}_i,\dot{\mathbf r}_j)
		+ U^{\mathrm R}(\mathbf r_i-\mathbf r_j,\dot{\mathbf r}_i-\dot{\mathbf r}_j)
		\Big]
		\nonumber\\
		&\quad
		+\sum_{j \in \langle ij \rangle}\frac{\mathrm{d}}{\mathrm{d}t}\partial_{\dot{\mathbf{r}}_i}   U^{\mathrm{R}} (\mathbf r_i-\mathbf r_j,\dot{\mathbf r}_i-\dot{\mathbf r}_j)
		\nonumber\\
		&\quad
		+\boldsymbol{\Xi}_i ,
	\end{align}
	where the unwanted correction terms are collected in
	\begin{align}
		\boldsymbol{\Xi}_i
		&=
		- \sum_{j\in\langle ij\rangle}
		\partial_{\mathbf r_i}
		\!\left[
		U^{\mathrm{NR}}_{j\rightarrow i}(\mathbf r_i,\mathbf r_j,\dot{\mathbf r}_i,\dot{\mathbf r}_j)
		-U^{\mathrm{NR}}_{j\rightarrow i}(\mathbf x_i,\mathbf r_j,\dot{\mathbf r}_i,\dot{\mathbf r}_j)
		\right]
		\nonumber\\
		&\quad
		+\sum_{j\in\langle ij\rangle}
		\frac{\mathrm d}{\mathrm dt}\,
		\partial_{\dot{\mathbf r}_i}
		\!\left[
		U^{\mathrm{NR}}_{j\rightarrow i}(\mathbf r_i,\mathbf r_j,\dot{\mathbf r}_i,\dot{\mathbf r}_j)
		+U^{\mathrm{NR}}_{i\rightarrow j}(\mathbf r_j,\mathbf r_i,\dot{\mathbf r}_j,\dot{\mathbf r}_i)
		\right.
		\nonumber\\
		&\left.\quad
		-U^{\mathrm{NR}}_{j\rightarrow i}(\mathbf x_i,\mathbf r_j,\dot{\mathbf r}_i,\dot{\mathbf r}_j)
		-U^{\mathrm{NR}}_{i\rightarrow j}(\mathbf x_j,\mathbf r_i,\dot{\mathbf r}_j,\dot{\mathbf r}_i)
		\right] .
	\end{align}
	The auxiliary degrees of freedom evolve according to
	\begin{align}
		\label{seq:EoM_auxiliary_velocity_2}
		m\ddot{\mathbf x}_i+\gamma\dot{\mathbf r}_i
		&=
		- \sum_{j \in \langle ij \rangle}
		\partial_{\mathbf{x}_i}
		\Big[
		U^{\mathrm{NR}}_{j\to i}(\mathbf x_i,\mathbf r_j,\dot{\mathbf r}_i,\dot{\mathbf r}_j)
		+ U^{\mathrm R}(\mathbf x_i-\mathbf x_j,\dot{\mathbf x}_i-\dot{\mathbf x}_j)
		\Big]
		\nonumber\\
		&\quad
		+\sum_{j \in \langle ij \rangle}\frac{\mathrm{d}}{\mathrm{d}t}\partial_{\dot{\mathbf{x}}_i}   U^{\mathrm{R}} (\mathbf x_i-\mathbf x_j,\dot{\mathbf x}_i-\dot{\mathbf x}_j) .
	\end{align}
	
	We impose the mirror constraint at the initial time:
	\begin{equation}
		\mathbf r_i(0)=\mathbf x_i(0),
		\qquad
		\dot{\mathbf r}_i(0)=\dot{\mathbf x}_i(0).
		\label{eq:constraint_velocity}
	\end{equation}
	As expected, on the constraint manifold, $\boldsymbol{\Xi}_i\equiv 0$ vanishes identically, and both the original and auxiliary equations reduce to the same dynamical equations. Once again, the constraint defines a closed submanifold: differentiating the constraint yields
	\begin{equation}
		\frac{\mathrm d}{\mathrm dt}\!\left[\mathbf r_i(t)-\mathbf x_i(t)\right]\Big|_{\mathrm{constr.}}=0,
		\qquad
		\frac{\mathrm d}{\mathrm dt}\!\left[\dot{\mathbf r}_i(t)-\dot{\mathbf x}_i(t)\right]\Big|_{\mathrm{constr.}}=0.
	\end{equation}
	This ensures that trajectories satisfying the constraint at $t=0$ remain on the constraint manifold for all times.
	
	The key advantage of this formulation is that the velocity dependence of the nonreciprocal potential $U^{\mathrm{NR}}$ is decoupled from the auxiliary degrees of freedom. Consequently, the embedding can accommodate nonreciprocal forces with polynomial or even transcendental velocity dependence, not merely linear terms. The force exerted by particle $j$ on particle $i$ is given by
	\begin{align}
		\mathbf{F}_{j \rightarrow i}^{\mathrm{NR}} 
		(\mathbf{r}_i, \mathbf{r}_j, \dot{\mathbf{r}}_i , \dot{\mathbf{r}}_j) = - \partial_{\mathbf{r}_i} U_{j \rightarrow i}^{\mathrm{NR}}
		(\mathbf{r}_i, \mathbf{r}_j, \dot{\mathbf{r}}_i , \dot{\mathbf{r}}_j) 
	\end{align}
	which is not restricted to linear velocity dependence. This demonstrates the generality and flexibility of the embedding framework for velocity-dependent nonreciprocal interactions.

	\subsubsection{Example: Lorentz-like forces}
	\label{subsec:Lorentz_like_force}
	
	As a concrete example of velocity-dependent nonreciprocal interactions, consider pairwise Lorentz-like forces of the form
	\begin{equation}
		\label{eq:Lorentz_like_eom}
		m\ddot{\mathbf r}_i+\gamma \dot{\mathbf r}_i
		=
		\sum_{j\in\langle ij\rangle}
		\mathbf F^{\mathrm{NR}}_{j\to i},
		\qquad
		\mathbf F^{\mathrm{NR}}_{j\to i}
		=
		\dot{\mathbf r}_i\times \mathbf B_{j\to i}(\mathbf r_i,\mathbf r_j),
	\end{equation}
	with
	\begin{equation}
		\mathbf B_{j\to i}(\mathbf r_i,\mathbf r_j)
		=
		\nabla_{\mathbf r_i}\times \mathbf A_{j\to i}(\mathbf r_i,\mathbf r_j).
	\end{equation}
	In general, $\mathbf A_{j\to i}\neq \mathbf A_{i\to j}$, so the force exerted by particle $j$ on particle $i$ need not be the opposite to the force exerted by $i$ on $j$; this interaction is therefore nonreciprocal. The overdamped limit is recovered by setting $m=0$.
	
	This example is useful because it allows us to make explicit the difference between the two velocity-dependent embeddings introduced above.
	
	\paragraph*{Direct generalized-potential embedding.}
	The construction of Sec.~\ref{sec:velocity_dependent_Lagrangian_embedding_1} applies provided the Lorentz-like force can be generated from a generalized (one-way) potential. A sufficient condition is that there exists a companion vector field $\tilde{\mathbf A}_{j\to i}$ such that
	\begin{equation}
		\label{eq:lorentz_integrability}
		\left(\partial_{\mathbf r_i}\tilde{\mathbf A}_{j\to i}\right)^T
		=
		\partial_{\mathbf r_j}\mathbf A_{j\to i}.
	\end{equation}
	Then the directed generalized potential
	\begin{equation}
		\label{eq:lorentz_direct_potential}
		U^{\mathrm{NR}}_{j\to i}
		=
		-
		\mathbf A_{j\to i}(\mathbf r_i,\mathbf r_j)\cdot \dot{\mathbf r}_i
		-
		\tilde{\mathbf A}_{j\to i}(\mathbf r_i,\mathbf r_j)\cdot \dot{\mathbf r}_j
	\end{equation}
	satisfies
	\begin{equation}
		-\partial_{\mathbf r_i}U^{\mathrm{NR}}_{j\to i}
		+
		\frac{\mathrm d}{\mathrm dt}\partial_{\dot{\mathbf r}_i}U^{\mathrm{NR}}_{j\to i}
		=
		\dot{\mathbf r}_i\times
		\big(\nabla_{\mathbf r_i}\times \mathbf A_{j\to i}\big),
	\end{equation}
	where Eq.~\eqref{eq:lorentz_integrability} ensures the cancellation of all terms proportional to $\dot{\mathbf r}_j$.
	Substituting Eq.~\eqref{eq:lorentz_direct_potential} into the Lagrangian embedding of Sec.~\ref{sec:velocity_dependent_Lagrangian_embedding_1}, Eq.~\eqref{Seq:Lagrangian_velocity}, yields an embedding of the form in Eq.~\eqref{eq:Lorentz_like_eom}. 
	More explicitly, if we set the reciprocal sector to zero for simplicity, the embedding Lagrangian becomes
	\begin{align}
		\label{eq:lorentz_direct_lagrangian}
		\mathcal L
		&=
		\sum_i
		\left[
		\frac{m}{2}\dot{\mathbf r}_i^2
		-
		\frac{m}{2}\dot{\mathbf x}_i^2
		+
		\frac{\gamma}{2}
		\left(
		\dot{\mathbf r}_i\cdot \mathbf x_i
		-
		\mathbf r_i\cdot \dot{\mathbf x}_i
		\right)
		\right]
		\nonumber\\
		&\quad
		+
		\sum_{\langle ij\rangle}
		\Big[
		\mathbf A_{j\to i}(\mathbf r_i,\mathbf r_j)\cdot \dot{\mathbf r}_i
		+
		\tilde{\mathbf A}_{j\to i}(\mathbf r_i,\mathbf r_j)\cdot \dot{\mathbf r}_j
		+
		\mathbf A_{i\to j}(\mathbf r_j,\mathbf r_i)\cdot \dot{\mathbf r}_j
		+
		\tilde{\mathbf A}_{i\to j}(\mathbf r_j,\mathbf r_i)\cdot \dot{\mathbf r}_i
		\nonumber\\
		&\qquad\qquad
		-
		\mathbf A_{j\to i}(\mathbf x_i,\mathbf r_j)\cdot \dot{\mathbf x}_i
		-
		\tilde{\mathbf A}_{j\to i}(\mathbf x_i,\mathbf r_j)\cdot \dot{\mathbf r}_j
		-
		\mathbf A_{i\to j}(\mathbf x_j,\mathbf r_i)\cdot \dot{\mathbf x}_j
		-
		\tilde{\mathbf A}_{i\to j}(\mathbf x_j,\mathbf r_i)\cdot \dot{\mathbf r}_i
		\Big].
	\end{align}
	The Euler--Lagrange equations derived from Eq.~\eqref{eq:lorentz_direct_lagrangian} reduce to Eq.~\eqref{eq:Lorentz_like_eom} on the constraint manifold. With the mirror constraint
	\begin{equation}
		\mathbf r_i(0)=\mathbf x_i(0),
		\qquad
		\dot{\mathbf r}_i(0)=\dot{\mathbf x}_i(0),
	\end{equation}
	the two copies remain locked for all times, and both follow the same Lorentz-like dynamics.

	\paragraph*{General embedding.}
	If no field $\tilde{\mathbf A}_{j\to i}$ satisfying Eq.~\eqref{eq:lorentz_integrability} exists, the direct route above is not viable. One can nevertheless use the broader construction of Sec.~\ref{sec:general_form}, which does not require the Lorentz-like force itself to arise from a generalized potential. Defining
	\begin{equation}
		\mathbf G_{j\to i}(\mathbf r_i,\mathbf r_j,\dot{\mathbf r}_i)
		=
		- \dot{\mathbf r}_i\times \mathbf B_{j\to i}(\mathbf r_i,\mathbf r_j),
	\end{equation}
	we introduce the embedding Lagrangian
	\begin{align}
		\label{eq:lorentz_general_embedding}
		\mathcal L
		&=
		\sum_i
		\left[
		\frac{m}{2}\dot{\mathbf r}_i^2
		-
		\frac{m}{2}\dot{\mathbf x}_i^2
		+
		\frac{\gamma}{2}
		\left(
		\dot{\mathbf r}_i\cdot \mathbf x_i
		-
		\mathbf r_i\cdot \dot{\mathbf x}_i
		\right)
		\right]
		\nonumber\\
		&\quad
		-
		\sum_{\langle ij\rangle}
		\Big[
		\mathbf G_{j\to i}(\mathbf r_i,\mathbf r_j,\dot{\mathbf r}_i)\cdot(\mathbf r_i-\mathbf x_i)
		+
		\mathbf G_{i\to j}(\mathbf r_j,\mathbf r_i,\dot{\mathbf r}_j)\cdot(\mathbf r_j-\mathbf x_j)
		\Big].
	\end{align}
	Away from the constraint manifold, the equations of motion for $\mathbf r_i$ and $\mathbf x_i$ contain extra terms proportional to $\mathbf r_i-\mathbf x_i$ and $\dot{\mathbf r}_i-\dot{\mathbf x}_i$. On the mirror-constraint manifold, these terms vanish identically. Hence, for initial conditions satisfying
	\begin{equation}
		\mathbf r_i(0)=\mathbf x_i(0),
		\qquad
		\dot{\mathbf r}_i(0)=\dot{\mathbf x}_i(0),
	\end{equation}
	the constraint is dynamically preserved and the reduced dynamics become
	\begin{equation}
		m\ddot{\mathbf r}_i+\gamma \dot{\mathbf r}_i
		=
		\sum_{j\in\langle ij\rangle}
		\dot{\mathbf r}_i\times \mathbf B_{j\to i}(\mathbf r_i,\mathbf r_j).
	\end{equation}
	
	In summary, Lorentz-like forces cleanly distinguish the scope of the two velocity-dependent embeddings. When the integrability condition~\eqref{eq:lorentz_integrability} holds, the interaction admits a bona fide generalized-potential representation and the direct construction of Sec.~\ref{sec:velocity_dependent_Lagrangian_embedding_1} is sufficient. When it does not, the more general framework of Sec.~\ref{sec:general_form} still provides a consistent embedding, with the mirror constraint selecting the physical nonreciprocal dynamics.

	\section{Examples of nonreciprocal systems treated within the Hamiltonian embedding}
	\label{sec:examples}
	
	Below, we discuss concrete examples of this general Hamiltonian embedding
	for nonreciprocal interactions, see Fig.~\ref{fig:illustration}.

	\begin{figure*}[t!]
		\centering\includegraphics[width=0.9\textwidth]{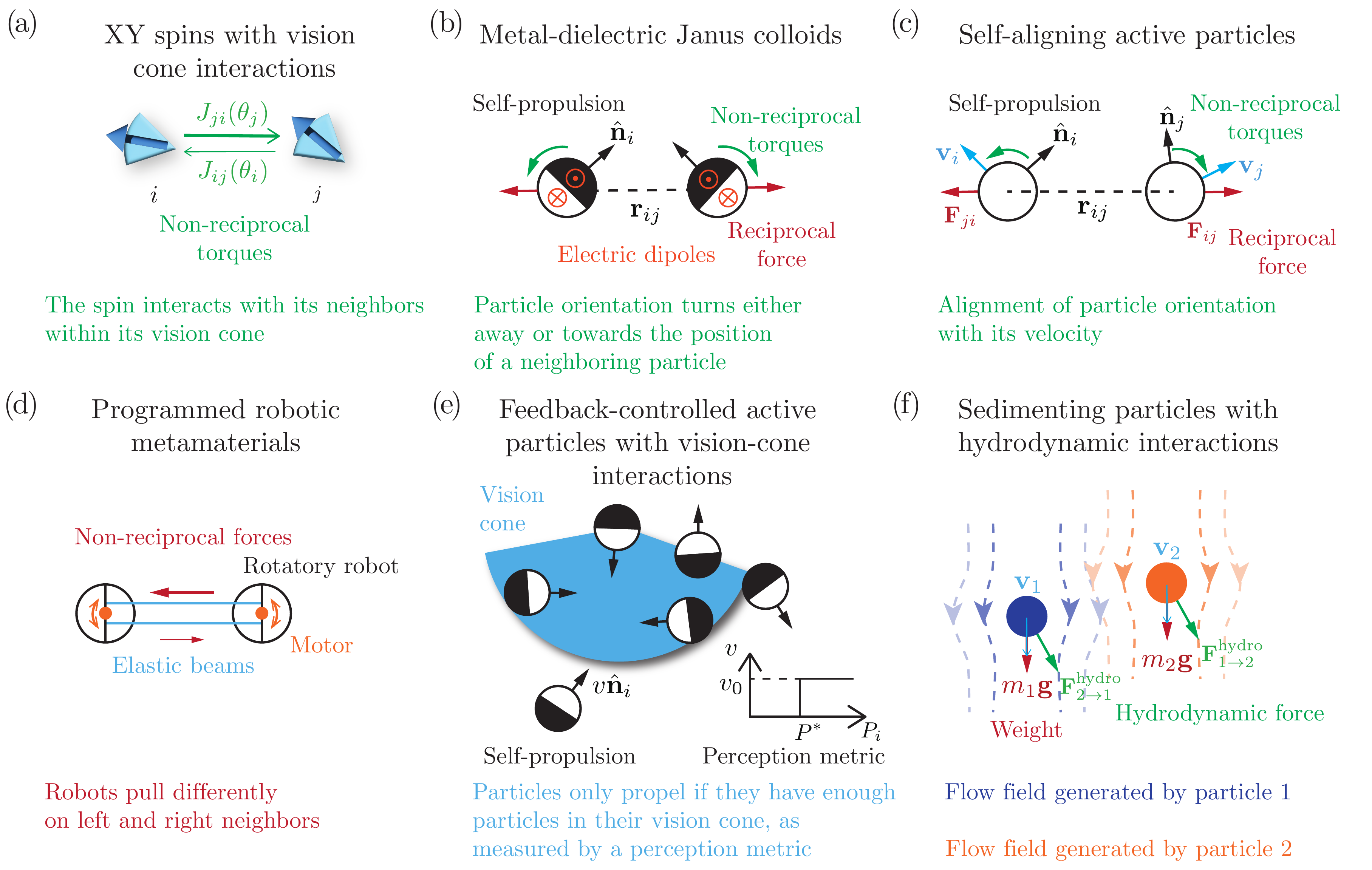}
		\caption{
			Schematic representation of various models exhibiting nonreciprocal equations of motion.
			(a) XY spins with vision-cone interactions (Sec.~\ref{subsec:XY_VC}) experiences torques from the neighboring spins within its vision cone.
			(b) Metal-dielectric self-propelled Janus colloidal particles (Sec.~\ref{subsec:Janus_particles}) experience self-propulsion, reciprocal repulsive forces and nonreciprocal torques.
			(c) Self-aligning active particles (Sec.~\ref{subsec:self-aligning active particle}) interact via reciprocal repulsive forces and exhibit self-propulsion. The nonreciprocal torques cause each particle to rotate toward its velocity direction.  
			(d) Programmed robotic metamaterials (Sec.~\ref{subsec:Programmed rotational robots (metamaterials)}) consist of robotic units connected by springs that exert asymmetric forces on the left and right sides.  
			(e) Feedback-controlled active particles (Sec.~\ref{subsec:Feedback-controlled active particles}) interact via reciprocal forces and self-propel when a sufficient number of neighboring particles enter their vision cone.
			(f) Sedimenting particles settling through a viscous fluid under gravity
			(Sec.~\ref{subsec:Hydrodynamic_couplings_sedimenting_systems}) 
			affect each other by generating flow fields. The resulting interactions are intrinsically nonreciprocal~\cite{poncet2022soft,guillet2025melting}.
		}
		\label{fig:illustration}
	\end{figure*}

	\subsection{XY spins with vision cone interactions}
	\label{subsec:XY_VC}
	
	A first illustrative example is the vision-cone XY spins, $\mathbf{S}_i=(\cos \theta_i,\sin \theta_i)$, on a square lattice discussed in the Main Text. Here, the nonreciprocal equations of motion take the form [cf.~Eq. (1) in the main text and~Fig.~\ref{fig:illustration}(a)]:
	\begin{equation}
		\label{Seq:NR_XY}
		\dot{\theta_i} = - \sum_{j\in\langle ij\rangle}J_{ij}(\theta_i)\sin(\theta_i-\theta_j) +\sqrt{2 T}\eta _{i}.
	\end{equation}
	To construct the Hamiltonian, we introduce the auxiliary spins $\mathbf{a}_i$ parametrized by the angle $\varphi_i$. Proceeding as explained in the Main Text, we obtain:
	\begin{eqnarray}
		\label{Seq:NR_Hamil}
		H &=& -\sum_{\langle ij\rangle}
		\left [ J_{ij}(\theta_i)(\mathbf{S}_i\cdot \mathbf{S}_j+\mathbf{a}_i\cdot \mathbf{S}_j)
		+
		J_{ji}(\theta_j)(\mathbf{S}_j \cdot \mathbf{S}_i+\mathbf{a}_j\cdot \mathbf{S}_i) \right ], \notag \\
		&=&-\sum_{\langle ij\rangle}
		\left \{ J_{ij}(\theta_i)[\cos(\theta_i-\theta_j)+\cos(\varphi_i-\theta_j)]
		+
		J_{ji}(\theta_j)[\cos(\theta_j-\theta_i)+\cos(\varphi_j-\theta_i)]   \right \} . 
	\end{eqnarray}
	The equations of motion are given in Eq.~(4) in the Main Text, ignoring the noise term.
	The origin and technical handling of the constraint are discussed in detail in Sec.~\ref{sec:Two ways of enforcing the constraint}. 
	
	As we now demonstrate, this way of constructing the Hamiltonian is generic and can be applied to other models with nonreciprocal interactions.

	\subsection{Metal-dielectric Janus particles}
	\label{subsec:Janus_particles}

	The next example is that of metal-dielectric self-propelled Janus colloidal particles. These particles are silica spheres coated with titanium on one hemisphere. Suspended in water, they sediment and form a monolayer on a conductive coverslip, separated from another conductive coverslip, which together form a capacitor~\cite{yan2016reconfiguring,zhang2021active,das2024flocking}. The particles are driven by an AC electric field produced by a voltage between the conductive coverslips. The electric field induces dipoles in each of the particles' hemispheres, which leads both to self-propulsion as well as to electrostatic interactions between the particles. The interactions consist of a repulsive force and a torque that tends to make the particles turn either towards or away from one another~\cite{zhang2021active,das2024flocking}. At low area fractions and self-propulsion speeds, the system behaves as an isotropic active gas. At higher area fractions and speeds, the system either undergoes phase separation when particles turn towards each other \cite{zhang2021active} or transitions to a polar-ordered flock when particles turn away from one another~\cite{das2024flocking}.
	
	To explain these phenomena, previous work proposed a two-dimensional microscopic model based on the dipolar interactions between the hemispheres of the particles (Fig.~\ref{fig:illustration}b)~\cite{zhang2021active}. The interaction torques, which tend to rotate a particle either away or towards the position of a second particle, are intrinsically nonreciprocal; they do not have a reciprocal limit~\cite{zhang2021active,das2024flocking}. The position and orientation of particle $i$ are denoted by $\mathbf{r}_i$ and $\mathbf{\hat{n}}_i$, respectively. The equations of motion are given by
	\begin{eqnarray}
		\label{Seq:Langevin_MDJP}
		\frac{\mathrm{d}\mathbf{r}_{i}}{\mathrm{d}t} &=&v_{0}\mathbf{\hat{n}}_{i}+\sum_{j\neq i} \frac{\mathbf{F}_{ji}}{\gamma_\text{t}} + \bm{\eta}_i^\text{t}, \notag \\
		\frac{\mathrm{d}\mathbf{\hat{n}}_{i}}{\mathrm{d}t} &=&\sum_{j\neq i}\frac{\mathbf{\Gamma }_{ji}}{\gamma_\text{r}} + \hat{\mathbf{n}}_i\times \bm{\eta}_i^\text{r}.
	\end{eqnarray}
	Here, $\bm{\eta}_i^\text{t}$ and $\bm{\eta}_i^\text{r}$ are translational and rotational Gaussian white noises with zero mean and with $\langle \eta^\text{t}_{i,\alpha}(t) \eta^\text{t}_{j,\beta}(t') \rangle = 2D_\text{t} \delta_{ij} \delta_{\alpha\beta} \delta(t-t')$ and $\langle \eta^\text{r}_{i,z}(t) \eta^\text{r}_{j,z}(t') \rangle = 2D_\text{r} \delta_{ij} \delta(t-t')$, where $D_\text{t}$ and $D_\text{r}$ are the translational and rotational diffusion coefficients of the particles. Here, Latin indices indicate particles, whereas Greek indices indicate spatial components. Note that the rotational noise has components only along the $\hat{\mathbf{z}}$ axis so that the particle orientation $\hat{\mathbf{n}}_i$ remains on the $xy$ plane.

	The particle positions change due to self-propulsion at a speed $v_0$, and due to the reciprocal repulsive force $\mathbf{F}_{ji}$ exerted by particle $j$ on particle $i$, with $\gamma_\text{t}$ being the translational friction coefficient. 
	This interaction force can be derived from an interaction potential $U({|\mathbf{r}_i - \mathbf{r}_j|})$ between particles $i$ and $j$. 
	In addition, $\mathbf{\Gamma }_{ji}$ denotes the interaction torque, which is intrinsically nonreciprocal, $\mathbf{\Gamma }_{ji} \neq - \mathbf{\Gamma }_{ij}$, and $\gamma_\text{r}$ is the associated rotational friction coefficient.

	Assuming turn-towards torques, the expressions for the force $\mathbf{F}_{ji}$ and the torque $\mathbf{\Gamma }_{ji}$ are given by~\cite{zhang2021active}
	\begin{eqnarray}
		\label{Seq:Langevin_explicit_form}
		\mathbf{F}_{ji} 
		&=&
		\frac{3}{4\pi \epsilon }    \frac{\left( d_\text{h}+d_\text{t}\right)^{2}}{r_{ji}^{4}}    
		\exp\left( -\frac{r_{ji}}{\lambda }\right)  \hat{\mathbf{r}}_{ji}, \notag \\ 
		\mathbf{\Gamma }_{ji} 
		&=& 
		\frac{3 \ell}{4\pi \epsilon }   \frac{d_\text{h}^{2}-d_\text{t}^{2}}{r_{ji}^{4}}  \exp\left( -\frac{r_{ji}}{\lambda }\right)  \mathbf{\hat{n}}_{i} \times\hat{\mathbf{r}}_{ji} 
		= 
		\bm{\tau}_{ji}\times\mathbf{\hat{n}}_{i}  ,
	\end{eqnarray}%
	where $\epsilon$ is the dielectric permittivity of the solvent, and $d_\text{h}$ and $d_\text{t}$ are the effective dipole strengths of the head and tail hemispheres. Here, to shorten the notation, we have rewritten the interaction torque in terms of the quantity
	\begin{equation}
		\bm{\tau}_{ji} \equiv  -\frac{3 \ell}{4\pi \epsilon }   \frac{d_\text{h}^{2}-d_\text{t}^{2}}{r_{ji}^{4}}  \exp\left( -\frac{r_{ji}}{\lambda }\right) \hat{\mathbf{r}}_{ji}.
	\end{equation}
	Here $\mathbf{r}_{ji} =\mathbf{r}_i-\mathbf{r}_j $ is the distance vector with its length and direction given by $r_{ji}= \left\vert \mathbf{r}_{ji}\right\vert$ and $\hat{\mathbf{r}}_{ji}=\mathbf{r}_{ji}/r_{ji}$, respectively.

	So far, we discussed all forces and torques. We now introduce the auxiliary particles with position and orientation denoted as $\mathbf{x}_i$ and $\mathbf{\hat{m}}_i$. The Hamiltonian embedding is given by
	\begin{eqnarray}
		\label{Seq:Hamiltonian_MDJP}
		H&=&\sum_{i} 
		-
		v_{0}\mathbf{\hat{n}}_{i}\cdot \left( \mathbf{r}_{i}+\mathbf{x}%
		_{i}\right) 
		+
		\sum_{i\neq j}\left [ \frac{U(|\mathbf{r}_i - \mathbf{r}_j|) }{{\gamma_\text{t}}}
		-
		\frac{U(|\mathbf{x}_i - \mathbf{x}_j|)}{{\gamma_\text{t}}}
		+ 
		\frac{\bm{\tau}_{ji}}{\gamma_\text{r}}\cdot \left( \mathbf{\hat{n}}_{i}+\mathbf{\hat{m}}_{i}\right)
		+ 
		\frac{\bm{\tau}_{ij}}{\gamma_\text{r}}\cdot \left( \mathbf{\hat{n}}_{j}+\mathbf{\hat{m}}_{j}\right) \right ].
	\end{eqnarray}
	We can derive the corresponding equations of motion from the Hamiltonian description. The translational motion of the particles is governed by
	\begin{eqnarray}
		\label{Seq:EOM_translation_MDJP}
		\frac{\mathrm{d}\mathbf{r}_{i}}{\mathrm{d}t} 
		=
		-   
		\frac{\partial H}{\partial 
			\mathbf{r}_{i}} 
		=
		v_{0}\mathbf{\hat{n}}_{i}   
		-    
		\sum_{j\neq i} 
		\left [   
		\frac{1 }{{\gamma_\text{t}}}\frac{\partial U(|\mathbf{r}_i - \mathbf{r}_j|)  }{\partial \mathbf{r}_{i}} 
		-
		\frac{\partial }{\partial \mathbf{r}_{i}}\frac{\bm{\tau}_{ij}}{\gamma_\text{r}}%
		\cdot \left( \mathbf{\hat{n}}_{i}+\mathbf{\hat{m}}_{i}\right)
		-
		\frac{\partial }{\partial \mathbf{r}_{i}}\frac{\bm{\tau}_{ji}}{\gamma_\text{r}}%
		\cdot \left( \mathbf{\hat{n}}_{j}+\mathbf{\hat{m}}_{j}\right)
		\right ]
		,
	\end{eqnarray}%
	and the orientational motion of the particle follows
	\begin{eqnarray}
		\label{Seq:EOM_orientation_MDJP}
		\frac{\mathrm{d}\mathbf{\hat{n}}_{i}}{\mathrm{d}t} 
		=
		\frac{\partial H}{\partial \mathbf{\hat{n}}_{i}}\times \mathbf{\hat{n}}_{i}
		=
		-
		v_{0}\left( \mathbf{r}_{i}+\mathbf{x}_{i}\right) \times \mathbf{\hat{n}}_{i}
		+
		\sum_{j \neq i} \frac{\bm{\tau}_{ji}}{\gamma_\text{r}}\times \mathbf{\hat{n}}_{i},
	\end{eqnarray}
	where we omit the corresponding equations for $\mathbf{x}_i$ and $\mathbf{m}_i$. 
	
	Applying the constraints
	\begin{eqnarray}
		\mathbf{r}_{i}(t)+\mathbf{x}_{i}(t)&=&0, \notag \\
		\mathbf{\hat{n}}_{i}(t)+\mathbf{\hat{m}}_{i}(t)&=&0
	\end{eqnarray}
	to the equations of motion yields
	\begin{eqnarray}
		\label{Seq:EOM_constraint_MDJP}
		\frac{\mathrm{d}\mathbf{r}_{i}}{\mathrm{d}t} 
		&=&
		v_{0}\mathbf{\hat{n}}_{i}+\sum_{j \neq i} \frac{\mathbf{F}_{ji}}{\gamma_\text{t}} ,\notag  \\
		\frac{\mathrm{d}\mathbf{\hat{n}}_{i}}{\mathrm{d}t} 
		&=&
		\sum_{j \neq i } \frac{\mathbf{\Gamma }_{ji}}{\gamma_\text{r}},
	\end{eqnarray}
	which is the same as Eq.~\eqref{Seq:Langevin_MDJP}, except for the noise terms $\bm{\eta}_i^\text{t}, \bm{\eta}_i^\text{r}$.

	\subsection{Self-aligning active particles}  
	\label{subsec:self-aligning active particle}
	
	Next we study systems of self-aligning active particles --- a model that has been proposed for cells, vibrated disks, and walker robots~\cite{szabo2006phase,deseigne2010collective,baconnier2022selective,baconnier2025self}. The interplay of self-alignment and central forces between the particles has been found to give rise to flocking or collective oscillations, depending on whether the system is confined, and whether it is liquid or solid~\cite{szabo2006phase,deseigne2010collective,henkes2011active,baconnier2022selective,baconnier2025self}.
	
	Self-aligning active particles, with positions $\mathbf{r}_i$ and orientations $\mathbf{\hat{n}}_i$, tend to align their orientation towards their velocity. They interact via central forces, arising for example from hard-disk collisions or elastic springs, resulting in nonreciprocal effective interactions. The equations of motion are given by \cite{baconnier2022selective,baconnier2025self}
	\begin{eqnarray}
		\label{Seq:Langevin_SAAP}
		\frac{\mathrm{d}\mathbf{r}_{i}}{\mathrm{d}t}
		&=&
		v_0 \mathbf{\hat{n}}_{i}
		+
		\sum_{j \neq i } \frac{\mathbf{F}_{ji}}{\gamma_\text{t}} + \bm{\eta}_i^\text{t}, \notag \\
		\frac{\mathrm{d}\mathbf{\hat{n}}_{i}}{\mathrm{d}t} 
		&=&
		\frac{1}{\ell_\text{a}}\left( \mathbf{\hat{n}}_{i}\times \dot{\mathbf{r}}_{i}\right) \times \mathbf{\hat{n}}_{i} + \mathbf{\hat{n}}_{i}\times\bm{\eta}_i^\text{r}.
	\end{eqnarray}
	Here, $v_0$ denotes the self-propulsion speed, $\gamma_\text{t}$ is the translational friction coefficient, and $\mathbf{F}_{ji}=-\mathbf{F}_{ij}$ represents the reciprocal repulsive force that particle $i$ feels from particle $j$. 
	Such interaction force can be derived from a potential $U(|\mathbf{r}_i - \mathbf{r}_j|)$ between particles $i$ and $j$.
	The second equation expresses that the orientation $\hat{\mathbf{n}}_i$ of a particle rotates towards its velocity $\dot{\mathbf{r}}_i$ over an alignment length $\ell_\text{a}$. A schematic illustration is presented in Fig.~\ref{fig:illustration}c.
	Ignoring the noise terms $\bm{\eta}_i^\text{t}$ and $\bm{\eta}_i^\text{r}$, we substitute the velocity in the second equation using the first equation, thereby obtaining the following expression for the equations of motion
	\begin{eqnarray}
		\label{Seq:Langevin_SAAP_replace}
		\frac{\mathrm{d}\mathbf{\hat{n}}_{i}}{\mathrm{d}t} = 
		\frac{1}{\ell_\text{a}\gamma_\text{t}}
		\sum_{j \neq i} \left( \mathbf{\hat{n}}_{i}\times \mathbf{F}_{ji}\right)
		\times  
		\mathbf{\hat{n}}_{i}.
	\end{eqnarray}
	This expression shows that the self-aligning torque tends to align the particle orientation with the interaction force. Such a torque is intrinsically nonreciprocal because, in general, $\left[ \mathbf{\hat{n}}_{i}\times \mathbf{F}_{ji}\right]
	\times  
	\mathbf{\hat{n}}_{i} \neq -\left[ \mathbf{\hat{n}}_{j}\times \mathbf{F}_{ij}\right]
	\times  
	\mathbf{\hat{n}}_{j} $ 
	as $\hat{\mathbf{n}}_i$, and $\hat{\mathbf{n}}_j$ are in general different.
	
	We now introduce auxiliary particles with position and orientation denoted by $\mathbf{x}_i$ and $\mathbf{\hat{m}}_i$. The Hamiltonian embedding is given by 
	\begin{eqnarray}
		\label{Seq:Hamiltonian_SAAP}
		H&= &- \sum_{i} v_0 \mathbf{\hat{n}}_{i}\cdot \left( \mathbf{r}_{i}+\mathbf{x}%
		_{i}\right)  \notag \\
		&&+
		\frac{1}{\gamma_\text{t}}  \sum_{i\neq j } \left\{
		U(|\mathbf{r}_i-\mathbf{r}_j|) 
		-
		U(|\mathbf{x}_i-\mathbf{x}_j|)
		+
		\frac{1}{\ell_\text{a}}
		\left[ \mathbf{\hat{n}}_{i}\times \mathbf{F}_{ji} \right] \cdot \left( \mathbf{\hat{n}}_{i}+\mathbf{\hat{m}}_{i}\right)
		+
		\frac{1}{\ell_\text{a}}
		\left[ \mathbf{\hat{n}}_{j}\times \mathbf{F}_{ij}\right] \cdot \left( \mathbf{\hat{n}}_{j}+\mathbf{\hat{m}}_{j}\right) \right\}.
	\end{eqnarray}%
	We can now derive the corresponding equations of motion from this Hamiltonian description. The translational motion of the particles obeys
	\begin{eqnarray}
		\label{Seq:EOM_T_SAAP}
		\frac{\mathrm{d}\mathbf{r}_{i}}{\mathrm{d}t} &=&
		- \frac{\partial H}{\partial  \mathbf{r}_{i}}  \notag \\
		&=&
		v_0 \mathbf{\hat{n}}_{i}
		+
		\sum_{j \neq i } \left \{-\frac{1}{\gamma_\text{t}}
		\frac{\partial U(|\mathbf{r}_i-\mathbf{r}_j|) }{\partial \mathbf{r}_i}
		+
		\frac{1}{\ell_\text{a}\gamma_\text{t}}
		\frac{\partial \mathbf{F}_{ ij } }{\partial \mathbf{r}_i} \left[ \left( \mathbf{\hat{n}}_{j}+\mathbf{\hat{m}}_{j}\right) \times \mathbf{\hat{n}}_{j}\right]
		+
		\frac{1}{\ell_\text{a}\gamma_\text{t}}
		\frac{\partial \mathbf{F}_{ ji } }{\partial \mathbf{r}_i} \left[ \left( \mathbf{\hat{n}}_{i}+\mathbf{\hat{m}}_{i}\right) \times \mathbf{\hat{n}}_{i}\right]\right \} ,
	\end{eqnarray}
	and the orientational motion of the particle follows
	\begin{eqnarray}
		\label{Seq:EOM_O_SAAP}
		\frac{\mathrm{d}\mathbf{\hat{n}}_{i}}{\mathrm{d}t} &=&     
		\frac{\partial H}{\partial \mathbf{\hat{n}}_{i}}\times \mathbf{\hat{n}}_{i}  \notag \\
		&=&
		-
		v_0 \left( \mathbf{r}_{i}+\mathbf{x}_{i}\right) \times \mathbf{\hat{n}}_{i}
		+
		\frac{1}{\ell_\text{a}\gamma_\text{t}}
		\sum_{j \neq i} \left\{\left[ \mathbf{\hat{n}}_{i}\times \mathbf{F}_{ji}\right] \times \mathbf{\hat{n}}_{i}
		+
		\left[\mathbf{F}_{ji}\times \left( \mathbf{\hat{n}}_{i}+\mathbf{\hat{m}}_{i}\right) \right] \times \mathbf{\hat{n}}_{i}\right\}.
	\end{eqnarray}%
	Finally, the constraints read as
	\begin{eqnarray}
		\mathbf{r}_{i}(t)+\mathbf{x}_{i}(t)=0, \notag \\
		\mathbf{\hat{n}}_{i}(t)+\mathbf{\hat{m}}_{i}(t)=0.
	\end{eqnarray}
	Applying the constraints to the equations of motion yields
	\begin{eqnarray}
		\label{Seq:EOM_constraint_SAAP}
		\frac{\mathrm{d}\mathbf{r}_{i}}{\mathrm{d}t}
		&=&
		v_{0}\mathbf{\hat{n}}_{i}+\frac{1}{\gamma_\text{t}}
		\sum_{j\neq i} \mathbf{F}_{ij}, \notag \\
		\frac{\mathrm{d}\mathbf{\hat{n}}_{i}}{\mathrm{d}t}
		&=&
		\frac{1}{\ell_\text{a}\gamma_\text{t}}
		\sum_{j\neq i} \left( \mathbf{\hat{n}}_{i}\times \mathbf{F}_{ij}\right) \times \mathbf{\hat{n}}_{i},
	\end{eqnarray}
	which is equivalent to Eq.~\eqref{Seq:Langevin_SAAP}, except for the noise terms.

	\subsection{Programmed robotic metamaterials} 
	\label{subsec:Programmed rotational robots (metamaterials)}

	Let us now look at mechanical metamaterials made of robots whose interactions are programmed to break reciprocity ~\cite{brandenbourger2019nonreciprocal,ghatak2020observation}.
	In such materials, nonreciprocal interactions give rise to spatially asymmetric standing waves and unidirectionally amplified propagating waves, realizing the mechanical analogue of the non-Hermitian skin effect~\cite{Okuma2020Topological,yao2018edge}. Specifically, the robots are programmed to turn more (less) depending on how their left (right) neighbors turn, as schematized in Fig.~\ref{fig:illustration}d. The dynamics can be mapped onto the following nonreciprocal discrete wave equation, which arises from a mass-spring model in which the forces exerted by the springs on the left and right sides differ \cite{brandenbourger2019nonreciprocal}:
	\begin{eqnarray}
		\label{Seq:Langevin_PRR}
		\ddot{u}_{i} 
		&=&
		-
		k\left( 1+\varepsilon \right) \left( u_{i-1}-u_{i}\right) 
		-
		k\left( 1-\varepsilon \right) \left(u_{i+1}-u_{i}\right).
	\end{eqnarray}
	Here $u_i$ denotes the displacement of robot $i$; $k\left( 1+\varepsilon \right)$ and $k\left( 1-\varepsilon \right)$ represent the stiffnesses of coupling springs from left to right and from right to left, respectively, with the dimensionless quantity $\varepsilon$ quantifying nonreciprocity. For convenience, we have set the mass to unity. 
	
	We now introduce the conjugate momentum $p_i^u$ for $u_i$, along with the auxiliary degrees of freedom $p_i^w$ and $w_i$ together with the Poisson bracket relations, $\left\{ u_i,p_j^u \right \} =\delta_{ij}$, and $\left\{ w_i,p_j^w \right \} =\delta_{ij}$. A Hamiltonian can then be constructed as follows:
	\begin{equation}
		\label{Seq:Hamiltonian_PRR}
		H
		=
		\sum_i 
		\left [\frac{\left( p_{i}^{u}\right) ^{2}}{2}-\frac{\left( p_{i}^{w}\right)^{2}}{2}
		+   
		k\left( 1+\varepsilon \right) \left( u_{i-1}-u_{i}\right) \left(u_{i}+w_{i}\right) 
		+   
		k\left( 1-\varepsilon \right) \left(u_{i+1}-u_{i}\right) \left( u_{i}+w_{i}\right)
		\right ].
	\end{equation}
	We observe that the kinetic energy of the auxiliary particles is negative. Such a negative kinetic energy term in the auxiliary system is necessary to enforce our constraints; a more detailed explanation will be provided in Sec.~\ref{subsec:Adding quadratic kinetic energy term to Hamiltonian}. 
	
	We can derive the corresponding equations of motion from the Hamiltonian description
	\begin{eqnarray}
		\label{Seq:EOM_PRR}
		\dot{u}_{i} &=& \{u_{i}, H \} = p_{i}^{u}, \notag \\
		\dot{w}_{i} &=& \{w_{i}, H \} = -p_{i}^{w},  \notag \\
		\dot{p}_{i}^{u} &=& \{p_{i}^{u}, H \} =
		-
		k\left( 1+\varepsilon \right) \left(u_{i-1}-u_{i}\right) 
		-   
		k\left( 1-\varepsilon \right) \left(u_{i+1}-u_{i}\right)  \notag \\
		&&+
		k\left( 1+\varepsilon \right) \left( u_{i}+w_{i}\right) 
		+
		k\left(1-\varepsilon \right) \left( u_{i}+w_{i}\right) 
		-   
		k\left( 1+\varepsilon
		\right) \left( u_{i+1}+w_{i+1}\right) -k\left( 1-\varepsilon \right) \left(
		u_{i-1}+w_{i-1}\right),  \notag \\
		\dot{p}_{i}^{w} &=& \{p_{i}^{w}, H \} = 
		-
		k\left( 1+\varepsilon \right) \left(
		u_{i-1}-u_{i}\right) -k\left( 1-\varepsilon \right) \left(
		u_{i+1}-u_{i}\right) .
	\end{eqnarray}
	The constraint here involves the momenta as well, and read as
	\begin{eqnarray}
		\label{Seq:constraint_PRR}
		u_{i}(t)+w_{i}(t)&=&0, \notag \\
		p_{i}^{u}(t)-p_{i}^{w}(t)&=& 0 .
	\end{eqnarray}
	Applying the constraint to the equations of motion yields
	\begin{eqnarray}
		\label{Seq:EOM_constraint_PRR}
		\ddot{u}_{i} &=&
		-
		k\left( 1+\varepsilon \right) \left( u_{i-1}-u_{i}\right)
		-
		k\left( 1-\varepsilon \right) \left( u_{i+1}-u_{i}\right).
	\end{eqnarray}
	The equation of motion of $u_i$ is the same as Eq.~\eqref{Seq:Langevin_PRR}.

	\subsection{Feedback-controlled active particles with vision-cone interactions}
	\label{subsec:Feedback-controlled active particles}
	
	We next consider light-activated active particles whose motility is controlled by an external feedback loop, consisting of a particle detection algorithm and a scanning laser system \cite{bauerle2018self,lavergne2019group}. Through this feedback system, particles can be programmed to self-propel only when there are enough other particles in their vision cone~\cite{lavergne2019group}, thus mimicking the behavior of animals such as midges. To decide whether a given particle $i$ should self-propel or not, a perception metric is computed as
	\begin{equation}
		P_i= \sum_{j\in V_i^\alpha} \frac{1}{2\pi r_{ij}},
	\end{equation}
	where $V_i^\alpha$ denotes the vision cone of angle $\alpha$ around the orientation of particle $i$ (Fig.~\ref{fig:illustration}e). Respectively, $r_{ij}$ is the distance between particles $i$ and $j$, so that the perception metric assigns less weight to more distant particles. If a given particle has a perception metric $P_i>P^*$, it is made to propel at speed $v_0$, which is achieved by shining light on it through the laser scanning system. Otherwise, the particle remains passive. Thus, particle motion can be modeled as
	\begin{eqnarray}
		\label{Seq:Langevin_FCAP}
		\frac{\mathrm{d}\mathbf{r}_{i}}{\mathrm{d}t}
		& =&
		v\left( P_{i}\right) \hat{\mathbf{n}}_{i}
		- \sum_{j\neq i}
		\frac{1}{\gamma_\text{t}}\frac{\partial U(r_{ij})}{ \partial \mathbf{r}_{i}} + \bm{\eta}_i^\text{t}, \notag \\
		\frac{\mathrm{d}\mathbf{\hat{n}}_{i}}{\mathrm{d}t} &=& \hat{\mathbf{n}}_i\times\bm{\eta}_i^\text{r},
	\end{eqnarray}
	where the self-propulsion speed $v$ is given by
	\begin{equation}
		v(P_i) = \left\{\begin{array}{ll} v_0; & P_i>P^*\\
			0; & P_i <P^*.
		\end{array}\right.
	\end{equation}
	Respectively, $\gamma_\text{t}$ is the translational friction coefficient, $\bm{\eta}_i^\text{t}$ and $\bm{\eta}_i^\text{r}$ are the translational and rotational noises, and $U(r_{ij})$ is an interaction potential to model excluded-volume interactions, which can be done through the repulsive Weeks-Chandler-Andersen (WCA) potential given by
	\begin{equation}
		U(r_{ij}) = \left\{
		\begin{array}{lr}
			4\epsilon\left[ \left(\frac{\sigma}{r_{ij}}\right)^{12} -  \left(\frac{\sigma}{r_{ij}}\right)^{6}  + \frac{1}{4} \right], & \text{for } r_{ij}\leq \sigma\sqrt[6]{2} \\
			0, & \ \ \text{for }  r_{ij}> \sigma\sqrt[6]{2},
		\end{array}
		\right.
	\end{equation}
	where $\epsilon$ is the potential strength and $\sigma$ is the particle diameter.

	We now introduce auxiliary particles with position and orientation denoted
	by $\mathbf{x}_{i}$ and $\mathbf{\hat{m}}_{i}$, respectively, to construct
	the Hamiltonian embedding 
	\begin{equation}
		H=
		-\sum_{i} 
		v\left( P_{i}\right) \hat{\mathbf{n}}_{i} \cdot \left( \mathbf{r}%
		_{i}+\mathbf{x}_{i}\right) 
		+
		\frac{1}
		{\gamma _{\text{t}}}
		\sum_{i<j}%
		\left[ U(r_{ij})-U(x_{ij})\right] .  \label{Seq:Hamiltonian_FCAP}
	\end{equation}%
	We can derive the equations of motion for the particle positions and orientations from the
	Hamiltonian description as follows
	\begin{eqnarray}
		\frac{\mathrm{d}\mathbf{r}_{i}}{\mathrm{d}t} 
		&=&
		-\frac{\partial H}{\partial  \mathbf{r}_{i}}
		=
		v\left( P_{i}\right) \hat{\mathbf{n}}_{i}
		+
		\sum_{j} \frac{\partial v\left( P_{j}\right) }{\partial \mathbf{r}_{i}}\hat{\mathbf{n}	}_{j}\cdot \left( \mathbf{r}_{j}+\mathbf{x}_{j}\right) 
		-
		\sum_{j\neq i}\frac{1%
		}{\gamma _{\text{t}}}\frac{\partial U(r_{ij})}{\partial \mathbf{r}_{i}}, 
		\notag  \label{Seq:EOM_translation_FCAP} \\
		\frac{\mathrm{d}\mathbf{\hat{n}}_{i}}{\mathrm{d}t} &=&
		\frac{\partial H}{\partial \mathbf{\hat{n}}_{i}}\times \mathbf{\hat{n}}_{i}
		=
		v\left(P_{i}\right) \left( \mathbf{r}_{i}+\mathbf{x}_{i}\right) \times \mathbf{\hat{n}}_{i}.
	\end{eqnarray}%
	Finally, the constraints read as 
	\begin{eqnarray}
		\mathbf{r}_{i}(t)+\mathbf{x}_{i}(t) &=&0,  \notag \\
		\mathbf{\hat{n}}_{i}(t)+\mathbf{\hat{m}}_{i}(t) &=&0.
	\end{eqnarray}%
	Applying the constraint to the equations of motion yields 
	\begin{eqnarray}
		\frac{\mathrm{d}\mathbf{r}_{i}}{\mathrm{d}t} &=&
		v\left( P_{i}\right) \hat{\mathbf{n}}_{i}
		-
		\frac{1}{\gamma _{\text{t}}}\sum_{j\neq i}\frac{\partial
			U(r_{ij})}{\partial \mathbf{r}_{i}},  \notag  \label{Seq:EOM_constraint_FCAP}
		\\
		\frac{\mathrm{d}\mathbf{\hat{n}}_{i}}{\mathrm{d}t} &=&0,
	\end{eqnarray}%
	which is the same as Eq.~\eqref{Seq:Langevin_FCAP}, except for the noise terms.

	\subsection{Sedimenting particles with hydrodynamic interactions}
	\label{subsec:Hydrodynamic_couplings_sedimenting_systems}
	
	Sedimenting particles provide a paradigmatic example of a system with velocity-dependent nonreciprocal interactions due to hydrodynamics. When particles settle through a viscous fluid under gravity, they generate flow fields that affect the motion of neighboring particles. In general, the flow generated by particle $j$ impacts particle $i$ differently than the flow generated by particle $i$ impacts particle $j$ (Fig.~\ref{fig:illustration}f). Therefore, the resulting interactions are intrinsically nonreciprocal. This nonreciprocity arises not from the particles themselves, but from the asymmetry of the hydrodynamic coupling through the surrounding fluid.
	
	Consider two spherical particles sedimenting under gravity in a Stokes flow regime, where inertial effects are negligible. The equation of motion for particle $i$ reads as
	\begin{align}
		\label{seq:eom_hydrodynamic_coupling}
		\gamma \dot{\mathbf{r}}_i = m_i \mathbf{g} + \sum_{j \neq i} \mathbf{F}_{j \to i}^{\text{hydro}}
		+ \bm{\eta}_i,
	\end{align}
	where $m_i$ is the particle mass, $\mathbf{g}$ is gravitational acceleration, $\gamma$ is the Stokes drag coefficient (proportional to the fluid viscosity), $\mathbf{F}_{j \to i}^{\text{hydro}}$ denotes the hydrodynamic force exerted on particle $i$ by the flow generated by particle $j$, and $\bm{\eta}_i$ is the noise term.
	
	The hydrodynamic interaction force can be expressed in terms of the Oseen tensor $\mathbf{G}$ (also called the Stokeslet):
	\begin{align}
		\mathbf{F}_{j \to i}^{\text{hydro}} = \gamma \mathbf{G}(\mathbf{r}_{ij}) \cdot m_j \mathbf{g},
	\end{align}
	where $\mathbf{r}_{ij} = \mathbf{r}_i - \mathbf{r}_j$ is the inter-particle separation, 
	and the Oseen tensor is given by
	\begin{align}
		\mathbf{G}(\mathbf{r}) = \frac{1}{8\pi\eta r}\left(\mathbf{I} + \frac{\mathbf{r}\otimes\mathbf{r}}{r^2}\right),
	\end{align}
	with $\eta$ the dynamic viscosity of the fluid, $\mathbf{I}$ the identity matrix, 
	and $[\mathbf{r}\otimes\mathbf{r}]_{\alpha\beta} = r_\alpha r_\beta$. The symmetry of this tensor implies that $\mathbf{G}(\mathbf{r}_{ij}) = \mathbf{G}(\mathbf{r}_{ji})$,
	reflecting the directional dependence of hydrodynamic coupling on the separation vector. Therefore, $\mathbf{F}_{j \to i}^{\text{hydro}}$ and $\mathbf{F}_{i \to j}^{\text{hydro}}$ point in the same direction since the masses of particles are all positive. 
	Hence, the interaction forces are not equal and opposite; the interactions are nonreciprocal. 
	Crucially, the nonreciprocal interactions in this system are velocity-\textit{independent}: they depend only on particle positions through the inter-particle distance.

	We can now introduce auxiliary particles with position denoted by $\mathbf{x}_i$. The Hamiltonian embedding is given by 
	\begin{align}
		H = - \sum_i m_i \mathbf{g}\cdot \mathbf{r}_i - \sum_{\langle  ij \rangle}\mathbf{F}_{j \to i}^{\text{hydro}} \cdot (\mathbf{x}_i + \mathbf{r}_i).
	\end{align}
	The corresponding equation of motion from this Hamiltonian description reads as
	\begin{align}
		\frac{\mathrm{d} \mathbf{r}_i}{\mathrm{d}t } 
		=
		- \frac{\partial H}{\partial \mathbf{r}_i} 
		&= m_i \mathbf{g} + \sum_{j \in \langle  ij \rangle} \mathbf{F}_{j \to i}^{\text{hydro}} + \sum_{j \in \langle  ij \rangle} \frac{\partial }{ \partial \mathbf{r}_i} \mathbf{F}_{j \to i}^{\text{hydro}} \cdot (\mathbf{x}_i + \mathbf{r}_i).
	\end{align}
	Moreover, the constraint is
	\begin{align}
		\mathbf{x}_i (t) + \mathbf{r}_i(t) = 0.
	\end{align}
	Applying the constraints to the equations of motion yields
	\begin{align}
		\frac{\mathrm{d} \mathbf{r}_i}{\mathrm{d}t } 
		&=m_i \mathbf{g} +   \sum_{j \in \langle  ij \rangle}\mathbf{F}_{j \to i}^{\text{hydro}}\; ,
	\end{align}
	which is equivalent to Eq.~\eqref{seq:eom_hydrodynamic_coupling}, up to the noise term.

	\section{Symplectic structure of the Hamiltonian embedding: Implementing the constraint for interactions with overdamped and undamped nonreciprocal dynamics}
	\label{sec:Two ways of enforcing the constraint}
	
	In Sec.~\ref{sec:embedding-proof}, we introduced Hamiltonian embeddings and outlined the appropriate constraints for several nonreciprocal systems, both for damped and undamped dynamics in Sec.~\ref{sec:examples}. Here, we elaborate on these two cases separately, discussing in detail the implementation of the constraint and the corresponding symplectic structure. Although the discussion below is general, for concreteness we shall illustrate the procedure using the vision-cone XY spins as an example, for which the interaction $J_{ij}(\theta _{i})$ takes the form
	\begin{equation}
		\label{Seq:VC_interaction}
		J_{ij}(\theta _{i})=\left\{ 
		\begin{array}{cc}
			J, & \text{for}\;  \min \left\{ 2\pi -\left\vert \theta _{i}-\psi _{ij}\right\vert
			,\left\vert \theta _{i}-\psi _{ij}\right\vert \right\} \leq \Psi \\ 
			0, & \mathrm{else.}%
		\end{array}
		\right. 
	\end{equation}%
	Below, we start by introducing the form of the Lagrangian and then derive the corresponding Hamiltonian, from which the dissipative and inertial nonreciprocal equations of motion can be obtained, along with the associated Poisson bracket relations. Subsequently, we impose the appropriate constraint and demonstrate that it is preserved at all times.

	\subsection{Overdamped nonreciprocal systems}
	\label{subsec:Dissipative nonreciprocal systems}
	
	Consider a dissipative system with nonreciprocal interactions in the equations of motion~\eqref{Seq:NR_XY}. Since the equation of motion is first-order in $\theta$, we consider a kinetic energy term linear in $\dot\theta$, expressed as $\mathcal{K} = \sum_{i}\varphi _{i}\dot{\theta}_{i}$. Such kinetic energy terms are commonly used to describe charged particles on a plane exposed to a perpendicular magnetic field~\cite{tong2016lectures}. 
	In addition, we denote by $\mathcal{U}$ the symmetrized interactions, cf.~Eq.~\eqref{Seq:NR_Hamil}. The explicit form of the Lagrangian is given by
	\begin{eqnarray}
		\label{Seq:Lagrangian_kinetic}
		\mathcal{L} &=& \mathcal{K}-\mathcal{U} \notag \\
		&=&
		\sum_{i}\varphi _{i}\dot{\theta}_{i}+\sum_{\langle ij \rangle}
		\left \{
		J_{ij}\left( \theta _{i}\right) \left[ \cos (\theta _{i}-\theta _{j})
		+
		\cos(\varphi _{i}-\theta _{j})\right] 
		+
		J_{ji}\left( \theta _{j}\right) \left[ \cos (\theta _{j}-\theta _{i})
		+
		\cos(\varphi _{j}-\theta _{i})\right] 
		\right \}.
	\end{eqnarray}
	From this Lagrangian, we can derive the canonical momentum associated with $\theta _{i}$ as
	\begin{equation}
		\label{Seq:Lagrangian_conjugate}
		p_{i}=\frac{\partial\mathcal{L}}{\partial \dot{\theta}_{i}}=\varphi _{i},
	\end{equation}%
	which leads to the Poisson bracket relation $\left\{ \theta _{i},\varphi _{j}\right \} =\delta_{ij}$. This implies that the original spins $\theta_i$ and the auxiliary spins $\varphi_i$ are conjugate degrees of freedom. Therefore, they define a phase space that is a torus (i.e., compact).
	
	Using the definition for the canonical momenta, we can construct the corresponding Hamiltonian from the Lagrangian $\mathcal{L}$ by performing a Legendre transform
	\begin{eqnarray}
		\label{Seq:Lagrangian_Hamiltonian}
		\mathcal{H}&=& \sum_i p_{i}\dot{\theta}_{i}-\mathcal{L}= \mathcal{U} \notag \\
		&=& -\sum_{\langle ij\rangle} \left\{ J_{ij}\left( \theta _{i}\right) \left[ \cos (\theta _{i}-\theta
		_{j})+\cos (\varphi _{i}-\theta _{j})\right] 
		+
		J_{ji}\left( \theta _{j}\right) \left[ \cos (\theta _{j}-\theta _{i})
		+
		\cos(\varphi _{j}-\theta _{i})\right] \right \}.
	\end{eqnarray}%
	Thus, we recover the Hamiltonian introduced in the Main Text, cf.~Eq.~(2). Based on the above result, we can derive the dynamics of $\theta _{i}$ and $\varphi_{i}$
	\begin{eqnarray}
		\label{Seq:Lagrangian_Hamiltonian_EOM}
		\dot{\theta}_{i} =
		\left\{ \theta _{i},\mathcal{H}\right\}
		&=&
		\sum_{j \in \langle ij \rangle} J_{ij}\left( \theta _{i}\right) \sin (\varphi _{i}-\theta_{j}), \notag \\
		\dot{\varphi}_{i}  = 
		\left\{ \varphi _{i},\mathcal{H}\right\}
		&=&
		\sum_{j \in \langle ij \rangle} \big\{-J_{ij}(\theta _{i})\sin (\theta _{i}-\theta _{j}) \notag \\
		&&+ 
		\partial _{\theta _{i}}J_{ij}(\theta _{i})\left[ \cos(\theta _{i}-\theta _{j})+\cos (\varphi _{i}- \theta _{j})\right]-J_{ji}(\theta _{j})\left[\sin (\theta _{i}-\theta _{j})+\sin (\theta _{i}-\varphi _{j})\right]\big\} .
	\end{eqnarray}
	Now, let us define the constraint $G_i$ in phase space as the difference between the degrees of freedom $\theta_i$ and $\varphi_i$,
	\begin{equation}
		\label{Seq:Dissipative_initial}
		G_i = \theta _{i} - \varphi _{i}.
	\end{equation}
	Using Eq.~\eqref{Seq:Lagrangian_Hamiltonian_EOM}, the equation of motion of the constraint is given by 
	\begin{eqnarray}
		\label{Seq:Dissipative_initial_dot}
		\frac{\mathrm{d} G_i}{\mathrm{d}t} &=& \{G_i,\mathcal{H}\} = \sum_j \partial_{\theta_j}G_i \partial_{\varphi_j}\mathcal{H} - \partial_{\varphi_j}G_i \partial_{\theta_j}\mathcal{H} = \dot\theta_i-\dot\varphi_i \notag \\
		&=&  \sum_{j \in \langle ij \rangle} J_{ij}\left( \theta _{i}\right) \sin (\varphi _{i}-\theta_{j}) - \sum_{j \in \langle ij \rangle} \big\{ J_{ij}(\theta _{i})\sin (\theta _{i}-\theta _{j}) \notag \\
		&&+ 
		\partial _{\theta _{i}}J_{ij}(\theta _{i})\left[ \cos(\theta _{i}-\theta _{j})+\cos (\varphi _{i}- \theta _{j})\right]-J_{ji}(\theta _{j})\left[\sin (\theta _{i}-\theta _{j})+\sin (\theta _{i}-\varphi _{j})\right] \big \}.
	\end{eqnarray}
	When the degrees of freedom lie on the constraint surface, i.e., $G_i = \pi$, the following relations hold
	\begin{eqnarray}
		\label{Seq:Hamiltonian_dissipative}
		\cos \left( \theta _{i}-\theta _{j}\right) +\cos \left( \varphi _{i}-\theta
		_{j}\right)  &=&0, \notag \\
		\sin \left( \theta _{i}-\theta _{j}\right) +\sin \left( \theta _{i}-\varphi_{j}\right)  &=&0 .
	\end{eqnarray} 
	This results in the value of Eq.~\eqref{Seq:Dissipative_initial_dot} being zero. Therefore, the constraint is preserved by the dynamics.
	Consequently, similar to the discussion in Eq.~\eqref{eq:constr_t}, if initially the system is under the constraint, $G_i(t=0)= \pi$, the constraint is preserved at all times: $G_i(t)= \pi$.

	Under the constraint, the equations of motion are given by
	\begin{eqnarray}
		\label{Seq:Lagrangian_Hamiltonian_EOM_constraint}
		\dot{\theta}_{i} 
		&=&
		\sum_{j \in \langle ij \rangle}
		J_{ij}\left( \theta _{i}\right) \sin(\varphi _{i}-\theta _{j})
		=
		-\sum_{j \in \langle ij \rangle}
		J_{ij}\left( \theta _{i}\right)\sin (\theta _{i}-\theta _{j}),  \notag \\
		\dot{\varphi}_{i} 
		&=&
		\sum_{j \in \langle ij \rangle}
		-J_{ij}(\theta _{i})\sin (\theta _{i}-\theta _{j})
		=
		-\sum_{j \in \langle ij \rangle}
		J_{ij}\left( \theta _{i}\right) \sin(\varphi _{i}-\varphi _{j}).
	\end{eqnarray}
	
	Thus, the application of the constraint to the Hamiltonian embedding allows us to retrieve the overdamped nonreciprocal dynamics for the original degrees of freedom $\theta_i$.

	\subsection{Undamped nonreciprocal systems}
	\label{subsec:Adding quadratic kinetic energy term to Hamiltonian}

	Inspired by the approach in the previous subsection, we introduce a general procedure to construct the Hamiltonian embedding for undamped systems with nonreciprocal interactions. We will provide the explicit form of the Hamiltonian for the vision-cone XY spins, and explain how the dynamics naturally emerge from the constraint.

	We first introduce the Lagrangian. It is composed of two terms: the first one is the quadratic kinetic energy $\mathcal{K} = \sum_{i} ( \frac{1}{2}\dot{\theta}_{i}^2-\frac{1}{2}\dot{\varphi}_{i}^2 )$ and the second one is the interaction energy $\mathcal{U}$ (see below for an explanation of the negative kinetic energy for $\varphi$). 
	The explicit form of the Lagrangian for undamped nonreciprocal systems is given by
	\begin{eqnarray}
		\label{Seq:Lagrangian_kinetic_I}
		\mathcal{L} &=& \mathcal{K}-\mathcal{U} \notag \\
		&=&
		\sum_{i}
		\left(\frac{1}{2}\dot{\theta}_{i}^2-\frac{1}{2}\dot{\varphi}_{i}^2 \right)
		+
		\sum_{\langle ij \rangle} \left\{ J_{ij}\left( \theta _{i}\right) \left[ \cos (\theta _{i}-\theta _{j})
		+
		\cos(\varphi _{i}-\theta _{j})\right] 
		+
		J_{ji}\left( \theta _{j}\right) \left[ \cos (\theta _{j}-\theta _{i})
		+
		\cos(\varphi _{j}-\theta _{i})\right] \right \} .
	\end{eqnarray}
	From this Lagrangian, we can derive the conjugate degree of freedom (i.e., the canonical momenta) associated with $\theta _{i}$ and $\varphi _{i}$:
	\begin{equation}
		\label{Seq:Lagrangian_conjugate_I}
		L_{i}^{\theta} = \frac{\partial\mathcal{L}}{\partial \dot{\theta}_{i}}=\dot{\theta}_{i},\qquad 
		L_{i}^{\varphi} = \frac{\partial\mathcal{L}}{\partial \dot{\varphi}_{i}}= - \dot{\varphi}_{i},
	\end{equation}
	which leads to the Poisson bracket relations $\{ \theta_i,L_{i}^{\theta} \} =\delta_{ij}$ and $\{ \varphi_i,L_{i}^{\varphi} \} =\delta_{ij}$. Unlike the overdamped case, here we end up with two spins; for each, the phase space is a semi-compact cylinder, since the original coordinate angle is compact but angular momentum is not. 
	
	Using the canonical momenta, we can construct the corresponding Hamiltonian $\mathcal{H}$ from the Lagrangian $\mathcal{L}$ using a Legendre transform:
	\begin{eqnarray}
		\label{Seq:Hamiltonian_quad_kinetic}
		\mathcal{H}&=&\sum_i \left(  L_{i}^{\theta} \dot{\theta}_i+L_{i}^{\varphi} \dot{\varphi}_i \right)-\mathcal{L} =\mathcal{K} + \mathcal{U}  \\
		&=&\frac{1}{2}\sum_{i} \left[\left( {L_{i}^{\theta }}\right)^{2}-\left( {L_{i}^{\varphi }} \right)^{2} \right ]
		-
		\sum_{\langle ij \rangle} 
		\left\{ J_{ij}\left( \theta _{i}\right) \left[ \cos (\theta_{i}-\theta _{j})+\cos (\varphi _{i}-\theta _{j})\right]
		+
		J_{ji}\left( \theta _{j}\right) \left[ \cos (\theta _{j}-\theta _{i})
		+
		\cos(\varphi _{j}-\theta _{i})\right]\right \} . \notag
	\end{eqnarray}
	An unusual feature of the kinetic energy $\mathcal{K}$ is that it is negative-definite for the auxiliary degree of freedom $\varphi_j$. As elaborated on above, mathematically this is required to maintain the constraint $\theta_i(t)=\varphi_i (t) + \pi$ at all times.

	Based on the above result, we derive the dynamics of $\theta _{i}$, $\varphi_{i}$, $L_{i}^{\theta }$, and $L_{i}^{\varphi }$ as follows:
	\begin{eqnarray}
		\label{Seq:Hamiltonian_quad_kinetic_EOM}
		\dot{\theta}_{i} =
		\left\{ \theta _{i},\mathcal{H}\right\} 
		&=&
		{L_{i}^{\theta}},  \notag \\
		\dot{\varphi}_{i} =
		\left\{ \varphi _{i},\mathcal{H}\right\} 
		&=&
		-{L_{i}^{\varphi}},  \notag \\
		{\dot{L}_{i}^{\theta }} =
		\left\{ {L_{i}^{\theta }},\mathcal{H}\right\}
		&=&
		-\sum_{j \in \langle ij \rangle}  \big \{J_{ij}(\theta _{i})\sin (\theta _{i}-\theta_{j}) \notag \\
		&&+
		\partial _{\theta _{i}}J_{ij}(\theta _{i})\left[ \cos (\theta_{i}-\varphi _{j})+\cos (\theta _{i}-\theta _{j})\right] +J_{ji}(\theta _{j})\left[ \sin (\theta _{i}-\theta _{j})+\sin (\theta _{i}-\varphi _{j})\right] \big \}, \notag \\
		{\dot{L}_{i}^{\varphi }} =
		\left\{ {L_{i}^{\varphi }},\mathcal{H}\right\}
		&=&
		-\sum_{j \in \langle ij \rangle} J_{ij}(\theta _{i})\sin (\varphi _{i}-\theta _{j}).
	\end{eqnarray}
	
	Now let us introduce two sets of constraints $G^{(1)}_i$ and $G^{(2)}_i$. The first set represents the difference between the degrees of freedom $\theta_i$ and $\varphi_i$,
	\begin{equation}
		\label{Seq:Hamiltonian_quad_kinetic_posi_initial}
		G^{(1)}_i = \theta _{i} - \varphi _{i},
	\end{equation}
	and the second concerns the sum of angular momenta
	\begin{equation}
		\label{Seq:Hamiltonian_quad_kinetic_mome_constraint}
		{G^{(2)}_i = L_{i}^{\theta }+L_{i}^{\varphi }.} 
	\end{equation}
	Using Eq.~\eqref{Seq:Hamiltonian_quad_kinetic_EOM}, the equations of motion of the constraints, $G^{(1)}_i$ and $G^{(2)}_i$, are given by 
	\begin{eqnarray}
		\frac{\mathrm{d}G^{(1)}_i}{\mathrm{d}t} &=& \left\{ G^{(1)}_i,\mathcal H \right\} = L_{i}^{\theta }+L_{i}^{\varphi } = G^{(2)}_i  \label{Seq:EOM_constraint_1} \\
		\frac{\mathrm{d}G^{(2)}_i}{\mathrm{d}t} &=& \left\{ G^{(2)}_i,\mathcal H \right\} = \dot{L}_{i}^{\theta }+\dot{L}_{i}^{\varphi } \notag \\
		&=& 
		-
		\sum_{j \in \langle ij \rangle} J_{ij}(\theta _{i})\sin (\theta _{i}-\theta_{j})-\sum_{j \in \langle ij \rangle} \big \{ J_{ij}(\theta _{i})\sin (\varphi _{i}-\theta _{j}) \notag \\
		&&+
		\partial _{\theta _{i}}J_{ij}(\theta _{i})\left[ \cos (\theta_{i}-\varphi _{j})+\cos (\theta _{i}-\theta _{j})\right] +J_{ji}(\theta _{j})\left[ \sin (\theta _{i}-\theta _{j})+\sin (\theta _{i}-\varphi _{j})\right] \big \}.\label{Seq:EOM_constraint_2}
	\end{eqnarray}
	When the system is on the constraint surface in phase space, defined by $G^{(1)}_i = \pi$ and $G^{(2)}_i = 0$, the time derivatives of the constraints are all zero. In particular, the constraint $G^{(2)}_i=0$, which represents the opposite direction of the angular momenta, will make Eq.~\eqref{Seq:EOM_constraint_1} evaluate to zero:
	\begin{equation}
		\label{Seq:Hamiltonian_quad_kinetic_condition_3}
		\frac{\mathrm{d}G^{(1)}_i}{\mathrm{d}t}  =0.
	\end{equation}
	Simultaneously, the constraint $G^{(1)}_i(0)=\pi$, which represents the opposite direction of the spins, leads to the following condition:
	\begin{eqnarray}
		\label{Seq:Hamiltonian_quad_kinetic_condition_1}
		\cos \left( \theta _{i}-\theta _{j}\right) +\cos \left( \varphi
		_{i}-\theta _{j}\right)  &=&0, \notag \\
		\sin \left( \theta _{i}-\theta _{j}\right) +\sin \left( \varphi
		_{i}-\theta _{j}\right)  &=&0.
	\end{eqnarray}
	This results in the value of Eq.~\eqref{Seq:EOM_constraint_2} being zero:
	\begin{equation}
		\label{Seq:Hamiltonian_quad_kinetic_condition_2}
		\frac{\mathrm{d}G^{(2)}_i}{\mathrm{d}t} =0.
	\end{equation}
	Similar to the overdamped case above, these constraints hold at all times.

	Under the constraints, the equations of motion are given by
	\begin{eqnarray}
		\label{Seq:Hamiltonian_quad_kinetic_EOM_constraint}
		\dot{\theta}_{i} =
		\left\{ \theta _{i},\mathcal{H}\right\} 
		&=&
		{L_{i}^{\theta}},  \notag \\
		\dot{\varphi}_{i} =
		\left\{ \varphi _{i},\mathcal{H}\right\} 
		&=&
		-{L_{i}^{\varphi}},  \notag \\
		{\dot{L}_{i}^{\theta }} =
		\left\{ {L_{i}^{\theta }},\mathcal{H}\right\}
		&=&
		-\sum_{j \in \langle ij \rangle} J_{ij}(\theta _{i})\sin (\theta _{i}-\theta_{j}) \notag \\
		{\dot{L}_{i}^{\varphi }} =
		\left\{ {L_{i}^{\varphi }},\mathcal{H}\right\}
		&=&
		\sum_{j \in \langle ij \rangle} J_{ij}(\theta _{i})\sin (\varphi _{i}-\varphi _{j}).
	\end{eqnarray}
	We can also write them in the form of the second-order time derivative of $\theta_i$ and $\varphi_i$,
	\begin{eqnarray}
		\label{Seq:Hamiltonian_EOM}
		\ddot{\theta}_{i} 
		&=&
		-\sum_{j \in \langle ij \rangle} J_{ij}(\theta _{i})\sin (\theta _{i}-\theta_{j}),  \notag \\
		\ddot{\varphi}_{i} 
		&=&
		- \sum_{j \in \langle ij \rangle} J_{ij}(\theta _{i})\sin (\varphi _{i}-\varphi _{j}). 
	\end{eqnarray}
	Thus, the application of the constraints to the Hamiltonian embedding allows us to retrieve the undamped nonreciprocal dynamics for the original degree of freedom $\theta_i$.

	The occurrence of negative kinetic energy for the auxiliary degrees of freedom can be explained as follows: Consider first the free system where the degrees of freedom are noninteracting, $\mathcal{U}=0$, and both terms in the kinetic energy are positive; then, by conservation of angular momentum $L^\varphi_j + L^\theta_j=0$, the $\theta$ and $\varphi$ degrees of freedom will rotate opposite to one another, thus violating the constraint that pins them to move together (i.e., in the same direction). To enforce the latter without over-constraining the dynamics of the resulting joint motion, a negative kinetic energy is required.
	Returning to the Hamiltonian~\eqref{Seq:Hamiltonian_quad_kinetic}, more formally, as discussed above, we need to construct a constraint as $\theta_i(t)=\varphi_i (t) + \pi $. Taking two time derivatives, we have $\ddot{\theta}_i(t) = \ddot{\varphi}_i (t) $. If the interactions between the original and the auxiliary degrees of freedom are defined as in Eq.~\eqref{Seq:NR_Hamil}, the torques experienced by the original spin and the auxiliary spin will be oppositely directed under the imposed constraint, as given by $-\sum_{j \in \langle ij \rangle } J_{ij} \sin(\theta_i - \theta_j) = -\big[-\sum_{j \in \langle ij \rangle} J_{ij} \sin(\varphi_i - \theta_j)\big]$. To maintain the opposite sign in the EOM of the angular momenta, the kinetic energy for the auxiliary degree of freedom has a negative sign, $-\frac{1}{2}\dot{\varphi}_{i}^2$.
	Such negative kinetic energy terms for the auxiliary degree of freedom have been discussed in Ref.~\cite{galley2013classical} where they arise as a result of reformulating the action principle as an initial- (rather than a boundary-) value problem. Similarly, in the programmed robotic metamaterials example from Sec.~\ref{subsec:Programmed rotational robots (metamaterials)}, we used the kinetic term $\frac{1}{2}\sum_i \left( p_{i}^{u}\right) ^{2}-\left( p_{i}^{w}\right)^{2}$ to construct the Hamiltonian. Such anisotropic kinetic energy terms have also been encountered before in the description of azimuthal curl forces~\cite{berry2015hamiltonian}.

	As we showed at the beginning of Appendix~\ref{sec:embedding-proof}, one can extend the arguments of this section to more general nonreciprocal systems whose equations of motion feature both inertial and dissipative terms.

	\section{Equivalence between nonreciprocal Langevin dynamics and Glauber Monte-Carlo dynamics}
	\label{sec:Langevin_MC_equivalence}

	This section demonstrates that the steady-state and non-stationary density distributions resulting from the original nonreciprocal Langevin dynamics and from the constrained Glauber dynamics, based on the Hamiltonian that we propose, are identical. 
	In Sec.~\ref{subsec:langevin_descr}, we first derive a Fokker-Planck-like equation for the phase-space density distribution obtained from Langevin dynamics; then, in Sec.~\ref{subsec:glauber}, we prove that it is equivalent to the corresponding distribution obtained using constrained Glauber dynamics.
	
	Below, we consider for concreteness the nonreciprocal Hamiltonian for the vision-cone interactions (see Main Text for the details)
	\begin{equation}
		\label{Seq:Hamiltonian}
		H=H_{SS}+H_{Sa} = - \sum_{\langle ij\rangle}\left\{ \left[ J_{ij}(\theta _{i})+J_{ji}(\theta _{j})\right] \cos (\theta _{i}-\theta _{j})  + J_{ij}(\theta _{i})\cos (\theta _{j}-\varphi_{i}) + J_{ji}(\theta _{j})\cos (\theta _{i}-\varphi _{j}) \right \},
	\end{equation}
	although the procedure is general and applies to other models (see Sec.~\ref{sec:examples}).
	
	Before presenting the detailed derivations, we briefly clarify the notions of detailed balance and global balance~\cite{landau2021guide} for Markov dynamics.
	Let $\mathcal{C}$ denote a microstate (configuration), e.g., the set of spin angles $\{\theta_i,\varphi_i\}$ of the Hamiltonian embedding in Eq.~\eqref{Seq:Hamiltonian}, and let $W(\mathcal{C}\to\mathcal{C}')$ be the transition rate of the Markov dynamics from $\mathcal{C}$ to $\mathcal{C}'$.
	
	\textit{Detailed balance} requires the probability current to vanish \emph{pairwise}:
	\begin{equation}
		\label{seq:detailed_balance}
		\rho(\mathcal{C})\, W(\mathcal{C}\to\mathcal{C}')=\rho(\mathcal{C}')\, W(\mathcal{C}'\to\mathcal{C})
		\qquad \forall\,\mathcal{C},\mathcal{C}' .
	\end{equation}
	This condition implies the absence of steady-state probability currents and is characteristic of equilibrium dynamics.
	
	\textit{Global balance} is weaker: it only requires that, for every microstate, the \emph{total} incoming probability flux equals the \emph{total} outgoing flux,
	\begin{equation}
		\label{seq:global_balance}
		\sum_{\mathcal{C}'} \rho(\mathcal{C}')\, W(\mathcal{C}'\to\mathcal{C})
		=
		\rho(\mathcal{C})\sum_{\mathcal{C}'} W(\mathcal{C}\to\mathcal{C}') ,
		\qquad \forall\,\mathcal{C},
	\end{equation}
	thereby ensuring stationarity while allowing for nonzero circulating currents.
	Since nonreciprocal interactions generically drive the system out of equilibrium, detailed balance is not expected to hold; accordingly, in the following, we take global balance to be the appropriate steady-state condition.
	We will state the global-balance condition in its most general form and characterize the class of stationary distributions consistent with it in our setting.
	At the same time, in specific models a stronger condition (intermediate between global and detailed balance) may hold; for instance, in Eq.~\eqref{seq:global_balance} the sum over $\mathcal{C}'$ may effectively be restricted to the subset of configurations that are dynamically connected to $\mathcal{C}$. Also, note that detailed balance implies global balance, but not vice versa.
	
	Finally, if global balance in Eq.~\eqref{seq:global_balance} is violated, the density distribution is \textit{non-stationary}.
	In that case, the time evolution of the density is governed by the master equation
	\begin{equation}
		\label{seq:global_balance_non_stationary}
		\partial_t \rho (\mathcal{C})=
		\sum_{\mathcal{C}'} \rho(\mathcal{C}')\,W(\mathcal{C}'\to\mathcal{C})
		-
		\rho(\mathcal{C})\sum_{\mathcal{C}'} W(\mathcal{C}\to\mathcal{C}') ,
		\qquad \forall\,\mathcal{C}.
	\end{equation}
	Although we do not use this relation directly in our proof, it can be recovered naturally from the derivations below.

	\subsection{Langevin steady-state distribution}
	\label{subsec:langevin_descr}

	Recall that imposing the constraint $\theta_i(t)-\varphi_i(t)=\pi$, and adding the same noise term for both the original and the auxiliary degrees of freedom results in the overdamped Langevin equations
	\begin{eqnarray}
		\label{Seq:Langevin_constraint}
		\dot{\theta _{i}}&=& \left\{ \theta_i,H\right\} +%
		\sqrt{2 T}\eta _{i}(t) = \partial _{\varphi _{i}}H|_\text{constr.}+
		\sqrt{2 T}\eta _{i}(t), \notag \\
		\label{eq:evolution_varphi}
		\dot{\varphi}_{i}&=& \left\{ \varphi_i,H\right\} +
		\sqrt{2 T}\eta _{i}(t) = - \partial _{\theta _{i}}H|_\text{constr.}+
		\sqrt{2 T}\eta _{i}(t),
	\end{eqnarray}
	with conjugate degrees of freedom $\{\theta_i, \varphi_j\}{=}\delta_{ij}$. To maintain consistency with the traditional form of Langevin dynamics and represent the equations of motion on the constraint surface, the notation $\partial
	_{\theta_{i}}H|_\text{constr.}$ is introduced to replace the Poisson bracket. Here $\eta_i(t)$ denotes white noise with $\langle\eta_i(t)\rangle=0$ and $\langle\eta_i(t)\eta_j(t')\rangle=\delta_{ij}\delta(t-t')$, with $T$ the temperature of the bath, where the average is taken over the phase-space density $\rho(\vec{\theta}(t),\vec{\varphi}(t))$. Note that the constraint implies that each pair $(\theta,\varphi_i)$ experiences the same noise realization. 
	
	Below, we first consider steady-state distributions (i.e., no explicit $t$-dependence in $\rho$, Sec.~\ref{subsec:stationary}), and then generalize the result to non-stationary distributions in Sec.~\ref{subsec:non-stationary}. 
	
	\subsubsection{Steady-state distribution}
	\label{subsec:stationary}
	To obtain the steady-state phase-space density distribution $\rho(\vec{\theta}(t),\vec{\varphi}(t))$ under the above dynamics, we follow the standard procedure~\cite{risken1996fokker,sekimoto1998langevin} and introduce a test function $f(\vec{\theta}(t),\vec{\varphi}(t))$ over phase space. The change of the test function within a small time window $\Delta t$ is to leading order
	\begin{eqnarray}
		\label{Seq:dynamics_text_function_total}
		&&f\left(\vec{\theta}(t+\Delta t),\vec{\varphi}(t+\Delta t) \right) - 
		f\left(\vec{\theta}(t),\vec{\varphi}(t) \right) \notag \\
		&=&\sum_{i}\frac{\partial f}{\partial \theta
			_{i}}\dot{\theta _{i}}\Delta t+\frac{\partial f}{\partial \varphi _{i}}\dot{%
			\varphi}_{i}\Delta t+\frac{\partial ^{2}f}{2\partial \theta _{i}\partial
			\theta _{i}}(\dot{\theta _{i}}\Delta t)^{2}+\frac{\partial ^{2}f}{2\partial
			\varphi _{i}\partial \varphi _{i}}(\dot{\varphi}_{i}\Delta t)^{2}+\frac{%
			\partial ^{2}f}{\partial \varphi _{i}\partial \theta _{i}}\dot{\varphi _{i}}%
		\dot{\theta _{i}}(\Delta t)^{2} \notag \\
		&&+\sum_{i\neq j}\frac{\partial ^{2}f}{\partial \theta _{i}\partial \theta
			_{j}}\dot{\theta _{i}}\dot{\theta _{j}}(\Delta t)^{2}+\frac{\partial ^{2}f}{%
			\partial \varphi _{i}\partial \varphi _{j}}\dot{\varphi _{i}}\dot{\varphi
			_{j}}(\Delta t)^{2} + \cdots\ .
	\end{eqnarray}
	For steady-state distributions, at long times the expectation value of the test function should be time-independent:
	\begin{equation}
		\langle f\left(t+\Delta t\right)\rangle  = \langle f\left(t \right)\rangle,
	\end{equation}
	which can be written infinitesimally as
	\begin{equation}
		\label{Seq:steady_state}
		\int \frac{\mathrm{d}f(\vec{\theta},\vec{\varphi} )}{\mathrm{d}t}\rho(\vec{\theta},\vec{\varphi}) \prod_j \mathrm{d}{\theta_j }\mathrm{d}{\varphi_j }  
		= 
		\lim_{\Delta t\to 0}
		\int 
		\frac{f\left(\vec{\theta}(t+\Delta t),\vec{\varphi}(t+\Delta t) \right) - f\left(\vec{\theta}(t),\vec{\varphi}(t) \right) }{\Delta t}
		\rho(\vec{\theta},\vec{\varphi}) \prod_j \mathrm{d}{\theta_j }\mathrm{d}{\varphi_j }  
		= 0.
	\end{equation}
	
	To proceed with the derivation, we write the noise terms in Eq.~\eqref{Seq:Langevin_constraint} as a Wiener process $W_i(t)$:
	\begin{eqnarray}
		\label{seq:Wiener_process}
		\dot{\theta}_i\Delta t =  \partial _{\varphi _{i}}H|_\text{constr.} \Delta t +
		\sqrt{2 T} \Delta W_i(t), \notag \\
		\dot{\varphi}_i\Delta t =  - \partial _{\theta _{i}}H|_\text{constr.} \Delta t +
		\sqrt{2 T} \Delta W_i(t),
	\end{eqnarray}
	where $\langle  \Delta W_i(t)\rangle=0$ and $\langle  \Delta W_i(t)\Delta W_j(t')\rangle=\Delta t \delta(t-t')\delta_{ij} $. 
	
	In the following, we substitute Eq.~\eqref{Seq:dynamics_text_function_total} in Eq.~\eqref{Seq:steady_state}. Focusing on the $\theta_i$ terms, the first-order contribution can be evaluated as 
	\begin{eqnarray}
		\int \frac{\partial f}{\partial \theta_{i}}\dot{\theta_{i}}\Delta t\; 
		\rho\left( \vec{\theta},\vec{\varphi}\right) \mathrm{d}\mathbf{\theta}_i\mathrm{d}\mathbf{\varphi}_i 
		&=& 
		\int \frac{\partial f}{\partial
			\theta _{i}}\left[ \frac{\partial  H}{  \partial \varphi _{i}}\Bigg|_\text{constr.}  \Delta t+
		\sqrt{2 T}\Delta W_i(t) \right]\rho \left(
		\vec{\theta},\vec{\varphi} \right) \mathrm{d}{\theta_i }\mathrm{d}%
		{\varphi_i } \nonumber \\
		&=& \Delta t\int \frac{\partial f}{\partial
			\theta _{i}}\frac{\partial H}{\partial \varphi _{i}}\Bigg|_\text{constr.}\rho
		\left( \vec{ \theta},\vec{\varphi } \right) \mathrm{d}{\theta_i }\mathrm{d}%
		{\varphi_i } ,
	\end{eqnarray}
	where we used the equation of motion and the property $\langle  \Delta W_i (t)\rangle=0$ in the last line. 
	Similarly, the second order in $\Delta t$ gives
	\begin{eqnarray}
		&&
		\int \frac{\partial ^{2}f}{\partial \theta _{i}\partial \theta _{j}}\dot{ \theta _{i}}\dot{\theta _{j}}\Delta t^{2}\rho \left( \vec{\theta},\vec{\varphi} \right) \mathrm{d}{\theta _{i}}\mathrm{d}{\varphi _{i}} =  \notag \\
		&=&
		\int \frac{\partial ^{2}f}{\partial \theta _{i}\partial\theta _{j}}\left[ \frac{\partial H}{\partial \varphi _{i}}\Bigg|_\text{constr.} \Delta t+\sqrt{2 T}\Delta W_i(t) \right] \left[ \frac{\partial H}{\partial \varphi _{j}}\Bigg|_\text{constr.} \Delta t
		+
		\sqrt{2 T}\Delta W_j(t)  \right] \rho \left( \vec{\theta},\vec{\varphi} \right) \mathrm{d}{\theta _{i}}\mathrm{d}{\varphi _{i}} \notag \\
		&=&
		\int 2T \frac{\partial ^{2}f}{\partial \theta _{i}\partial \theta _{j}}\Delta W _{i}(t)\Delta W _{j}(t)
		\rho \left( \vec{\theta},\vec{\varphi} \right)
		\mathrm{d}\theta _{i}\mathrm{d}\varphi _{i}
		+
		\sqrt{2 T}\int \frac{\partial ^{2}f}{\partial \theta _{i}\partial \theta _{j}}\frac{ \partial H}{\partial \varphi _{j}}\Bigg|_\text{constr.}\Delta W_{i}(t)
		\rho \left( \vec{\theta},\vec{\varphi} \right)
		\Delta t\mathrm{d}{\theta _{i}}\mathrm{d}{\varphi_{i}}  \notag \\
		&&+
		\sqrt{2 T}\int \frac{\partial ^{2}f}{\partial \theta _{i}\partial \theta _{j}}\frac{\partial H}{\partial \varphi _{i}}\Bigg|_\text{constr.} \Delta W_{j}(t)
		\rho \left( \vec{\theta},\vec{\varphi} \right) \Delta t \mathrm{d}{\theta _{i}}\mathrm{d}{\varphi _{i}}
		+
		\Delta t^{2}\int \frac{\partial ^{2}f}{\partial\theta _{i}\partial \theta _{j}}\left( \frac{\partial H}{\partial \varphi _{i}}\Bigg|_\text{constr.} \frac{\partial H}{\partial \varphi _{j}}\Bigg|_\text{constr.} \right) \rho \left( \vec{\theta},\vec{\varphi} \right) \mathrm{d}{\theta_{i}}\mathrm{d}{\varphi _{i}} \notag \\
		&=&
		\Delta t \delta _{ij}2 T\int \frac{\partial ^{2}f}{\partial \theta_{i}\partial \theta _{j}}\rho \left( \vec{\theta},\vec{\varphi} \right) \mathrm{d}\theta _{i}\mathrm{d}\varphi _{i}
		+
		O\left[ \left( \Delta t\right)^{2}\right], 
	\end{eqnarray}
	where we neglect terms of order $\Delta t^2$ since they vanish in the infinitesimal limit $\Delta t\to 0$. 
	The expressions for the other terms in Eq.~\eqref{Seq:dynamics_text_function_total} can be derived using the same reasoning. 
	
	Collecting all $\Delta t$ contributions, and plugging them back to Eq.~\eqref{Seq:steady_state} leads to the following condition
	\begin{eqnarray}
		\sum_{i}\int\prod_{j\neq i}\mathrm{d}\theta_j \mathrm{d}\varphi_j \int \mathrm{d}\theta_i \mathrm{d}\varphi_i \Bigg[ 
		\frac{\partial f}{\partial \theta _{i}}
		\left(\frac{\partial H}{\partial \varphi _{i}}\Bigg|_\text{constr.}\right)-\frac{\partial f}{\partial \varphi _{i}}
		\left(\frac{\partial H}{\partial \theta _{i}}\Bigg|_\text{constr.}\right)&& \notag \\
		+ T\left( \frac{\partial ^{2}f}{\partial \theta _{i}\partial \theta _{i}}+\frac{\partial
			^{2}f}{\partial \varphi _{i}\partial \varphi _{i}}+2\frac{\partial ^{2}f}{%
			\partial \varphi _{i}\partial \theta _{i}}\right) \Bigg] \rho \left( \vec{\theta},\vec{\varphi} \right) &=&0\, .
	\end{eqnarray}
	Noticing that the term proportional to $ T$ is a perfect square, we have
	\begin{equation}
		\label{Seq:FlokkerPlanck}
		\sum_{i}\int\prod_{j\neq i}\mathrm{d}\theta_j \mathrm{d}\varphi_j \int \mathrm{d}\theta_i \mathrm{d}\varphi_i \left[ 
		\frac{\partial f}{\partial \theta _{i}}
		\left(\frac{\partial H}{\partial \varphi _{i}}\Bigg|_\text{constr.}\right)-\frac{\partial f}{\partial \varphi _{i}}%
		\left(\frac{\partial H}{\partial \theta _{i}}\Bigg|_\text{constr.}\right)
		+  T \left(\frac{\partial }{\partial \theta _{i}} +\frac{\partial }{\partial \varphi _{i}}  \right)^2 f\right] \rho \left( \vec{\theta},\vec{\varphi} \right) =0.
	\end{equation}

	Under the constraint $\theta _{i}-\varphi _{i}=\pi $, the dynamics
	of the $\varphi _{i}$ spins mirrors that of the $\theta _{i}$ spins. Hence, we have 
	$\partial _{\varphi _{i}}H|_{\text{constr.}}=-\partial _{\theta _{i}}H|_{%
		\text{constr.}}$. Plugging this into Eq.~\eqref{Seq:FlokkerPlanck} leads to%
	\begin{equation}
		\sum_{i}\int \prod_{j\neq i}\mathrm{d}\theta _{j}\mathrm{d}\varphi _{j}\int 
		\mathrm{d}\theta _{i}\mathrm{d}\varphi _{i}\left[ -\left( \frac{\partial }{%
			\partial \theta _{i}}+\frac{\partial }{\partial \varphi _{i}}\right) f\frac{%
			\partial H}{\partial \theta _{i}}\Bigg|_{\text{constr.}}+T\left( \frac{%
			\partial }{\partial \theta _{i}}+\frac{\partial }{\partial \varphi _{i}}%
		\right) ^{2}f\right] \rho \left( \vec{\theta},\vec{\varphi} \right) =0.
	\end{equation}
	If we integrate the $T$ term by parts and, recalling there is no boundary
	term for a compact phase space, we obtain 
	\begin{equation}
		\sum_{i}\int \prod_{j\neq i}\mathrm{d}\theta _{j}\mathrm{d}\varphi _{j}\int 
		\mathrm{d}\theta _{i}\mathrm{d}\varphi _{i}f\left[ \left( \frac{\partial }{%
			\partial \theta _{i}}+\frac{\partial }{\partial \varphi _{i}}\right) \left( 
		\frac{\partial H}{\partial \theta _{i}}\Bigg|_{\text{constr.}}
		\rho \left( \vec{\theta},\vec{\varphi} \right) \right) +T\left( \frac{\partial }{\partial
			\theta _{i}}+\frac{\partial }{\partial \varphi _{i}}\right) ^{2}\rho \left( \vec{\theta},\vec{\varphi} \right) \right] =0.  \label{Seq:FlokkerPlanck_copies}
	\end{equation}%
	Since this equation for the steady state is valid for all test functions $f$, we find that the phase-space density in the steady state satisfies 
	\begin{equation}
		\sum_i \left( \frac{\partial }{\partial \theta _{i}}+\frac{\partial }{\partial
			\varphi _{i}}\right) \left[ \frac{\partial H}{\partial \theta _{i}}\Bigg|_{%
			\text{constr.}}\ \rho \left( \vec{\theta},\vec{\varphi} \right) +T\left( \frac{%
			\partial }{\partial \theta _{i}}+\frac{\partial }{\partial \varphi _{i}}%
		\right) \rho \left( \vec{\theta},\vec{\varphi} \right) \right] =0. \label{Seq:Langevin_distribution}
	\end{equation}
	
	So far the derivation was generic since we did not make use of the precise
	form of the equations of motion. We will show in Sec.~\ref{subsec:glauber} that Eq.~
	\eqref{Seq:Langevin_distribution} governs also the constrained Glauber
	dynamics.
	
	\subsubsection{Time-dependent non-stationary distribution}
	\label{subsec:non-stationary}
	Before we do that, let us complete the analysis and examine the case where the density distribution exhibits explicit time dependence, $\rho (\vec{\theta }(t),\vec{\varphi}(t),t)$. Such situations can occur away from equilibrium where oscillating states arise. 
	
	First, we allow the test function to include explicit time dependence. Building on the same analytical framework as above, the evolution of the expectation value of the test function takes the form
	\begin{equation}
		\frac{\mathrm{d}\left\langle f\right\rangle }{\mathrm{d}t}
		=
		\left\langle 
		\frac{\partial f}{\partial t}\right\rangle 
		+
		\sum_{i} \left(
		\left\langle \frac{\partial f}{
			\partial \theta _{i}}\frac{\partial H}{\partial \varphi _{i}}\Bigg|_{\text{%
				constr.}}\right\rangle -\left\langle \frac{\partial f}{\partial \varphi _{i}}%
		\frac{\partial H}{\partial \theta _{i}}\Bigg|_{\text{constr.}}\right\rangle
		+T\left\langle \left( \frac{\partial }{\partial \theta _{i}}+\frac{\partial 
		}{\partial \varphi _{i}}\right) ^{2}f\right\rangle \right) ,
	\end{equation} 
	which can be written as 
	\begin{eqnarray}
		\label{eq:time_denp_test_function_1}
		&&\frac{\mathrm{d}}{\mathrm{d}t}\int \prod_j
		\mathrm{d}\theta _{j}\mathrm{d}\varphi _{j} f\rho 
		= \\
		&=&\int \prod_j
		\mathrm{d}\theta _{j}\mathrm{d}\varphi _{j} \frac{\partial f}{\partial t}\rho
		+
		\sum_{i}\int \prod_{j\neq i}\mathrm{d}\theta _{j}\mathrm{d}\varphi _{j}\int 
		\mathrm{d}\theta _{i}\mathrm{d}\varphi _{i} \Bigg[ 
		\frac{\partial f}{\partial \theta _{i}}\frac{%
			\partial H}{\partial \varphi _{i}}\Bigg|_{\text{constr.}}\rho 
		-
		\frac{\partial f}{\partial \varphi
			_{i}}\frac{\partial H}{\partial \theta _{i}}\Bigg|_{\text{constr.}}\rho 
		+
		T\int \left( \frac{\partial }{%
			\partial \theta _{i}}+\frac{\partial }{\partial \varphi _{i}}\right)
		^{2}f\rho 
		\Bigg]
		. \notag 
	\end{eqnarray} 
	On the other hand, directly from the left-hand side of the equation above, we have
	\begin{equation}
		\frac{\mathrm{d}}{\mathrm{d}t}\int \prod_j
		\mathrm{d}\theta _{j}\mathrm{d}\varphi _{j}  f\rho 
		=
		\int 
		\prod_j
		\mathrm{d}\theta _{j}\mathrm{d}\varphi _{j}  \left(
		\frac{\partial f}{\partial t}\rho + f\frac{\partial \rho }{\partial t} \right ) .  \label{eq:time_denp_test_function_2}
	\end{equation}
	Setting Eq.\eqref{eq:time_denp_test_function_1} equal to Eq.%
	\eqref{eq:time_denp_test_function_2} yields
	\begin{eqnarray}
		&&\int \prod_j
		\mathrm{d}\theta _{j}\mathrm{d}\varphi _{j}
		f\frac{\partial \rho }{\partial t}  
		=
		\sum_{i}\int \prod_{j\neq i}\mathrm{d}\theta _{j}\mathrm{d}\varphi _{j}\int 
		\mathrm{d}\theta _{i}\mathrm{d}\varphi _{i} \Bigg[ 
		\frac{\partial f}{\partial t}\rho 
		+
		\frac{\partial f}{\partial \theta _{i}}\frac{%
			\partial H}{\partial \varphi _{i}}\Bigg|_{\text{constr.}}\rho 
		-
		\frac{\partial f}{\partial \varphi
			_{i}}\frac{\partial H}{\partial \theta _{i}}\Bigg|_{\text{constr.}}\rho 
		+
		T\int \left( \frac{\partial }{%
			\partial \theta _{i}}+\frac{\partial }{\partial \varphi _{i}}\right)
		^{2}f\rho 
		\Bigg]\, . \notag
	\end{eqnarray} 
	After integrating the right-hand side of this equation by parts and plugging $%
	\partial _{\varphi _{i}}H|_{\text{constr.}}=-\partial _{\theta _{i}}H|_{%
		\text{constr.}}$ into the above equation, we have 
	\begin{eqnarray}
		&&\int \prod_j
		\mathrm{d}\theta _{j}\mathrm{d}\varphi _{j}
		f\frac{\partial \rho }{\partial t} 
		=
		\int \prod_j
		\mathrm{d}\theta _{j}\mathrm{d}\varphi _{j}
		f  \sum_{i} 
		\left[ \left( \frac{\partial }{\partial \theta _{i}}+\frac{\partial }{\partial
			\varphi _{i}} \right) \left( \frac{\partial H}{\partial \theta _{i}}\Bigg|_{\text{%
				constr.}}\rho \right) 
		+
		T\left( \frac{\partial }{\partial \theta _{i}}+\frac{\partial }{\partial
			\varphi _{i}}\right) ^{2}\rho 
		\right].
		\label{seq:FP_equation_Langevin}
	\end{eqnarray} 
	Since this result has to hold for all test functions $f\left( \theta
	_{i}(t),\varphi _{i}(t),t\right) $, we find 
	\begin{equation}
		\label{seq:Fokker_Planck_Langevin}
		\frac{\partial \rho }{\partial t}=\sum_{i}\left[
		\left(  \frac{\partial }{\partial \theta _{i}}+\frac{\partial }{\partial
			\varphi _{i}} \right) \left( \frac{\partial H}{\partial \theta _{i}}\Bigg|_{%
			\text{constr.}}\rho \right) +T\left( \frac{\partial }{\partial \theta _{i}}+%
		\frac{\partial }{\partial \varphi _{i}}\right) ^{2}\rho \right ].
	\end{equation} 
	This Fokker-Planck equation, consistent with Eq.~\eqref{seq:global_balance_non_stationary}, governs the dynamics of the phase-space distribution. As expected, for a stationary distribution, $\partial_t\rho=0$, it reduces to Eq.~\eqref{Seq:Langevin_distribution}.

	\subsubsection{Eliminating the auxiliary degree of freedom from Langevin dynamics}
	When studying the nonreciprocal Langevin dynamics, it is often convenient to keep only the original angles $\vec{\theta}$ and not track the auxiliary variables $\vec{\varphi}$ explicitly. In this reduced description, the dynamics is written directly in terms of the (generally nonreciprocal) drift force $F_i(\vec{\theta})$ acting on $\theta_i$.
	
	We start from the overdamped Langevin equation (It\^o convention)
	\begin{align}
		\theta_i(t+\Delta t)-\theta_i(t)
		=
		F_i(\vec{\theta})\,\Delta t + \sqrt{2T}\,\Delta W_i(t),
	\end{align}
	where $\Delta W_i$ are Wiener increments with $\langle \Delta W_i\rangle=0$ and
	$\langle \Delta W_i\,\Delta W_j\rangle=\delta_{ij}\Delta t$. The corresponding configuration-space probability density $\rho(\theta_i,t)$ obeys the standard Fokker--Planck equation
	\begin{align}
		\partial_t \rho(\vec{\theta} ,t)
		=
		-\sum_i \partial_{\theta_i}\left[F_i(\vec{\theta})\,\rho(\vec{\theta} ,t)\right]
		+T\sum_i \partial_{\theta_i}^2 \rho(\vec{\theta} ,t).
		\label{seq:wo_dynamics}
	\end{align}
	In a stationary state, $\partial_t\rho=0$, and we obtain
	\begin{align}
		0=
		\sum_i \partial_{\theta_i}\left[
		F_i(\vec{\theta})\,\rho(\vec{\theta})
		+
		T\,\partial_{\theta_i}\rho(\vec{\theta})
		\right].
		\label{seq:wo_steady_distribution}
	\end{align}
	For the vision-cone XY model in Eq.~\eqref{Seq:Hamiltonian}, the drift force is
	\begin{align}
		F_i(\vec{\theta})
		=
		-\sum_{j\in\langle ij\rangle} J_{ij}(\theta_i)\,\sin(\theta_i-\theta_j).
	\end{align}
	This drift force is precisely the one obtained from the Hamiltonian embedding once the constraint $\varphi_i=\theta_i+\pi$ is imposed:
	\begin{align}
		F_i(\vec{\theta})
		=
		\partial_{\varphi_i}H\Big|_{\mathrm{constr.}}
		=
		-\partial_{\theta_i}H\Big|_{\mathrm{constr.}}.
	\end{align}
	Moreover, the higher-order derivatives of $F_i(\vec{\theta})$ with respect to $\theta_i$ can be expressed directly in terms of the constrained Hamiltonian:
	\begin{align}
		\partial_{\theta_i}^n F_i(\vec{\theta})
		=
		-\partial_{\theta_i}\big(\partial_{\theta_i}+\partial_{\varphi_i}\big)^n H \Big|_{\mathrm{constr.}},
		\qquad n=1,2,\ldots .
	\end{align}
	Thus, we can always describe the force $F_i$ and its higher-order derivatives with respect to $\theta_i$ using the Hamiltonian embedding.
	The steady-state distribution and non-stationary dynamics of the density introduced in Eq.~\eqref{Seq:Langevin_distribution} and Eq.~\eqref{seq:Fokker_Planck_Langevin} 
	is recovered by expressing all derivatives of $\rho (\vec{\theta})$ in Eq.~\eqref{seq:wo_dynamics} and Eq.~\eqref{seq:wo_steady_distribution}, $\partial_{\theta_i}^n \rho (\vec{\theta})$, in terms of $(\partial_{\theta_i} +\partial_{\varphi_i} )^n \rho (\vec{\theta},\vec{\varphi})$.
	This shows that the Langevin dynamics obtained from Hamiltonian embedding on the constrained manifold and the conventional nonreciprocal Langevin description give the same evolution for the physical degrees of freedom.

	\subsection{Glauber dynamics description}
	\label{subsec:glauber}
	
	We now present a Glauber algorithm that samples the steady state and the nonstationary distribution of the Langevin dynamics using the Hamiltonian embedding. 
	In particular, we focus on defining transition rates that are consistent with the constraint.
	
	When the Hamiltonian is evaluated on the constraint, it vanishes identically, $H(\vec{\theta},\vec{\varphi})\big|_\text{constr.}=0$. Thus, it is nontrivial to define the Glauber update rule based on energy differences alone.
	To address this, we consider a Markov process in which every state $\{\vec{\theta},\vec{\varphi} \}$ satisfies the constraint $\theta_i = \varphi_i + \pi$ while the transition rates are defined based on the energy difference between two states that differ by a small increment $\delta$ in the angle of the original spin and the auxiliary spin, respectively.
	
	We show that taking $\delta$ sufficiently small guarantees that the Glauber and Langevin steady states coincide. 
	We emphasize that, in the strict limit $\delta \to 0$, the Hamiltonian vanishes identically under the constraint, making the update rule ill-defined. However, for any small but finite $\delta$, the Glauber dynamics correctly reproduces the physics of the nonreciprocal Langevin equation.
	
	We start from the probability current from state $\{ \vec{\theta},\vec{\varphi}\}$ to state $\{\theta_i+\delta, \{\vec{\theta} \}_{j \neq i }, \varphi_i+\delta ,\{\vec{\varphi} \}_{j \neq i } \}$, which is given by
	\begin{align}
		\label{seq:joint_detailed_balance}
		\mathcal{J}(\theta _{i},\varphi _{i}\rightarrow
		\theta _{i}+\delta ,\varphi _{i}+\delta) &=  
		\rho (\vec{\theta} ,\vec{\varphi} )
		w(\theta _{i},\varphi _{i}\rightarrow
		\theta _{i}+\delta ,\varphi _{i}+\delta ) 
		\\
		&-  \rho (\theta _{i}+\delta, \{\vec{\theta} \}_{j \neq i }  ,\varphi _{i}+\delta
		, \{\vec{\varphi} \}_{j \neq i })
		w(\theta _{i}+\delta ,\varphi _{i}+\delta \rightarrow \theta _{i},\varphi
		_{i}). \notag
	\end{align}
	Here the increment $\delta$ is restricted to the interval $[-\Delta, \Delta]$. The Glauber transition rates read as
	\begin{equation}
		\label{seq:joint_transition_rate}
		w\left( \theta _{i}\rightarrow \theta _{i}^{\prime }, \varphi_i\rightarrow\varphi_i'
		\right) =\frac{1}{2}%
		\left( 1-\tanh \left( \frac{ H\left(
			\theta ^{\prime },\varphi \right) -H\left( \theta ,\varphi ^{\prime }\right)}{4T}\right) \right)\, ,
	\end{equation}%
	where the constraint is built-in. 
	In other words, we compare the change in energy resulting from the update of each variable $\theta, \varphi$, separately. 
	In particular, we change the angle of the original spin as $\theta \rightarrow \theta^\prime{=}\theta+\delta$. The resulting energy difference between the two states is given by $ H(\theta^{\prime } _{i},\varphi_i)-H(\theta
	_{i},\varphi_{i})$. 
	Next, we compute the energy difference when the corresponding auxiliary spin is changed under the constraint, i.e., $\varphi \rightarrow \varphi ^\prime{=}\varphi+\delta$, with the energy difference $ H(\theta _{i},\varphi^{\prime }_i)-H(\theta
	_{i},\varphi_{i})$. 
	This gives a total change of $H\left(
	\theta ^{\prime },\varphi \right) -H\left( \theta ,\varphi ^{\prime }\right)$.
	
	Notice the additional factor of $1/2$ in the argument of the $\tanh$-function, which accounts for the update of both degrees of freedom; it serves (i) to derive the correct equation for the phase-space distribution and time-dependent dynamics, cf.~Eq.~\eqref{eq:Hamitonian_distribution2} and~Eq.~\eqref{seq:MC_rho_vs_time}, respectively, and (ii) to reproduce the correct rates in the reciprocal limit (cf.~Sec.~\ref{subsec:Eliminating_auxiliary} below). 
	
	The detailed balance condition in Eq.~\eqref{seq:detailed_balance} requires that the probability current vanishes on every channel, i.e., 
	$ \mathcal{J}(\theta _{i},\varphi _{i}\rightarrow
	\theta _{i}+\delta ,\varphi _{i}+\delta )=0$ in Eq.~\eqref{seq:joint_detailed_balance} for every $\delta \in [-\Delta, \Delta]$. 
	However, this condition is generally not satisfied for nonreciprocal systems, which break time-reversal symmetry. 
	Therefore, in Sec~\ref{subsec:stationary_MC}, we derive the steady-state distribution of the Glauber dynamics based on the global balance and then consider the time-dependent distribution in Sec.~\ref{subsec:non-stationary_MC} when the global balance condition is not satisfied.

	\subsubsection{Steady-state distribution from global balance}
	\label{subsec:stationary_MC}
	The steady-state distribution is derived from the balance of total probability flux into and out of a given configuration, which constitutes the so-called global balance~\cite{landau2021guide}
	\begin{align}
		\sum_i \int_{-\pi}^{\pi} \mathrm{d} \delta \left[  \mathcal{J}(\theta _{i},\varphi _{i}\rightarrow
		\theta _{i}+\delta ,\varphi _{i}+\delta ) 
		-
		\mathcal{J}(\theta _{i} +\delta,\varphi _{i} +\delta\rightarrow
		\theta _{i} ,\varphi _{i}  )
		\right] \Big|_\mathrm{constr.} = 0.
	\end{align}
	In detail, it reads as
	\begin{align}
		\label{eq:global_balance}
		&\sum_i \int_{-\pi}^{\pi} \mathrm{d} \delta  
		\big[\rho (\theta _{i}+\delta, \{\vec{\theta} \}_{j \neq i }  ,\varphi _{i}+\delta
		, \{\vec{\varphi} \}_{j \neq i }) 
		W (\theta _{i}+\delta, \varphi _{i}+\delta
		\rightarrow \theta _{i}, \varphi_{i})  \notag \\
		& - \rho (\vec{\theta},\vec{\varphi})W (\theta _{i} , \varphi_i \rightarrow \theta 
		_{i}+\delta, \varphi_i + \delta) \big]\Big|_\text{constr.}
		= 0 .
	\end{align}
	Here, we use $W (\theta _{i}+\delta, \varphi _{i}+\delta
	\rightarrow \theta _{i}, \varphi_{i}) =w (\theta _{i}+\delta, \varphi _{i}+\delta
	\rightarrow \theta _{i}, \varphi_{i}) \,\mathbf{1}_{[-\Delta ,\Delta ]}$, where $\mathbf{1}$ is the indicator function which restricts the range of the shift $\delta$ to the interval $[-\Delta ,\Delta ]$.
	In the Glauber scheme, the proposed angle increment $\delta$ is drawn independently at each update step. 
	Consequently, to enforce global balance, one must sum over all lattice sites $i$ and integrate over all proposed shifts $\delta$ in Eq.~\eqref{eq:global_balance}. This guarantees that we have considered all probability flux channels.
	
	For small $\delta \in \lbrack -\Delta ,\Delta \rbrack$ we can approximate
	Eq.~\eqref{eq:global_balance} as
	\begin{align}
		\label{seq:balance_Glauber}
		&\sum_i
		\int_{-\Delta }^{\Delta }\Bigg\{\left[ \rho (\theta _{i},\varphi
		_{i})+\delta \times \left( \partial _{\theta _{i}}+\partial _{\varphi
			_{i}}\right)
		\rho (\vec{\theta},\vec{\varphi})
		+
		\frac{1}{2}\delta ^{2}\times
		\left( \partial _{\theta _{i}}+\partial _{\varphi _{i}}\right) ^{2}
		\rho (\vec{\theta},\vec{\varphi})
		\right]   \notag \\
		&\times \frac{1}{2}\left( 1-\tanh \frac{-2\left( \partial _{\theta
				_{i}}H-\partial _{\varphi _{i}}H\right) \times \delta -\left( \partial
			_{\theta _{i}}^{2}H-\partial _{\varphi _{i}}^{2}H\right) \times \delta ^{2}}{%
			8T}\right)   \notag \\
		&-
		\rho (\vec{\theta},\vec{\varphi})
		\frac{1}{2}\left( 1+\tanh \frac{-2\left(
			\partial _{\theta _{i}}H-\partial _{\varphi _{i}}H\right) \times \delta
			-\left( \partial _{\theta _{i}}^{2}H-\partial _{\varphi _{i}}^{2}H\right)
			\times \delta ^{2}}{8T}\right) \Bigg\}\Bigg|_{\text{constr.}}\ \mathrm{d} \delta =0.
	\end{align}%
	Next, we perform a Taylor expansion in $\delta$ of the above equation. The first-order contribution vanishes immediately due to the symmetry of the integrand under $\delta \to -\delta$:
	\begin{equation}
		\sum_i \frac{1}{2}\Big\{T\left( \partial _{\theta _{i}}+\partial
		_{\varphi _{i}}\right) 
		\rho (\vec{\theta},\vec{\varphi})
		+
		\frac{1}{2}
		\rho (\vec{\theta},\vec{\varphi})
		\left( \partial _{\theta _{i}}H-\partial _{\varphi
			_{i}}H\right) \Big\}\Big|_{\text{constr.}} \; \int_{-\Delta }^{\Delta }\delta\; \mathrm{d}\delta  =0.
	\end{equation}
	Hence, the steady-state distribution always satisfies global balance to order $\mathcal{O}(\delta)$.
	
	The second-order contribution in $\delta $ does not vanish. Multiplying by $T$, we have 
	\begin{eqnarray}
		\sum_i
		\frac{1}{4}\Big\{\left( \partial _{\theta _{i}}+\partial
		_{\varphi _{i}}\right) \left[ T\left( \partial _{\theta _{i}}+\partial
		_{\varphi _{i}}\right) \rho (\vec{\theta},\vec{\varphi})+\frac{1}{2}\rho (\vec{\theta},\vec{\varphi})
		\left( \partial _{\theta _{i}}H-
		\partial _{\varphi
			_{i}}H\right) \right] \Big\}\Big|_{\text{constr.}}\ \int_{-\Delta }^{\Delta } \delta ^{2}\mathrm{d}%
		\delta  &=& 0.  \label{Seq:MC_taylor_2}
	\end{eqnarray}%
	Rearranging and cancelling the constants, we arrive at the following equation for the phase-space
	distribution function $\rho $ 
	\begin{equation}
		\sum_i
		\left( \partial _{\theta _{i}}+\partial _{\varphi _{i}}\right) \left[
		T\left( \partial _{\theta _{i}}+\partial _{\varphi _{i}}\right) \rho (\vec{\theta},\vec{\varphi})
		+\rho (\vec{\theta},\vec{\varphi})
		\partial
		_{\theta _{i}}H \right] =0.  
		\label{eq:Hamitonian_distribution2}
	\end{equation}
	We find that Eq.~\eqref{eq:Hamitonian_distribution2} is equivalent to Eq.~\eqref{Seq:Langevin_distribution}, after using the relation $-\partial _{\varphi _{i}}H=\partial _{\theta _{i}}H$ which is valid under the constraint.

	\subsubsection{Non-stationary distributions from Monte-Carlo Glauber dynamics}
	\label{subsec:non-stationary_MC}
	
	As we have seen, the constrained Glauber dynamics can generate nonequilibrium steady-state distributions through the constraints imposed at each update step. Hence, it is natural to ask whether one can also use Glauber dynamics to obtain the non-stationary states of the original Langevin dynamics. This will enable the application of the Hamiltonian embedding to study non-stationary states of nonreciprocal systems, which can feature configuration-space entropy currents and time-translation symmetry breaking~\cite{loos2020irreversibility}.
	In particular, we now show that the Glauber dynamics defined by our Hamiltonian embedding reproduces the Fokker–Planck equation derived for the Langevin dynamics in Eq.~\eqref{seq:Fokker_Planck_Langevin}
	
	We denote the density at iteration step $n$ by $\rho (\vec{\theta} , \vec{\varphi}, n)$, and recall the corresponding transition rate $w$, defined in Eq.~\eqref{seq:joint_transition_rate}. 
	In each update step, the probability of selecting site $j$ is uniformly given by $1/L^2$ with $L$ the linear system size; the shift angle $\delta$ is drawn uniformly from the interval $[-\Delta,\Delta]$, with probability density $1/(2\Delta)$.
	Sampling under the constraint is reminiscent of directed loop algorithms, which are used to explore the extensive discrete set of ground states in spin models supporting an emergent gauge theory~\cite{syljuaasen2002quantum}, and more efficient algorithms may exist.
	
	After a single Glauber update $n \rightarrow n+1$, the change in the density is given by
	\begin{align}
		\rho (\vec{\theta},\vec{\varphi}, n+1) - \rho (\vec{\theta},\vec{\varphi}, n) &= 
		\frac{1}{2\Delta L^2} \sum_i
		\int_{-\Delta}^{\Delta} \mathrm d \delta \Big[
		\rho(\theta_i+\delta,\{\vec{\theta}\}_{i \neq j},\varphi_i+\delta,\{\vec{\varphi}\}_{i \neq j},n)\; 
		w (\theta_i + \delta \to \theta_i , \varphi_i + \delta \to \varphi_i)  \notag \\
		&- \rho (\theta_i ,\{\vec{\theta}\}_{i \neq j}, \varphi_i,\{\vec{\varphi}\}_{i \neq j}, n) \; 
		w (\theta_i \to \theta_i + \delta , \varphi_i \to \varphi_i + \delta) \Big].
	\end{align}
	The right-hand side represents the difference between the total inflow and outflow of density current for the Glauber update. As before, in the limit of small $\Delta$, it can be approximated by expanding both the transition rates and the density up to second order in $\delta$:
	\begin{align}
		\label{seq:density_current}
		\rho (\vec{\theta},\vec{\varphi}, n+1) - \rho (\vec{\theta},\vec{\varphi}, n) 
		&= 
		\frac{1}{2\Delta L^2} \sum_i
		\times  \\
		&\int_{-\Delta}^{\Delta} \mathrm d \delta 
		\Bigg\{\left[ \rho (\vec{\theta},\vec{\varphi}, n) 
		+\delta \times \left( \partial _{\theta _{i}}+\partial _{\varphi
			_{i}}\right)\rho (\vec{\theta},\vec{\varphi}, n) 
		+\frac{1}{2}\delta ^{2}\times
		\left( \partial _{\theta _{i}}+\partial _{\varphi _{i}}\right) ^{2}
		\rho (\vec{\theta},\vec{\varphi}, n) 
		\right]   \notag \\
		&\times \frac{1}{2}\left( 1-\tanh \frac{-2\left( \partial _{\theta
				_{i}}H-\partial _{\varphi _{i}}H\right) \times \delta -\left( \partial
			_{\theta _{i}}^{2}H-\partial _{\varphi _{i}}^{2}H\right) \times \delta ^{2}}{%
			8T}\right)   \notag \\
		&-\rho (\vec{\theta},\vec{\varphi}, n) \frac{1}{2}\left( 1+\tanh \frac{-2\left(
			\partial _{\theta _{i}}H-\partial _{\varphi _{i}}H\right) \times \delta
			-\left( \partial _{\theta _{i}}^{2}H-\partial _{\varphi _{i}}^{2}H\right)
			\times \delta ^{2}}{8T}\right) \Bigg\}\Bigg|_{\text{constr.}}. \notag 
	\end{align}
	As expected, for stationary distributions where $\rho (\vec{\theta},\vec{\varphi}, n) $ is time independent, the equation reduces to Eq.~\eqref{seq:balance_Glauber} discussed in Sec.~\ref{subsec:Glauber}, except for the overall prefactor $1/(2\Delta L^2)$. 
	Similar to the previous section, we perform a Taylor expansion of the above equation and retain terms up to second order in $\delta$. This yields
	\begin{align}
		\rho (\vec{\theta},\vec{\varphi}, n+1)  &-\rho (\vec{\theta},\vec{\varphi}, n) 
		= \sum_i \frac{1}{2\Delta L^2} \int_{-\Delta}^{\Delta} \mathrm d\delta 
		\Bigg\{ \delta \Big[ \frac{1}{2}(\partial_{\theta_i} + \partial_{\varphi_i}) \rho (\vec{\theta},\vec{\varphi}, n)  + \frac{1}{4T}\rho (\vec{\theta},\vec{\varphi}, n)  (\partial_{\theta_i} - \partial_{\varphi_i}) H \Big] \notag \\ 
		& 
		+ \delta^2 \frac{1}{4T}
		(\partial_{\theta_i}+\partial_{\varphi_i}) 
		\Big[
		\frac{1}{2}\rho (\vec{\theta},\vec{\varphi}, n)  (\partial_{\theta_i} - \partial_{\varphi_i}) H 
		+ T (\partial_{\theta_i}+\partial_{\varphi_i}) \rho (\vec{\theta},\vec{\varphi}, n) 
		\Big]
		\Bigg\}\Bigg|_{\text{constr.}}.
	\end{align}
	Upon integration, the first-order term in $\delta$ vanishes owing to the symmetry of the integrand. 
	Hence, performing the integration over $\delta$ for the second-order term, we obtain
	\begin{align}
		\label{seq:FP_Glauber}
		\rho (\vec{\theta},\vec{\varphi}, n+1)  &-\rho (\vec{\theta},\vec{\varphi}, n) 
		= 
		\sum_i
		\frac{\Delta^2}{12 L^2T }   
		(\partial_{\theta_i}+\partial_{\varphi_i})  
		\Big[
		\rho (\vec{\theta},\vec{\varphi}, n)  \partial_{\theta_i} H  \big |_\text{constr.} 
		+ T (\partial_{\theta_i}+\partial_{\varphi_i}) \rho (\vec{\theta},\vec{\varphi}, n) 
		\Big] + \mathcal{O}(\Delta^4),
	\end{align}
	where we used the equality $\partial_{\theta_i} H\big|_{\text{constr.}} = -\partial_{\varphi_i} H\big|_{\text{constr.}}$ valid on the constraint manifold.
	By the same symmetry arguments, the leading order correction is of the order $\mathcal{O}(\Delta^4)$.
	
	In order to compare the Glauber simulations to real-time Langevin dynamics in practice, we now introduce the notion of Monte-Carlo time, $t_{\mathrm{MC}}$, for the constrained Glauber dynamics. To avoid confusion, the time variable $t$ used outside this section is identified as the Langevin time and is here relabeled as $t_{\mathrm{Lan}}$; this is the proper physical time from the equations of motion of the original nonreciprocal system.  
	We associate one unit of (dimensionless) Monte-Carlo time with updating each spin once (on average), and hence define
	\begin{equation}
		\label{seq:MC_time}
		t_{\textrm{MC}}=\frac{n}{L^2} .
	\end{equation}
	Accordingly, we identify the discrete difference with
	\begin{equation}
		\rho (\vec{\theta},\vec{\varphi}, n+1) - \rho (\vec{\theta},\vec{\varphi}, n) 
		= \partial _ {t_\mathrm{MC}}\rho(\theta_i,\varphi_i , t_{\mathrm{MC}})\; \mathrm{d} t_\mathrm{MC},
	\end{equation}
	with $\mathrm{d} t_\mathrm{MC} = 1/L^2$. This allows us to formally define a continuum limit for the Monte-Carlo time.
	
	Following the derivation above, the equation of motion for the density distribution then reads as
	\begin{align}
		\label{seq:MC_rho_vs_time}
		\frac{\partial \rho(\vec{\theta},\vec{\varphi}, t_{\textrm{MC}})}{\partial t_{\textrm{MC}}} 
		= \frac{\Delta^2}{12T}    \sum_i 
		(\partial_{\theta_i}+\partial_{\varphi_i})
		\Big[
		\rho \partial_{\theta_i} H  \big |_\text{constr.}
		+ T (\partial_{\theta_i}+\partial_{\varphi_i}) \rho
		\Big]  + \mathcal{O}(\Delta^4) .
	\end{align}
	This equation is equivalent to the Fokker-Planck equation derived for the Langevin dynamics in Eq.~\eqref{seq:Fokker_Planck_Langevin}, up to the prefactor $\Delta^2/(12T)$. In units where the friction coefficient is set to unity, it follows from the equations of motion in Eq.~\eqref{Seq:NR_XY} that the interaction strength $J$, and hence energy, has the same unit as frequency (inverse time). Since the only model-independent energy scale in the Glauber rate $w$ is the temperature $T$ (which also has units of energy), we find that the time unit is set by the inverse bath temperature $T$. The dimensionless factor $\Delta^2/12$ is a property of the Glauber update rule, and is most likely not universal.
	
	These results demonstrate that the constrained Glauber dynamics reproduce the same non-stationary evolution as the Langevin dynamics as $\Delta \to 0$, provided the Monte-Carlo time $t_{\mathrm{MC}}$ is related to the physical Langevin time $t_{\mathrm{Lan}}$ through a proportionality constant, namely
	\begin{align}
		\frac{\Delta^2}{12T} t_{\mathrm{MC}} = t_{\mathrm{Lan}}.
		\label{seq:time_relation}
	\end{align}
	Physically, this relation implies that the lower the bath temperature $T$, the fewer Monte-Carlo updates are necessary to reach the same unit of physical Langevin time at fixed $\Delta$. Moreover, since the constrained Glauber update rule reproduces the nonreciprocal physics in the limit $\Delta\ll 1$ and because the next-order correction in Eq.~\eqref{seq:MC_rho_vs_time} is $\Delta^4$, there is a natural timescale set by $1/\Delta^2$ (at a fixed temperature $T$), beyond which Glauber dynamics is expected to disagree with Langevin dynamics; 
	we emphasize that this scale can be controllably extended to an arbitrarily long Langevin time by decreasing $\Delta$ which, however, requires increasing the number of Glauber iterations $n$. 
	
	We provide numerical evidence in support of Eq.~\eqref{seq:time_relation} in Fig.~\ref{fig:comparison_period} and Sec.~\ref{sec:nonstationary}.

	\subsubsection{Eliminating the auxiliary degree of freedom from the Glauber transition rates}
	\label{subsec:Eliminating_auxiliary}

	Whereas the nonreciprocal equations of motion of interest depend only on the original degrees of freedom $\theta$, the Glauber update rules derived in the preceding subsections appear to depend on both the original and the auxiliary degrees of freedom: $w\left( \theta _{i}\rightarrow \theta _{i}^{\prime }, \varphi_i\rightarrow\varphi_i'\right)$; this is because the Glauber transition rate $w$ was derived using the Hamiltonian embedding. As a result, one might be tempted to argue that the constrained Glauber Monte-Carlo simulations are computationally more expensive than the corresponding original Langevin dynamics. 
	
	Here, we show that this is not the case: one can eliminate the auxiliary field $\varphi$ completely from the expression for the constrained Glauber transition rate: $w = w(\theta_i\to\theta_i')$. 
	To get an intuition for why this is the case, notice that, in constrained Glauber dynamics, observables depend only on the original spins $\theta$, while the values of the auxiliary spins $\varphi$ are fixed by the constraint. As a result, the Monte-Carlo fluctuations of $\varphi$ are slaved to those of $\theta$ at each Glauber step.
	
	For clarity, we apply the procedure to the vision-cone XY model using the Hamiltonian in Eq.~\eqref{Seq:Hamiltonian}.
	First, we initialize the simulation in a $(\theta_i,\varphi_i)$-configuration that obeys the constraint. 
	Hence, at the start of each Glauber update step, the constraint $\varphi_i=\theta_i+\pi$ holds, so the auxiliary angle can be expressed in terms of $\theta_i$.  
	We then update $\theta_i\to\theta_i'=\theta_i+\delta$ and $\varphi_i\to\varphi_i'=\varphi_i+\delta$, which preserves the constraint manifold: $\varphi_i'=\theta_i'+\pi$.  
	The key observation is that $\varphi_i'$ may be replaced by $\theta_i'+\pi$ when evaluating the energy differences which define the Glauber transition rate.  
	
	To see this in practice, we will revisit the expression for the transition rate from Eq.~\eqref{seq:joint_transition_rate}. The energy difference in the argument of the hyperbolic tangent can be expressed as 
	\begin{eqnarray}
		\label{eq:joint_E-diff}
		\Delta E &=& \frac{1}{2}\big(H(\theta',\varphi)-H(\theta,\varphi')\big)\big|_{\varphi'=\theta'+\pi}^{\varphi=\theta+\pi} \nonumber\\
		&=& 
		\frac{1}{2}\big(H(\theta',\varphi)-H(\theta,\varphi) + H(\theta,\varphi) - H(\theta,\varphi')\big)\big|_{\varphi'=\theta'+\pi}^{\varphi=\theta+\pi} \nonumber\\
		&=& \frac{1}{2}\left( \Delta E_1\big|_{\varphi=\theta+\pi} - \Delta E_2\big|_{\varphi'=\theta'+\pi}^{\varphi=\theta+\pi} \right),
	\end{eqnarray}
	where, crucially, the constraint is applied only after the differences are evaluated.
	
	Now consider the term $\Delta E_1$, and note that the energy change associated with updating the original spin $\theta_i$, while keeping $\varphi_i$ fixed, is given by
	\begin{align}
		\Delta E_1 &= H(\theta',\varphi)-H(\theta,\varphi) \notag\\
		&= -\sum_{j\in\langle ij\rangle}\Big\{[J_{ij}(\theta_i')+J_{ji}(\theta_j)]\cos(\theta_i'-\theta_j)
		-[J_{ij}(\theta_i)+J_{ji}(\theta_j)]\cos(\theta_i-\theta_j)\Big\}\notag\\
		&\quad -\sum_{j\in\langle ij\rangle}\Big\{[J_{ij}(\theta_i')-J_{ij}(\theta_i)]\cos(\theta_j-\varphi_i)
		+J_{ji}(\theta_j)[\cos(\theta_i'-\varphi_j)-\cos(\theta_i-\varphi_j)]\Big\}.
	\end{align}
	Since the Glauber update depends only on the energy difference, the auxiliary angles can be eliminated using the constraint $\varphi=\theta+\pi$, without changing the transition rate. Substituting $\varphi=\theta+\pi$ gives
	\begin{align}
		\Delta E_1\big|_{\varphi=\theta+\pi}
		&=H(\theta',\theta+\pi)-H(\theta,\theta+\pi)\notag\\
		&=-\sum_{j\in\langle ij\rangle}\Big\{[J_{ij}(\theta_i')+J_{ji}(\theta_j)]\cos(\theta_i'-\theta_j)
		-[J_{ij}(\theta_i)+J_{ji}(\theta_j)]\cos(\theta_i-\theta_j)\Big\} \notag \\
		&\quad  +\sum_{j\in\langle ij\rangle}[J_{ij}(\theta_i')-J_{ij}(\theta_i)]\cos(\theta_j-\theta_i)
		+\sum_{j\in\langle ij\rangle}J_{ji}(\theta_j)[\cos(\theta_i'-\theta_j)-\cos(\theta_i-\theta_j)]\notag\\
		&= -\sum_{j\in\langle ij\rangle}J_{ij}(\theta_i')[\cos(\theta_i'-\theta_j)-\cos(\theta_i-\theta_j)].
	\end{align}

	Likewise, the energy difference associated with updating the auxiliary spin $\varphi_i$ while keeping $\theta_i$ fixed reads as
	\begin{align}
		\Delta E_2 = H(\theta,\varphi')-H(\theta,\varphi)
		= -\sum_{j\in\langle ij\rangle}J_{ij}(\theta_i)[\cos(\theta_j-\varphi_i')-\cos(\theta_j-\varphi_i)].
	\end{align}
	Applying the constraint by substituting $\varphi'=\theta'+\pi$ and $\varphi=\theta+\pi$ yields
	\begin{align}
		\Delta E_2\big|_{\varphi'=\theta'+\pi}^{\varphi=\theta+\pi}
		=H(\theta,\theta'+\pi)-H(\theta,\theta+\pi)
		=\sum_{j\in\langle ij\rangle}J_{ij}(\theta_i)[\cos(\theta_i'-\theta_j)-\cos(\theta_i-\theta_j)].
	\end{align}

	Therefore, the total energy difference entering the transition rate in Eq.~\eqref{seq:joint_transition_rate} simplifies to
	\begin{align}
		\Delta E &=
		\frac{1}{2}\left( \Delta E_1\big|_{\varphi=\theta+\pi} - \Delta E_2\big|_{\varphi'=\theta'+\pi}^{\varphi=\theta+\pi} \right)\notag\\
		&= -\frac{1}{2}\sum_{j\in\langle ij\rangle}[J_{ij}(\theta_i')+J_{ij}(\theta_i)] [\cos(\theta_i'-\theta_j)-\cos(\theta_i-\theta_j)].
		\label{seq:energy_difference}
	\end{align}
	Consequently, the joint-update constrained Glauber dynamics can be implemented using only the original variables $\theta_i$, and the corresponding transition rate is
	\begin{align}
		w(\theta_i\to\theta_i')=\frac{1}{2}\left(1-\tanh\frac{\Delta E}{2T}\right),
	\end{align}
	with $\Delta E$ given by Eq.~\eqref{seq:energy_difference}. 
	
	Notably, in the reciprocal limit $\Psi{=}\pi$, the interaction becomes orientation-independent: $J_{ij}(\theta_i')=J_{ij}(\theta_i)=J_0$. Then
	$\Delta E=-J_0\sum_{j\in\langle ij\rangle}\big[\cos(\theta_i'-\theta_j)-\cos(\theta_i-\theta_j)\big]$,
	which matches exactly the energy difference obtained from the conventional XY Hamiltonian
	$H_\mathrm{XY}=-J_0\sum_{\langle ij\rangle}\cos(\theta_i-\theta_j)$ (and hence justifies the factor of $1/2$ in the definition of $\Delta E$ above a posteriori).
	Thus, although imposing the constraint breaks detailed balance in general, the latter is restored in the reciprocal limit, as expected.
	
	As a consequence of eliminating the auxiliary degrees of freedom $\varphi$, the total number of XY-spins to be simulated in the constrained Glauber algorithm is equivalent to that used in the original Langevin dynamics.
	Note that we proved this elimination procedure only for the constrained Glauber dynamics considered in this work; further research is required to determine whether, and under which conditions, it extends to other energy-difference-based algorithms.
	
	Finally, let us emphasize that the role of the Hamiltonian embedding here is to provide the correct Glauber transition rate (which we have proved to give rise to the correct Fokker-Planck equation in the limit of $\delta \to 0$, cf.~Sec.~\ref{subsec:Glauber}).

	\section{Determining the critical temperature for the vision-cone model}
	\label{sec:Tc}
	
	To determine the critical temperature within the statistical uncertainty, $T_c\pm\Delta T_c$, for a given algorithm, we first conduct $N$ independent simulations (runs) of the dynamics for fixed values $T$ of the reservoir temperature. For each run $i$, we determine the transition temperature $T_{c,i}$ from the peak of the specific heat. The mean and standard deviation over the $N$ independent runs are subsequently computed from
	\begin{eqnarray}
		T_c &=& \frac{1}{N} \sum_{i=1}^{N} T_{c,i}, \qquad\qquad
		\Delta T_c=  \frac{1}{N}\sqrt{\sum_{i=1}^{N}\left(T_{c,i} - T_c \right)^2}.
	\end{eqnarray}

	\section{Supplementary results for the Monte-Carlo simulations}
	\label{sec:MC_results}

	This section presents supplementary results for the Monte-Carlo simulations presented in the Main Text. In Sec.~\ref{subsec:MC_algo}, we detail the implementation of the Glauber algorithm under the proposed constraints, derived in Sec.~\ref{subsec:Glauber}. In Sec.~\ref{subsec:Unconst}, we present numerical results for the individual-update Glauber dynamics that we propose, for Glauber dynamics with selfish energy as proposed in previous works~\cite{avni2023non,loos2023long,blom2025local}, and for Langevin dynamics across various system sizes. Additionally, we examine the outcomes of Glauber dynamics with the Hamiltonian dropping the auxiliary spin (Eq.~\eqref{Seq:Hamiltonian_ss}), and highlight the significant differences from the behavior observed in the constrained Glauber dynamics. Finally, in Sec.~\ref{subsec:MC_selfish},  we explore a different vision-cone interaction strength proposed in 
	Ref.~\cite{rouzaire2025nonreciprocal} and compare the constrained Glauber dynamics with selfish energy dynamics~\cite{rouzaire2025nonreciprocal} and Langevin dynamics: We find numerically identical behavior obtained from the constrained Glauber dynamics and the Langevin dynamics, whereas the selfish energy approach results in a different behavior.

	\subsection{Constrained Glauber algorithm}
	\label{subsec:MC_algo}
	
	In this subsection, we introduce the details of the constrained Glauber algorithm from Sec.~\ref{subsec:glauber} that we use in the Main Text. The Hamiltonian is introduced in Eq.~\eqref{Seq:Hamiltonian}. We use $\Psi^{(n)}$ to represent the system configuration $\Psi^{(n)} =(\theta_{1}^{(n)},\theta_2^{(n)},...\theta_i^{(n)},...\theta_N^{(n)},\varphi_1^{(n)},\varphi_2^{(n)},...\varphi_i^{(n)},...\varphi_N^{(n)})$, where $n$ denotes the iteration step of the Monte-Carlo simulation. The algorithm proceeds as follows:
	\begin{enumerate}
		\item \textbf{Initial Configuration:} Set initial values for the original and auxiliary spins in a uniform configuration, with $\theta_i^{(0)} = 0$ and $\varphi_i^{(0)} = \pi$. This initial configuration satisfies the embedding constraint. 
		\item \textbf{Original Spin Update:} Propose a new configuration by selecting an \textit{original} spin uniformly at random and rotating it to a new angle $\theta_i^{(0)} \rightarrow \theta_i^{(0)}+\delta$, where $\delta$ is drawn from a uniform distribution between $[-\Delta,\Delta]$. As anticipated in Sec.~\ref{subsec:glauber}, we find that the smaller $\Delta$, the better the agreement with the Langevin steady state. The energy of the new configuration reads as $H(\theta_i^{(0)} + \delta, \varphi ^{(0)}_i )$ where $H$ is the embedding Hamiltonian.
		\item \textbf{Auxiliary Spin Update:} We also update the the auxiliary spin $\varphi_i$ by the same angle $\delta$ to ensure that the constraint is preserved at each step, i.e., $\varphi_i^{(0)} \rightarrow \varphi_i^{(0)}+\delta$. The energy of this configuration is $H(\theta_i^{(0)}, \varphi_i^{(0)} + \delta)$.
		\item \textbf{Energy Change Calculation:} The energy difference between the initial and the proposed configurations is given by $\Delta E = \frac{1}{2}[ H(\theta_i^{(0)}+\delta, \varphi_i^{(0)})-H(\theta_i^{(0)}, \varphi_i^{(0)} + \delta )]$. For the XY spins with vision-cone interactions, it is given by
		\begin{eqnarray}
			\label{Seq:}
			\Delta E  &=& 
			-\frac{1}{2}
			\sum_{j \in \langle ij \rangle} \bigg \{
			J_{ij}(\theta_i^{(0)}+\delta)\left[\cos(\theta_i^{(0)}+\delta-\theta_j^{(0)}) + \cos(\varphi_i^{(0)}-\theta_j^{(0)})\right] 
			\\
			&& \qquad\qquad  +  
			J_{ji}(\theta_j^{(0)})\left[\cos(\theta_j^{(0)}-\theta_i^{(0)}-\delta) + \cos(\varphi_j^{(0)}-\theta_i^{(0)}-\delta)\right]  \bigg \} \notag \\
			&&+  \frac{1}{2}
			\sum_{j \in \langle ij \rangle} \bigg \{
			J_{ij}(\theta_i^{(0)})
			\left[\cos(\theta_i^{(0)}-\theta_j^{(0)}) + \cos(\varphi_i^{(0)} + \delta - \theta_j^{(0)})\right] 
			+  
			J_{ji}(\theta_j^{(0)})\left[\cos(\theta_j^{(0)}-\theta_i^{(0)}) + \cos(\varphi_j^{(0)}  -\theta_i^{(0)})\right] \bigg \} \notag.
		\end{eqnarray}
		\item \textbf{Transition Probability:} Calculate the transition probability $w$ using the Glauber transition rates at temperature $T$:
		\begin{equation}
			w =\frac{1}{2}\left(1-\tanh \frac{\Delta E}{2T} \right).
		\end{equation}
		
		\item \textbf{Acceptance/Rejection Step:} To accept or reject the proposed update, draw a random number $r$ from a uniform distribution between $0$ and $1$. If it is less than the transition probability, i.e., $r<w$, accept the new configuration by updating the spins on the selected site as $\theta_i^{(1)} = \theta_i^{(0)}+\delta$, and $\varphi_i^{(1)} = \varphi_i^{(0)}+\delta$. Otherwise, reject the update and leave the spin unchanged. Note that all other spins remain unchanged $\theta_j^{(1)} = \theta_j^{(0)}$, for all $j \neq i$. 
		\item \textbf{Repeat:} Iterate the procedure for a large number of steps. In our numerical simulations, we perform $200L^4$ iterations where $L$ is the linear dimension.
	\end{enumerate}
	
	\subsection{Constrained Glauber algorithm after eliminating the auxiliary spins}
	\label{subsec:MC_algo_eliminating}
	
	In Sec.~\ref{subsec:Eliminating_auxiliary}, we have shown that the auxiliary spins can be eliminated from the expression for the transition rate, and hence we can simulate the constrained Glauber dynamics using only the original spins $\theta_i$. The system configuration is now $\Psi^{(n)} = (\theta_1^{(n)},\theta_2,...\theta_N^{(n)})$. 
	
	The algorithm proceeds as follows:
	\begin{enumerate}
		\item \textbf{Initial Configuration:} Set initial values for the original spins in a uniform configuration, with $\theta_i^{(0)} = 0$. This initial configuration satisfies the constraint. 
		\item \textbf{Spin Update:} Propose a new configuration by selecting a spin uniformly at random and rotating it to a new angle $\theta_i^{(0)} \rightarrow \theta_i^{(0)}+\delta$, where $\delta$ is drawn from a uniform distribution between $[-\Delta,\Delta]$.
		\item \textbf{Energy Change Calculation:} The energy difference between the initial and the proposed configurations is given by Eq.~\eqref{seq:energy_difference}, which reads as
		\begin{align}
			\Delta E  = - \frac{1}{2}\sum_{j\in\langle ij\rangle}\left[J_{ij}(\theta_i^{(0)}+\delta)+J_{ij}(\theta_i^{(0)})\right] \, \left[\cos(\theta_i^{(0)}+\delta-\theta_j^{(0)})-\cos(\theta_i^{(0)}-\theta_j^{(0)})\right].     
		\end{align}
		
		\item \textbf{Transition Probability:} Calculate the transition probability $w$ using the Glauber transition rates at temperature $T$:
		\begin{equation}
			w =\frac{1}{2}\left(1-\tanh \frac{\Delta E }{2T} \right).
		\end{equation}
		
		\item \textbf{Acceptance/Rejection Step:} To accept or reject the proposed new configuration, draw a random number $r$ from a uniform distribution between $0$ and $1$. If it is less than the transition probability, i.e., $r<w$, accept the new configuration by updating the selected spin as $\theta_i^{(1)} = \theta_i^{(0)}+\delta$. Otherwise, reject the update and leave the spin unchanged. Note that all other spins remain unchanged $\theta_j^{(1)} = \theta_j^{(0)}$, for all $j \neq i$. 
		
		\item \textbf{Repeat:} Iterate the procedure for a large number of steps. In our numerical simulations, we perform $200L^4$ iterations where $L$ is the linear dimension.
	\end{enumerate}

	\begin{figure}[t!]
		\centering\includegraphics[width=0.98\textwidth]{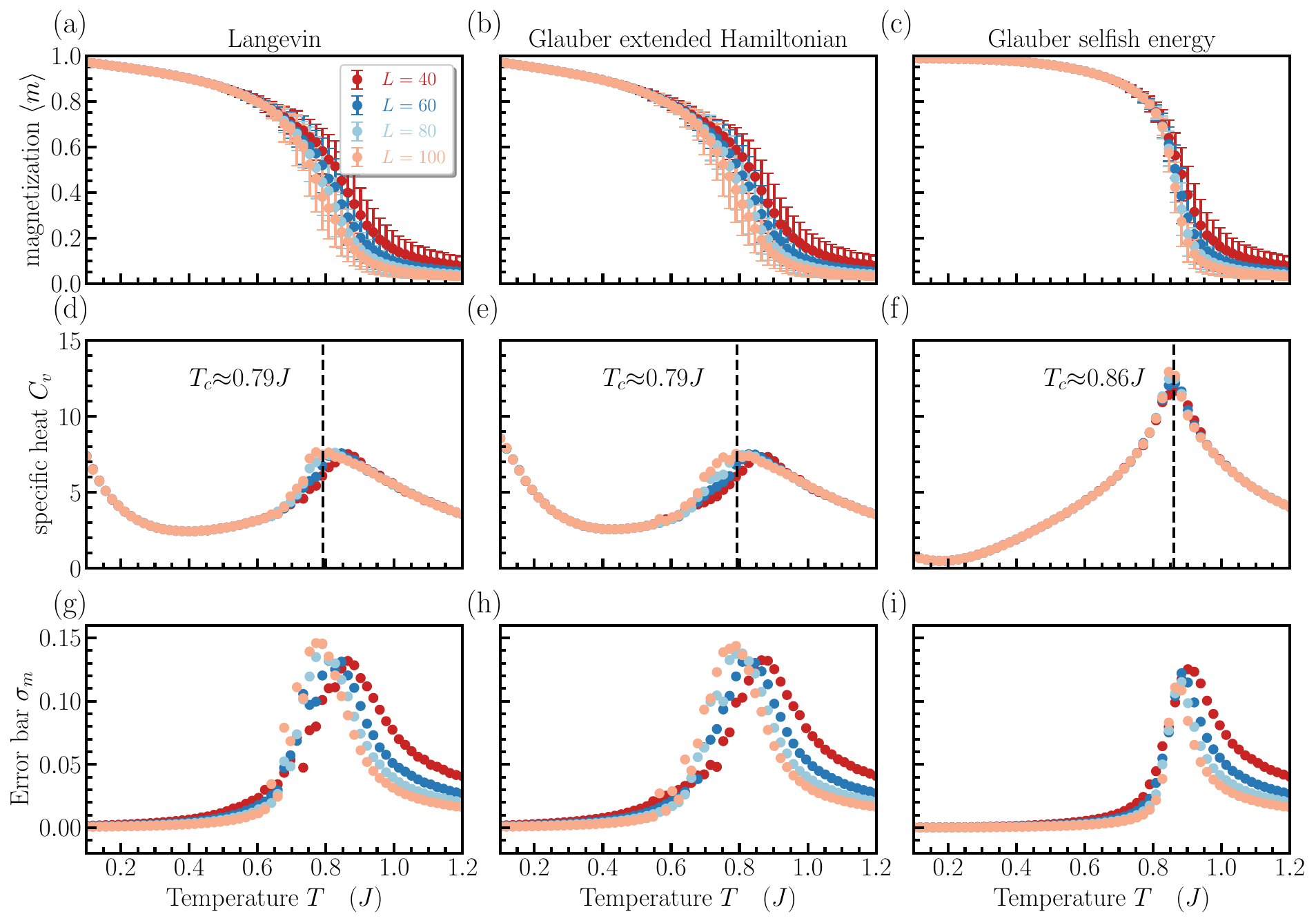}
		\caption{
			Average magnetization $\langle m \rangle$ (a,b,c), specific heat $C_v$ (d,e,f), and error bars 
			of average magnetization $\sigma_m$ (g,h,i) obtained from numerical simulations of Langevin dynamics, constrained Glauber dynamics, and selfish-energy Glauber dynamics as functions of temperature $T$.
			Away from the transition, the magnetization fluctuations are self-averaging, so the standard deviation (error bars) decreases with increasing linear size $L$ for all three dynamics.
			Close to the critical point, the peak of error bars increases with increasing $L$ for both Langevin and constrained Glauber dynamics, whereas the selfish-energy Glauber dynamics shows decreasing error bars upon increasing $L$ (at least up to $L=100$). The results indicate that the selfish-energy Glauber dynamics samples a different stationary distribution rather than reproducing the critical behavior of the Langevin dynamics.
			The remaining parameter values are the same as in Fig.~3 in the Main Text.
		}
		\label{fig:Langevin_Hamiltonian_system-size}
	\end{figure}

	\subsection{Finite-size effects}
	\label{subsec:Unconst}
	
	In this subsection, we investigate the effects of system size as shown in Fig.~\ref{fig:Langevin_Hamiltonian_system-size}. The average magnetization and specific heat from both Langevin dynamics and the constrained Glauber dynamics that we propose here exhibit similar behavior across ordered and disordered phases at different system sizes, differing significantly from the results of selfish-energy Glauber dynamics proposed in previous works~\cite{avni2023non,loos2023long,blom2025local}. 
	
	To better characterize the fluctuations in the magnetization, we also examine the standard deviation $\sigma_m =  \sqrt{\langle m^2 \rangle  - \langle m\rangle^2}$ across different system sizes (shown as error bars in the panels of the first row in Fig.~\ref{fig:Langevin_Hamiltonian_system-size}). 
	Both Langevin and constrained Glauber dynamics exhibit similar scaling behavior, which differs markedly from that of the selfish-energy Glauber dynamics. 
	Away from the critical temperature, the error bars decrease with increasing system size $L$ in all three methods, as expected due to self-averaging in large systems. 
	However, near the critical temperature, Langevin and constrained Glauber dynamics show increasing error bars with system size, reflecting critical fluctuations that grow in the thermodynamic limit. 
	By contrast, the selfish-energy Glauber dynamics exhibits the opposite trend, with decreasing error bars near the transition, suggesting a different critical behavior. 
	This discrepancy further supports the usefulness of our constrained Hamiltonian embedding approach in capturing the correct nonequilibrium physics of nonreciprocal systems.

	\begin{figure*}[t!]
		\centering\includegraphics[width=0.8\textwidth]{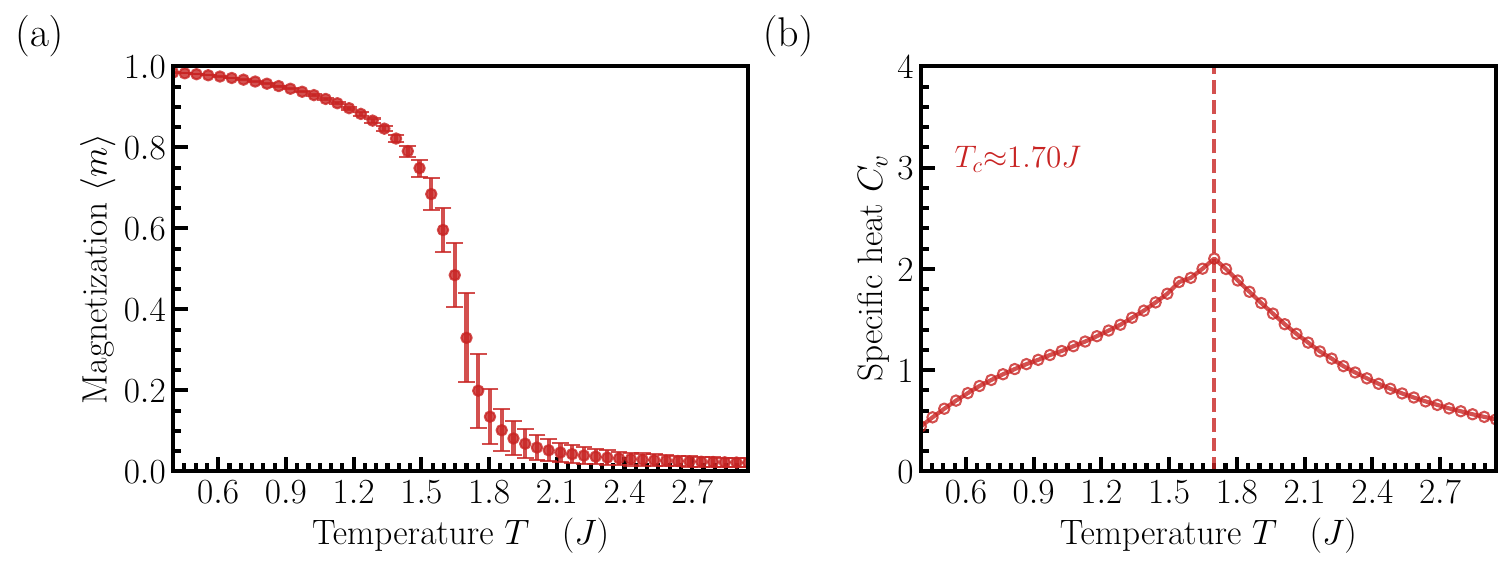}
		\caption{Average magnetization $\langle m \rangle$ (a) and specific heat $C_v$ (b) obtained from Monte-Carlo simulations of Glauber dynamics without auxiliary spins as functions of temperature $T$.
			The critical temperature is $T_c = 1.70J$, which differs significantly from the value $T_c=0.79J$ obtained through the constrained Glauber dynamics. 
			Parameter values are the same as in Fig.~3 in the Main Text.
		}
		\label{fig:magnet_and_specific_heat_total}
	\end{figure*}
	
	\subsection{Glauber dynamics without auxiliary spins}
	\label{subsec:Glauber}
	To show the key role of the auxiliary spins in our Hamiltonian embedding construction, we also consider the Glauber dynamics for the Hamiltonian without auxiliary DOF
	\begin{equation}
		\label{Seq:Hamiltonian_ss}
		H_{SS} = - \sum_{\langle ij\rangle}\left[ J_{ij}(\theta _{i})+J_{ji}(\theta _{j})\right] \cos (\theta _{i}-\theta _{j}) .
	\end{equation} 
	Here, $J_{ij}(\theta _{i})$ is the vision-cone interaction introduced in the Main Text.
	
	The average magnetization and the specific heat are plotted in Fig.~\ref{fig:magnet_and_specific_heat_total}. The critical temperature is found at $T_c = 1.70J$, which differs significantly from the value of $T_c=0.79J$ obtained from either the constrained Glauber dynamics or the original Langevin dynamics. This result exemplifies the crucial role played by the auxiliary degrees of freedom and the constraint in the Hamiltonian construction proposed in this work. Finally, consistent with the results of Ref.~\cite{bandini2024xy}, the critical temperature for the Hamiltonian in Eq.~\eqref{Seq:Hamiltonian_ss} is nearly twice that of the value $T_c=0.86J$ found for the selfish-energy Glauber dynamics.

	\begin{figure*}[t!]
		\centering\includegraphics[width=0.8\textwidth]{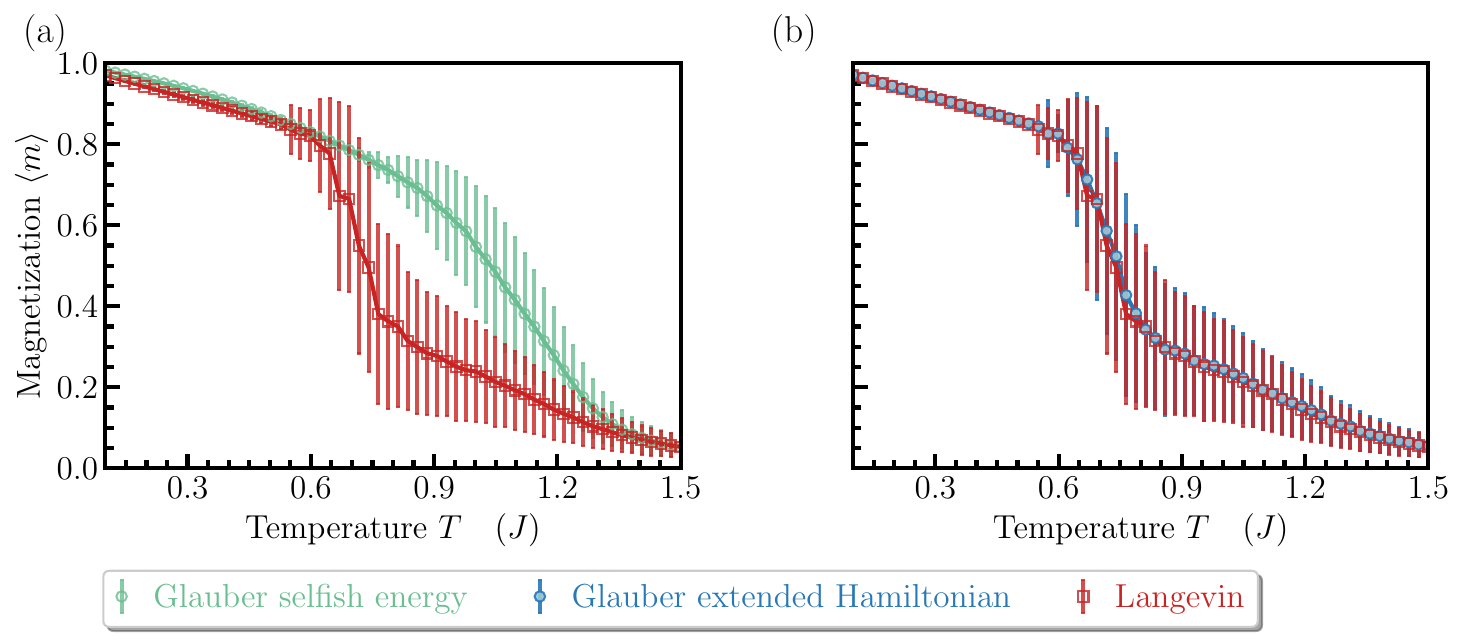}
		\caption{
			Average magnetization $\langle m \rangle$ in the steady state on a square lattice of linear dimension $L=100$, as a function of temperature $T$ for constrained Glauber (blue dots), selfish-energy Glauber (green dots), and Langevin (red squares) dynamics. 
			The results show that in the long-time steady state, the constrained Glauber dynamics accurately reproduces the behavior of the Langevin dynamics and captures the same critical temperature, $T_c /J \approx 0.74$ (a), whereas the selfish-energy Glauber dynamics predicts a significantly different critical temperature, $T_c /J \approx 1.07$ (b). 
			We use the vision-cone interaction strength in Eq.~\eqref{Seq:NR_Force_rou} with $\sigma=1$, the semi-width $\Delta=0.2$ for constrained Glauber dynamics, and a time step of $\delta t = 0.01$ for Langevin dynamics.
			We perform $200L^4$ iterations ($100 L^2$ time steps) for constrained Glauber (Langevin) dynamics and $70L^4$ iterations for selfish-energy Glauber dynamics. The ensemble consists of $\mathcal{N}=100$ independent trajectories.
		}
		\label{fig:comparison}
	\end{figure*}

	\subsection{Soft-vision-cone interactions: Comparison between selfish energy and constrained Glauber dynamics}
	\label{subsec:MC_selfish}
	
	To demonstrate that the agreement between the constrained Glauber dynamics and the Langevin dynamics is not a coincidence, we now adopt the vision-cone interaction proposed in Ref.~\cite{rouzaire2025nonreciprocal} and apply it to the XY spins on the square lattice. Specifically, we consider the interaction strength
	\begin{equation}
		\label{Seq:NR_Force_rou}
		J_{ij}(\theta_i) = \exp{[\sigma \cos{(\theta_i-\psi_{ij})}]},
	\end{equation}
	where $\psi_{ij}$ is defined in the Main Text. 
	The explicit form of the selfish energy~\cite{avni2023non,loos2023long,blom2025local} for this interaction reads as
	\begin{equation}
		E_i = 
		- \sum_{j \in \langle ij \rangle} \exp{[\sigma \cos{(\theta_i-\psi_{ij})}]} \cos(\theta_i-\theta_j).
	\end{equation}
	The form of the Hamiltonian embedding remains identical to Eq.~\eqref{Seq:Hamiltonian} with the interaction replaced by Eq.~\eqref{Seq:NR_Force_rou}. The Langevin equations of motion derived from the Hamiltonian under the constraint $\theta_i(t) - \varphi_i(t)  = \pi$ are given by
	\begin{eqnarray}
		\label{Seq:Langevin_CB}
		\dot{\theta_i} = -\sum_{j \in \langle ij \rangle} \exp{[\sigma \cos{(\theta_i-\psi_{ij})}]} \sin(\theta_i-\theta_j) +
		\sqrt{2 T}\eta _{i}(t), \notag \\
		\dot{\varphi_i} = \sum_{j \in \langle ij \rangle} \exp{[\sigma \cos{(\theta_i-\psi_{ij})}]} \sin(\varphi_i-\varphi_j) +%
		\sqrt{2 T}\eta _{i}(t).
	\end{eqnarray}
	
	The average steady-state magnetization for the three alternative approaches at system size $L=100$ is shown in Fig.~\ref{fig:comparison}, for $\sigma=1$. For the constrained Glauber dynamics, we use a semi-width $\Delta = 0.2$ and $200\times L^4$ iterations; a time step $\delta t = 0.01$ is used for Langevin dynamics.

	As shown in Fig.~\ref{fig:comparison}, the average steady-state magnetization from selfish-energy Glauber dynamics (green dots) is consistently higher than that obtained from constrained Glauber (blue dots) and Langevin (red squares) dynamics. 
	Importantly, the transition temperature from the ordered to the disordered state in the selfish-energy Glauber dynamics differs significantly from that of the other two dynamics for the chosen parameters.
	However, we observe that in both the ordered and disordered phases, and across the critical region, the constrained Glauber dynamics accurately reproduce the average magnetization obtained from Langevin dynamics.

	\section{Chase-and-run dynamics for nonreciprocal XY spins}
	\label{sec:nonstationary}

	\begin{figure*}[t!]
		\centering\includegraphics[width=0.95\textwidth]{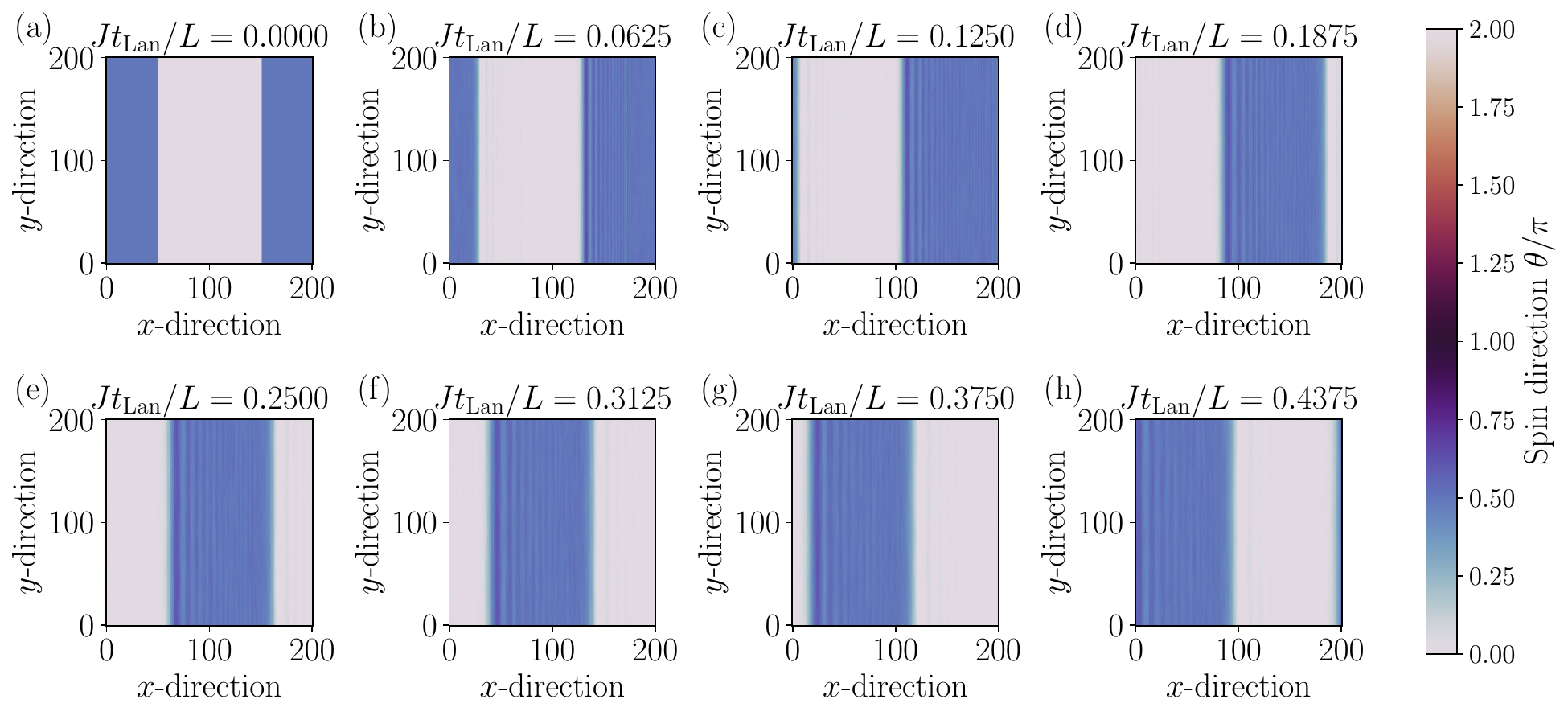}
		\caption{
			Snapshots of the spin configuration of nonreciprocal XY spins on a square lattice exhibiting chase-and-run dynamics, generated by solving the Langevin equation, Eq.~\eqref{seq:Langevin_chiral}.
			Colors encode the spin angle $\theta_{x,y}$.
			Starting from a stripe initial condition, the pattern briefly relaxes into a stable shape and then propagates leftward at approximately a constant speed.
			Parameters are set as $J_\rightarrow = J$, $J_\leftarrow/J = -0.99$, $J_\mathrm{rec}/J = 10$, $T/J=0.08$; the linear dimension is $L=200$. The Langevin time-step is $\delta t J = 0.01$.
		}
		\label{fig:snapshot}
	\end{figure*}

	In Sec.~\ref{sec:Langevin_MC_equivalence}, we provided a general derivation of the equivalence of Langevin and constrained Glauber dynamics based on the Hamiltonian embedding defined in the Main Text, in capturing both nonequilibrium steady states and nonstationary states of nonreciprocal systems. Whereas we presented numerical evidence for this equivalence in the case of a nonequilibrium steady state for vision-cone XY spins in the Main Text, we now turn our attention to a model whose long-time dynamics exhibit persistent oscillations.

	In this section, we consider nonreciprocal XY spins $\theta_{x,y}$ on a square $L\times L$ lattice which feature persistent chase-and-run dynamics.
	The corresponding Langevin equation of motion reads as
	\begin{align}
		\dot{\theta}_{x,y} = & - J_{\rightarrow} \sin(\theta_{x,y} - \theta_{x+1,y}) - J_{\leftarrow} \sin(\theta_{x,y} - \theta_{x-1,y}) 
		- J_\mathrm{rec} [\sin(\theta_{x,y} - \theta_{x,y+1})+ \sin (\theta_{x,y} - \theta_{x,y-1})] + \sqrt{2T} \eta_{x,y}(t).  \label{seq:Langevin_chiral}
	\end{align}
	Explicitly, the model has constant reciprocal interactions in the $y$-direction of strength $J_\mathrm{rec}$, and constant \textit{nonreciprocal} interactions in the $x$-direction of strength $J_\rightarrow$ (with the spin to the right) and $J_\leftarrow$ (with the spin to the left). 
	Whereas the reciprocal interactions along the $y$-direction can be derived from the potential energy $-J_\textrm{rec} \cos (\theta_{x,y} - \theta_{x,y+1})$, there is no potential energy that captures the nonreciprocal interactions along the $x$-direction for $J_{\rightarrow}\neq J_{\leftarrow}$ since the nonreciprocal force is nonconservative.
	The spins are coupled to an external reservoir at temperature $T$ and, as before, we set the damping factor to unity. 
	We consider periodic boundary conditions.
	
	To understand the underlying physics, consider first only two spins at two neighboring positions $1$ and $2$ in the $x$-direction interacting via strengths $J_{\rightarrow}$ and $J_{\leftarrow}$, respectively, and ignore the other spins. When their coupling strengths satisfy the antisymmetric relation $J_{\rightarrow} = -J_{\leftarrow} = J$, where $J>0$, the two spins exhibit chase-and-run phase dynamics: spin $1$ tends to align with $2$ while spin $2$ tends to anti-align with spin $1$.
	As a consequence, they will both rotate at a constant speed, maintaining a constant angle difference. 
	These time-dependent chiral dynamics do not have an equilibrium analogue. 
	In the full 2D nonreciprocal model described above, this chiral state can be stable at finite temperature, as a consequence of a subtle interplay between noise and many-body effects~\cite{fruchart2021non}. In the following, we regard $J$ as the unit of energy.

	We consider an initial state defined by
	\begin{align}
		\theta_{x,y} =\left\{  \begin{array}{cc}
			0, & \text{for}\;  x \in (L/4,3L/4) \\ \pi/2, & \mathrm{else,}
		\end{array}
		\right. 
	\end{align}
	It represents a stripe pattern along the $y$-axis, where spins in the central region point in the positive $x$-direction, while those in the outer region (connected by the periodic boundary condition) point in the positive $y$-direction.
	Snapshots of the initial dynamics until time $J t_\mathrm{Lan} \approx 0.44 L$ are presented in Fig.~\ref{fig:snapshot} [we label the time variable by $t_\text{Lan}$ to explicitly denote that this is the time associated with the solution of the Langevin equation].
	This configuration exhibits chase-and-run dynamics characteristic of the model.
	We find that, during the transient dynamics, the configuration reshapes slightly and the pattern then moves to the left at a constant speed $v$.
	In Sec.~\ref{subsec:scaling_effect_langevin}, we show that the speed remains constant as the linear dimension $L$ is increased, while the period $T$ scales linearly with $L$ since $T=L/v$; therefore, we use the rescaled time $Jt_\mathrm{Lan}/L$ to eliminate the trivial scaling effect in the dynamical snapshots.
	
	To characterize this oscillatory dynamics, we consider an ensemble of $\mathcal{N}$ trajectories, each initialized in the same initial state. We introduce the order parameter~\cite{fruchart2021non}
	\begin{align}
		\langle O(t) \rangle =  \frac{1}{\mathcal{N}}\sum_{\ell = 1}^{\mathcal{N}} \frac{1}{L} \sum_{y = 1}^{L} e^{i \theta^{(\ell)} _{1,y}(t)},
	\end{align}
	where $\ell$ labels the trajectory index and $x=1$ indexes spins in the first column of the square lattice.
	This order parameter characterizes the average orientation of the spins in the first column. 
	The real and imaginary parts of $\langle O(t) \rangle$ exhibit temporal oscillations that reflect the collective rotation of the stripe pattern.
	
	During the evolution, the pattern of spins propagates collectively at a constant velocity, thereby breaking continuous time-translation symmetry. Thus, the system never relaxes to a steady state, and the state exhibits perpetual oscillatory motion. Hence, it represents a perfect testbed for the Glauber Monte-Carlo dynamics defined with the help of our Hamiltonian embedding. 
	
	In the following, we first investigate the scaling effect of this oscillatory dynamics with respect to the system size $L$. We also compare the Langevin dynamics to constrained Glauber dynamics. To avoid confusion in the comparison, we use $t_\mathrm{Lan}$ to represent the time in Langevin dynamics and $t_\mathrm{MC}$ to denote the Monte-Carlo time in the constrained Glauber dynamics. The analytical derivation of the equivalence between the two methods is presented in Sec.~\ref{sec:Langevin_MC_equivalence}.

	\subsection{Langevin dynamics}
	\label{subsec:scaling_effect_langevin}
	
	To enhance the numerical accuracy of the Langevin simulations, we adopt Heun's method~\cite{ruemelin1982numerical}. This method provides second-order accuracy in the time step, i.e., $\mathcal{O}(\delta t^2)$, which significantly reduces numerical errors compared to the Euler method. 
	The integration algorithm reads as
	\begin{align}
		\theta_{x,y}(t + \delta t) & = \theta_{x,y}(t) + \frac{1}{2}[F_{x,y}(\{\theta_{x,y}(t)\}) + F_{x,y}(\{\tilde{\theta}_{x,y}(t+\delta t)\})]\delta t + \sqrt{2T\delta t}\;  \eta_{x,y}(t), \label{seq:Heun_dynamics_1}\\
		\tilde{\theta}_{x,y}(t + \delta t) & = \theta_{x,y}(t) + F_{x,y}(\{\theta_{x,y}(t)\})\delta t + \sqrt{2T\delta t} \; \eta_{x,y}(t), \label{seq:Heun_dynamics_2} 
	\end{align}
	where $F_{x,y}(\{\theta_{x,y}(t)\})$ denotes the force acting on spin $\theta_{x,y}$ at time $t$, as defined in Eq.~\eqref{seq:Langevin_chiral}. 
	
	The update procedure proceeds as follows: at each step $t$, we first calculate the force $F_{x,y}(\{\theta_{x,y}(t)\})$ for all spins. We then propose the provisional state $\tilde{\theta}_{x,y}(t + \delta t)$ using Eq.~\eqref{seq:Heun_dynamics_2} above. Next, we calculate the force $F_i(\{\tilde{\theta}_{x,y}(t+\delta t)\})$ based on this provisional state. Finally, we update the spin configuration $\theta_{x,y}(t + \delta t)$ according to Eq.~\eqref{seq:Heun_dynamics_1}.
	The stochastic force $\eta_{x,y}(t)$ is the same in Eq.~\eqref{seq:Heun_dynamics_1} and Eq.~\eqref{seq:Heun_dynamics_2} in one update step. 
	
	\begin{figure*}[t!]
		\centering\includegraphics[width=0.95\textwidth]{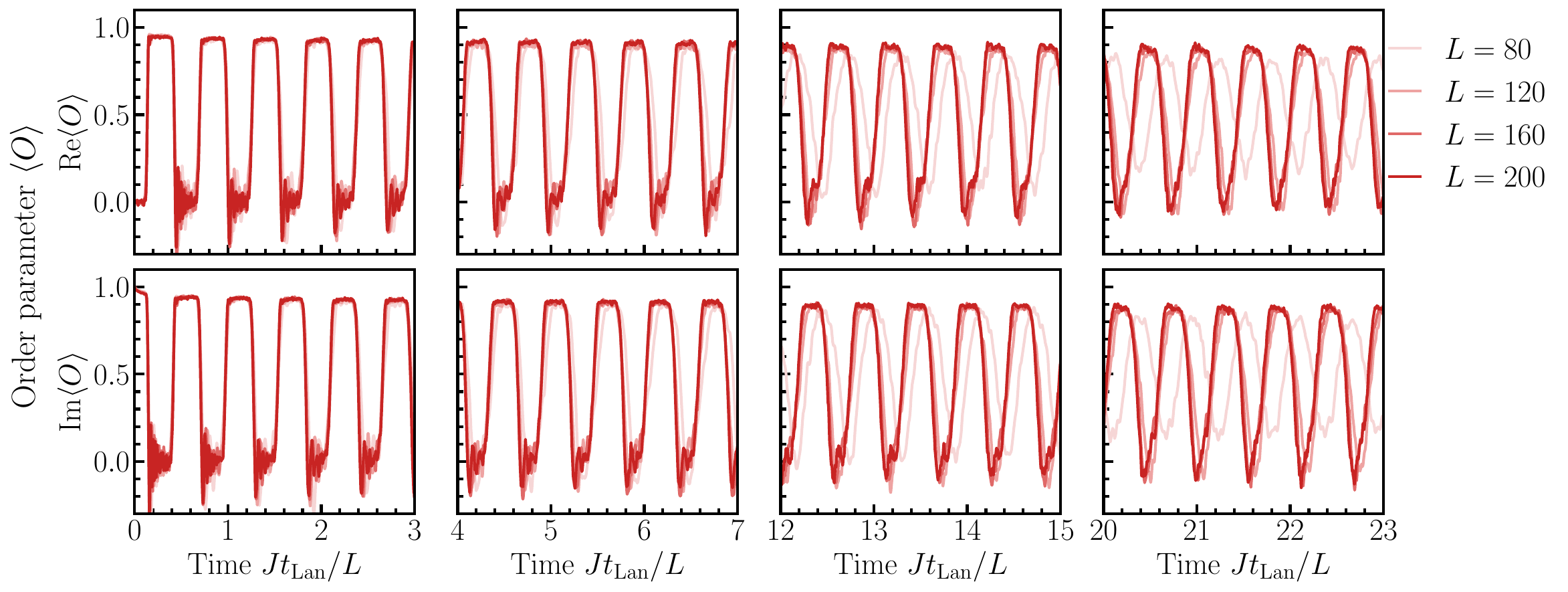}
		\caption{
			System size scaling of the oscillatory pattern obtained from Langevin dynamics. 
			The real and imaginary parts of the order parameter $\langle O(t) \rangle$ for the nonequilibrium chiral state are plotted against time for various system sizes $L$. The color intensity increases as the system size grows from $L=80$ to $L=200$ (see legend). The oscillation period exhibits a linear scaling with the system size $L$: when the time is rescaled by the system size, $Jt_\mathrm{Lan}/L$, the oscillation curves nearly overlap for different system sizes. 
			The order-parameter amplitude decay decreases as the system size increases, indicating stable oscillatory dynamics in the thermodynamic limit. 
			The parameters are $J_{\rightarrow}/J = 1$, $J_{\leftarrow}/J = -0.99$, $J_\mathrm{rec}/J = 10$, $T/J=0.08$; the number of trajectories in the ensemble is $\mathcal{N} = 100$. We use Heun's method with time step $J \delta t_\mathrm{Lan} = 0.01$ [see text].
		}
		\label{fig:scaling_effect_oscillation_Langevin}
	\end{figure*}
	
	Let us now examine the scaling behavior of the Langevin dynamics with respect to system size.
	Figure~\ref{fig:scaling_effect_oscillation_Langevin} shows the real and imaginary parts of $\langle O(t) \rangle$ (red curves). We show the initial dynamics, the transients, and the late-time dynamics, respectively. 
	The $L$-rescaled time axis confirms that the oscillation period scales linearly with the system size $L$, indicating that the propagation velocity of the stripe pattern remains constant. 
	
	To analyze the scaling behavior of the oscillation amplitude more clearly, we plot the results for different system sizes. For the smallest system size $L=80$ (lightest color), the oscillation amplitude decays noticeably over time. However, as the system size increases, the amplitude stabilizes when plotted against the rescaled time $Jt_\mathrm{Lan}/L$, demonstrating that the oscillatory chiral state is stabilized in the thermodynamic limit, underscoring the importance of interactions.
	Note that the oscillation period also stabilizes as the system size increases.
	
	It is important to emphasize that the stability of the oscillations imposes strict requirements on numerical accuracy. When employing the Euler method for integrating the stochastic differential equations with a time step of $J\delta t_\mathrm{Lan} = 0.01$, we observe a decay in the oscillation amplitude, even for system sizes as large as $L=300$. Eventually, this numerical imprecision would lead to a complete loss of the oscillatory dynamics [not shown]. This effect is, however, avoided here by using the second-order Heun's method.

	\begin{figure*}[t!]
		\centering\includegraphics[width=0.95\textwidth]{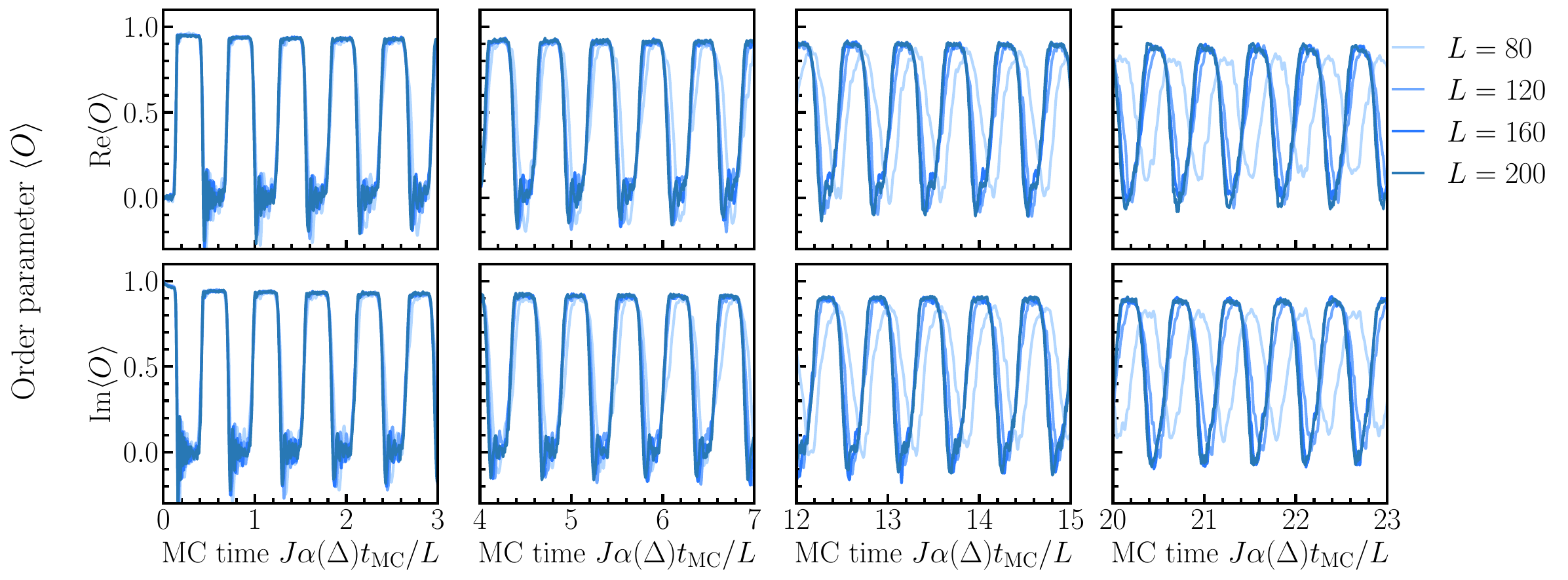}
		\caption{System size scaling of the oscillatory pattern obtained from constrained Glauber dynamics. The real and imaginary parts of the order parameter $\langle O(t) \rangle$ are plotted as functions of the rescaled Monte-Carlo time $J \alpha(\Delta) t_\mathrm{MC}/L$ for various system sizes $L$.
			The scaling behavior of constrained Glauber dynamics is in excellent agreement with the Langevin dynamics, see Fig.~\ref{fig:comparison_oscillation}.
			The time rescaling factor $\alpha(\Delta) \approx 2.37\times 10^{-3}/J$ is determined by matching the first oscillation period of $\langle O(t) \rangle$ from the constrained Glauber dynamics to the same period in the Langevin dynamics for $L=200$. [cf.~Fig.~\ref{fig:scaling_effect_oscillation_Langevin}].
			The parameters are $J_\rightarrow /J= 1$ $J_{\leftarrow}/J = -0.99$, $J_\mathrm{rec}/J = 10$, $T/J=0.08$, $\Delta = 0.05$, and the number of trajectories is $\mathcal{N}=100$.
		}
		\label{fig:scaling_effect_oscillation_MC}
	\end{figure*}

	\subsection{Constrained Glauber dynamics}
	
	The Hamiltonian embedding for the nonreciprocal XY spins from Eq.~\eqref{seq:Langevin_chiral} reads as
	\begin{align}
		H_\text{tot} & = H_{SS} + H_{Sa} - H_{aa}, \notag \\
		H_{SS} & = -   \sum_{ x,y}(J_{\rightarrow} + J_{\leftarrow})[ \cos(\theta_{x,y} - \theta_{x+1,y}) +\cos(\theta_{x,y} - \theta_{x-1,y})  ]  + J_\mathrm{rec}[\cos(\theta_{x,y} - \theta_{x,y+1}) + \cos(\theta_{x,y} - \theta_{x,y-1})] ,  \notag \\
		H_{Sa} & = - \sum_{ x,y} J_{\leftarrow} \cos(\theta_{x,y} - \varphi_{x+1,y}) + J_{\rightarrow} \cos(\theta_{x,y} - \varphi_{x-1,y}), \notag \\
		H_{aa} &= - \sum_{ x,y} J_\mathrm{rec}[\cos(\varphi_{x,y} - \varphi_{x,y+1}) + \cos(\varphi_{x,y} - \varphi_{x,y-1})] .
	\end{align}
	We implement constrained Glauber dynamics in which the auxiliary spins $\varphi_{x,y}$ are eliminated by imposing the constraint $\theta_{x,y} - \varphi_{x,y} = \pi$, see Sec.~\ref{subsec:Eliminating_auxiliary}.  
	Consequently, the Glauber algorithm evolves only the original spins $\theta_{x,y}$ and follows the procedure described in Sec.~\ref{subsec:MC_algo_eliminating}.  
	The corresponding energy difference $\Delta E$ is given in Eq.~(27) in the Main Text.

	For the numerical simulation, we use the same initial state as the Langevin dynamics. We also consider the ensembles with the same number of trajectories. 
	For the uniform distribution we sample the Glauber updates from, we choose the semi-width $\Delta = 0.05$ to reduce the difference between the constrained Glauber dynamics and Langevin dynamics: we have proven in Sec.~\ref{sec:Langevin_MC_equivalence} that the difference between the two vanishes in the limit $\Delta \to 0$.
	
	\subsubsection{Monte-Carlo time vs.~physical time}
	
	As discussed in Sec.~\ref{subsec:non-stationary_MC}, we can define a (dimensionless) Monte-Carlo time as $t_{\mathrm{MC}} = n/L^2$, where $n$ is the iteration step. This MC time is then proportional to the Langevin time (which is the original proper time of the dynamics) through $t_\mathrm{Lan}=\alpha(\Delta) t_\mathrm{MC}$. Leading-order perturbation theory in $\Delta$ predicts $\alpha(\Delta) = \Delta^2/(12T) \approx 2.60\times 10^{-3}/J$, cf.~Eq.~\eqref{seq:time_relation}. 
	
	In practice, we find that a slightly better agreement between Langevin and Monte-Carlo simulations is obtained by matching the oscillation period of the order parameter $\langle O(t) \rangle$ from the constrained Glauber dynamics to that of Langevin dynamics for the linear dimension $L=200$, leading to $\alpha(\Delta ) \approx 2.37\times10^{-3}/J$.
	The main discrepancy between the numerically fitted $\alpha(\Delta)$ and the analytical prediction arises from the limited validity of the small-$\Delta$ expansion, which assumes $\Delta \ll T/J$. Since in our simulations the temperature strength, $T/J=0.08$, is comparable to the semi-width $\Delta=0.05$, truncating the analysis in Sec.~\ref{subsec:non-stationary_MC} at second order becomes inaccurate. In practice, this leads primarily to a mismatch between the oscillation period extracted numerically and that predicted by the theory.
	
	\begin{figure*}[t!]
		\centering\includegraphics[width=0.8\textwidth]{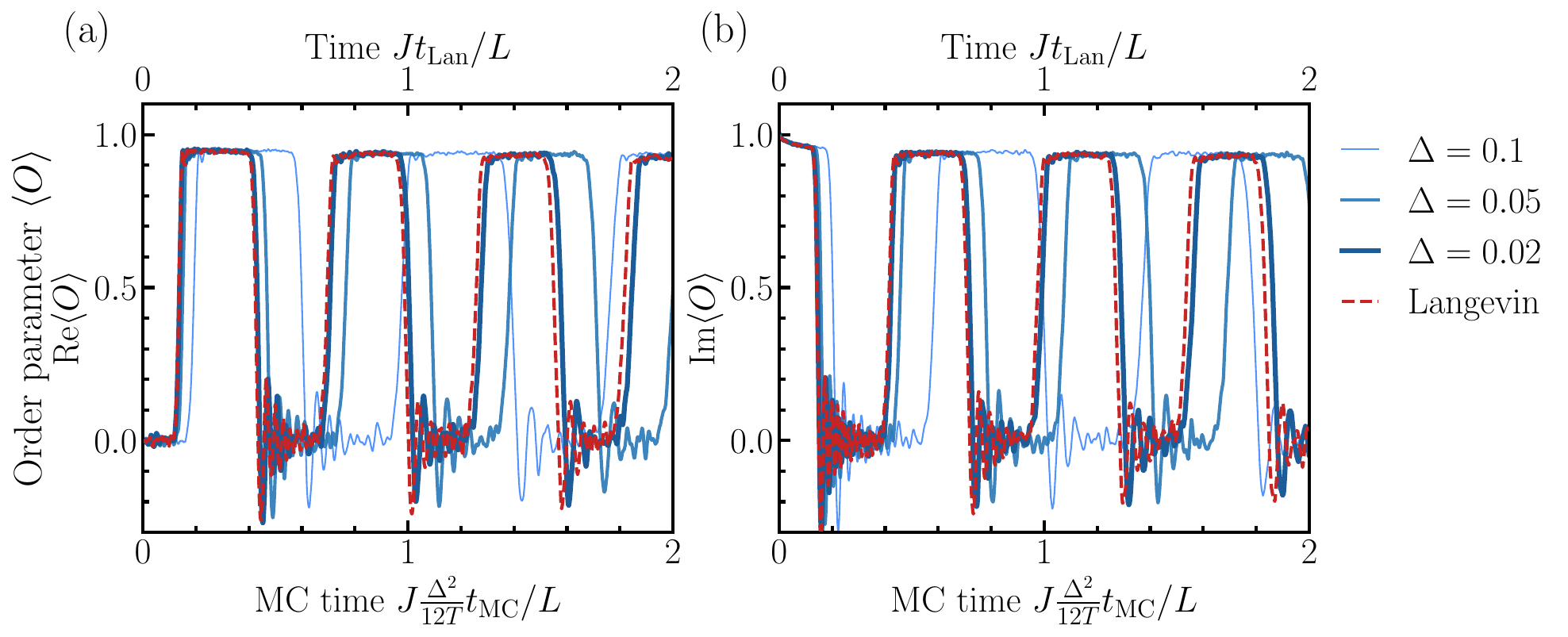}
		\caption{
			Comparison between physical Langevin time and rescaled Monte-Carlo time for different semi-widths $\Delta$ [see legend]. We use the analytical rescaling factor $\alpha (\Delta) = \Delta^2/(12T)$ to match the Langevin (top $x$-axis) and Monte-Carlo (bottom $x$-axis) times [see text]. The difference between the constrained Glauber and the Langevin order-parameter curves decreases in the first three oscillations as the semi-width $\Delta$ decreases. The parameters are $J_\rightarrow = J$, $J_\leftarrow/J = -0.99$, $J_\mathrm{rec}/J = 10$, $T/J=0.08$, and the number of trajectories is $\mathcal{N} = 100$. The linear system dimension is $L=200$. 
		}
		\label{fig:comparison_period}
	\end{figure*}

	Figure~\ref{fig:comparison_period} compares the short-time oscillations obtained from Heun-integrated (physical) Langevin dynamics and constrained Glauber dynamics, using the analytical rescaling factor $\alpha(\Delta) = \Delta^{2}/(12T)$.
	As $\Delta$ decreases, the order parameter curves agree increasingly well over the first few periods, in line with the above analysis.
	Note that, formally taking $\Delta \to 0$ requires an infinite number of Monte-Carlo steps to reach finite physical times, making the simulations costly; developing more efficient Hamiltonian-embedding schemes is an interesting direction for future work.
	
	\subsubsection{Order-parameter dynamics}
	
	The dynamics of the real part and imaginary part of order parameter $\langle O(t) \rangle$ from the constrained Glauber dynamics are shown in Fig.~\ref{fig:scaling_effect_oscillation_MC} using blue curves.  
	We find that the oscillation amplitude stabilizes as the system size increases, consistent with the results of Langevin dynamics. 
	The oscillation period also exhibits linear scaling with the system size, in agreement with Langevin dynamics.
	
	\begin{figure*}[t!]
		\centering\includegraphics[width=0.95\textwidth]{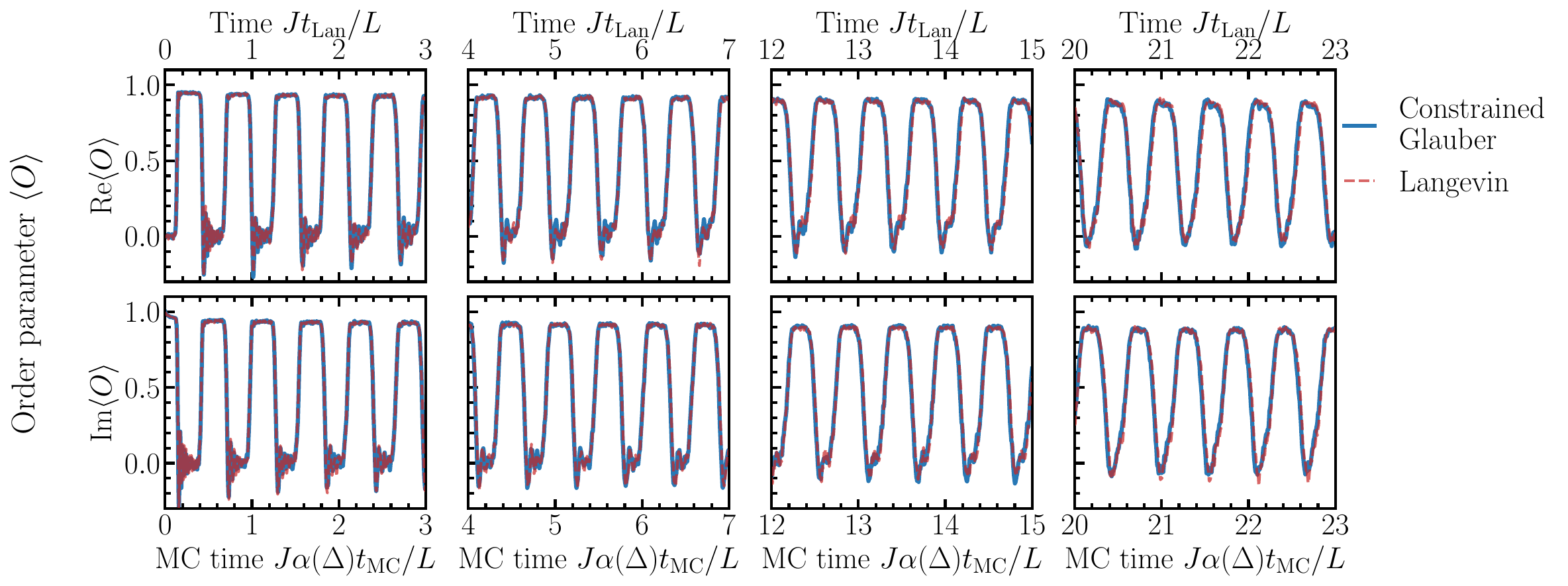}
		\caption{
			Comparison of time evolution obtained from Langevin and constrained Glauber dynamics. The real and imaginary parts of the order parameter $\langle O(t) \rangle$ from constrained Glauber dynamics (blue curves) are plotted as functions of the rescaled Monte-Carlo time $J \alpha(\Delta) t_\mathrm{MC}/L$ [bottom $x$-axis], while those from Langevin dynamics (blue and green curves) are plotted as functions of Langevin time $J t_\mathrm{Lan}/L$ [top $x$-axis]. The order-parameter curves coincide to an excellent precision, in agreement with the theory analysis in Sec.~\ref{subsec:non-stationary_MC}. 
			The parameters are $J_\rightarrow = J$, $J_\leftarrow/J = -0.99$, $J_\mathrm{rec}/J = 10$, $T/J=0.08$, and the number of trajectories is $\mathcal{N} = 100$. The linear system dimension is $L=200$.
			We use $\Delta = 0.05$ for the constrained Glauber dynamics, which is sufficiently small to ensure good agreement with Langevin dynamics. 
		}
		\label{fig:comparison_oscillation}
	\end{figure*}

	We also compare numerically the oscillatory dynamics produced by the Langevin and the constrained Glauber approaches. For a direct comparison, we plot the order parameter $\langle O(t)\rangle$ side by side in Fig.~\ref{fig:comparison_oscillation}. This confirms the validity of the analytical derivation from Sec.~\ref{subsec:non-stationary_MC}.

	Finally, we note that constrained Glauber dynamics can reproduce the correct long-time dynamics at time steps (e.g., $J \delta t_{\mathrm{Lan}} = 0.01$) where Euler-integration of the corresponding Langevin dynamics exhibits inaccurate amplitude damping of the order parameter oscillations. 
	This indicates that numerical artifacts caused by integrating the Langevin dynamics can lead to potential inaccuracies when determining the phase diagram. 
	By contrast, we find that periodic oscillations remain undamped in the constrained Glauber dynamics even at relatively coarse angle updates (e.g., $\Delta = 0.3$) [not shown], highlighting one potential advantage of the Hamiltonian embedding method. 
	Thus, for capturing nonequilibrium properties of some nonreciprocal systems, constrained Glauber dynamics can be more robust and efficient than direct Langevin integration.

	\section{Entropy production in the Hamiltonian embedding}
	
	In this section, we discuss entropy production from the perspective of the Hamiltonian embedding. 
	Let us denote by $w(\sigma\to\sigma')$ the transition rate from state $\sigma$ to state $\sigma'$ and by $p(\sigma)$ the probability of being in state $\sigma$. Then, the total entropy production rate $\dot\Sigma$ is defined as
	\begin{eqnarray}
		\dot\Sigma &=& \dot S_\text{sys} + \dot S_\text{sys-env},\nonumber\\
		\dot S_\text{sys} &=&  \frac{1}{2}\sum_{\sigma,\sigma'}\mathcal{J}(\sigma\to\sigma')\log\frac{p(\sigma)}{p(\sigma')} = \sum_{\sigma,\sigma'} \mathcal{J}(\sigma\to\sigma')\log p(\sigma)
		= - \sum_{\sigma,\sigma'} \dot p(\sigma)\log p(\sigma), \nonumber\\
		\dot S_\text{sys-env} &=& \frac{1}{2}\sum_{\sigma,\sigma'} \mathcal{J}(\sigma\to\sigma')\log\frac{w(\sigma\to\sigma')}{w(\sigma'\to\sigma)}  ,\nonumber\\
	\end{eqnarray}
	with the probability current $\mathcal{J}(\sigma\to\sigma')= p(\sigma)w(\sigma\to\sigma')-p(\sigma')w(\sigma'\to\sigma)$. In a nonequilibrium steady state, $\dot p(\sigma) = -\sum_{\sigma'} \mathcal{J}(\sigma\to\sigma') = 0$, and hence $\dot S_\text{sys}=0$ (even though the configuration space currents $\mathcal{J}(\sigma\to\sigma')$ need not vanish). Thus, we only need to consider the entropy flow between the system and its environment $\dot S_\text{sys-env}$. 
	
	For XY variables, using the Glauber expression for the rates, we have $\log(w(\theta\to\theta')/w(\theta'\to\theta)) =\beta\Delta E(\theta,\theta')$, and hence
	\begin{equation}
		\dot S_\text{sys-env} = \frac{\beta}{2}\int_0^{2\pi}\mathrm d\theta\mathrm d\theta'\;  \mathcal{J}(\theta\to\theta')\; \Delta E(\theta,\theta').
	\end{equation}
	In the nonreciprocal case, $\Delta E$ cannot be derived from the gradient of a Hamiltonian --- a direct manifestation of the violation of detailed balance. Instead, it can be calculated from the embedding Hamiltonian
	\begin{equation}
		\label{eq:Delta_E}
		\Delta E(\theta,\theta') = \frac{1}{2}\big\{H(\theta',\varphi)-H(\theta,\varphi')\big\}\big|_\text{constr.}\; .
	\end{equation}  
	We can then write the entropy flow between the system and its environment as
	\begin{equation}
		\label{seq:ent_flow_nr}
		\dot S_\text{sys-env} = \frac{\beta}{4}\int_0^{2\pi}\mathrm d\theta\mathrm d\theta'\; \mathcal{J}(\theta\to\theta')\; \big\{H(\theta',\varphi)-H(\theta,\varphi')\big\}\big|_\text{constr.},
	\end{equation}
	where, as stated, one first evaluates the expression in the curly bracket and only then applies the constraint to eliminate the $\varphi$ and $\varphi'$ degrees of freedom. 
	
	We now write down the same expression for the system-environment entropy flow of a ``conventional'' Hamiltonian system on an enlarged phase space $(\theta,\varphi)$ (and without a constraint):
	\begin{equation}
		\label{seq:ent_flow_H}
		\dot S_\text{sys-env} = \frac{\beta}{2}\int_0^{2\pi}\mathrm d\theta\mathrm d\theta'\mathrm d\varphi\mathrm d\varphi'\; \mathcal{J}(\theta,\varphi\to\theta',\varphi')\; \big\{H(\theta',\varphi')-H(\theta,\varphi)\big\}\big. .
	\end{equation}
	Compared to Eq.~\eqref{seq:ent_flow_nr}, in Eq.~\eqref{seq:ent_flow_H} there are three essential and one minor differences: 
	\begin{itemize}
		\item[(i)] The energy difference in Eq.~\eqref{seq:ent_flow_H} differs from that in Eq.~\eqref{seq:ent_flow_nr} (cf.~the position of the primed variables). 
		\item[(ii)] Eq.~\eqref{seq:ent_flow_nr} contains in addition the constraint which reduces the number of degrees of freedom (rather than having integrals over all $\theta,\theta',\varphi,\varphi'$ as in Eq.~\eqref{seq:ent_flow_H}). This is an effective change in the phase-space measure; it ensures that the entropy flow is measured by the correct measure on the configuration space of the original $\theta$-degree of freedom. 
		\item[(iii)] The probability current in Eq.~\eqref{seq:ent_flow_H} differs from that in Eq.~\eqref{seq:ent_flow_nr}; in particular, when one considers a ``conventional'' Hamiltonian system without a constraint and coupled to a thermal bath, detailed balance stipulates that $\mathcal{J}(\theta,\varphi\to\theta',\varphi')=0$. This need not be the case for a system with a constraint: $\mathcal{J}(\theta\to\theta')\neq 0$ (and, in fact, it is not, for nonreciprocal systems). 
		\item[(iv)] There is an additional factor of $1/2$ in Eq.~\eqref{seq:ent_flow_nr}; it ensures that the reciprocal limit is reproduced correctly (see SM, Sec.~\ref{subsec:Eliminating_auxiliary} where we derive the expression in Eq.~\eqref{eq:Delta_E}). 
	\end{itemize}
	Ultimately, the entropy production is caused by the non-vanishing probability currents $\mathcal{J}(\theta\to\theta')$ in the original degrees of freedom on the constraint manifold. Therefore, it is the application of the constraint that breaks time-reversal symmetry for the original degrees of freedom starting from the Hamiltonian dynamics of the extended, unconstrained system.

\end{widetext}

\fi 

\bibliography{bibfile}
	
\end{document}